\documentclass[aps,reprint,superscriptaddress,nofootinbib,longbibliography]{revtex4-2}
\usepackage[utf8]{inputenc}
\usepackage[T1]{fontenc}
\usepackage{amsmath,amssymb,bm}
\usepackage{graphicx}
\usepackage[colorlinks=true,citecolor=blue,linkcolor=blue,urlcolor=blue]{hyperref}

\newcommand{\rev}[1]{#1}

\begin{document}

\title{\emph{Quo vadis}, stochastic thermodynamics?}

\author{Jan Korbel}
\thanks{All authors contributed equally to this perspective. Authors AK, SAML, GM, RG-M, OMM, and ER are arranged in alphabetical order.}
\affiliation{Complexity Science Hub, Metternichgasse 8, 1030, Vienna, Austria}
\email{korbel@csh.ac.at}

\author{Artemy Kolchinsky}
\thanks{All authors contributed equally to this perspective. Authors AK, SAML, GM, RG-M, OMM, and ER are arranged in alphabetical order.}
\affiliation{Pompeu Fabra University, 08003 Barcelona, Spain}

\author{Sarah A.M. Loos}
\thanks{All authors contributed equally to this perspective. Authors AK, SAML, GM, RG-M, OMM, and ER are arranged in alphabetical order.}
\affiliation{Max Planck Institute for Dynamics and Self-Organization, G\"ottingen 37077, Germany}

\author{Gonzalo Manzano}
\thanks{All authors contributed equally to this perspective. Authors AK, SAML, GM, RG-M, OMM, and ER are arranged in alphabetical order.}
\affiliation{Institute for Cross-Disciplinary Physics and Complex Systems IFISC (UIB-CSIC), Campus Universitat Illes Balears, E-07122 Palma de Mallorca, Spain}

\author{Rosalba Garcia-Millan}
\thanks{All authors contributed equally to this perspective. Authors AK, SAML, GM, RG-M, OMM, and ER are arranged in alphabetical order.}
\affiliation{Department of Mathematics, King's College London, Strand, London WC2R 2LS, United Kingdom}

\author{Olga Movilla Miangolarra}
\thanks{All authors contributed equally to this perspective. Authors AK, SAML, GM, RG-M, OMM, and ER are arranged in alphabetical order.}
\affiliation{Departamento de F\'isica, Universidad de La Laguna, 38203 La Laguna, Spain}

\author{\'Edgar Rold\'an}
\thanks{All authors contributed equally to this perspective. Authors AK, SAML, GM, RG-M, OMM, and ER are arranged in alphabetical order.}
\affiliation{ICTP --- The Abdus Salam International Centre for Theoretical Physics, Strada Costiera 11, Trieste 34151, Italy}

\begin{abstract}
Stochastic thermodynamics is a framework for describing non-equilibrium processes at the level of fluctuating trajectories, where the state of a system evolves as a stochastic time series, allowing thermodynamic quantities such as work, heat, and entropy production to be defined along individual realizations rather than at the ensemble level only. Over the past three decades, the field has yielded fundamental results, including fluctuation theorems and several universal bounds, such as thermodynamic uncertainty relations, speed limit theorems, and many others. Many of them have been tested on a range of experimental platforms. This Perspective reviews recent developments in stochastic thermodynamics that extend its scope beyond its traditional domains, including systems with memory and hidden degrees of freedom, microscopic approaches to interacting and active matter, and geometric formulations based on optimal transport. Next, the Perspective surveys the challenges that arise when applying these ideas to macroscopic and complex systems, where the link between statistical irreversibility and thermodynamic dissipation becomes less direct. Finally,   emerging applications in non-physical contexts are highlighted, including computation, biological systems, and social dynamics. Transcending the traditional boundaries of physics, these developments catalyze an unorthodox framework to tackle the thermodynamics of complex systems.
\end{abstract}


\maketitle

Stochastic thermodynamics is a rapidly growing subfield of statistical physics. Its main goal is to extend concepts from non-equilibrium thermodynamics --- such as heat, work, and entropy production --- to microscopic fluctuating systems.
A key achievement has been to describe the thermodynamics of small-scale fluctuations associated with individual stochastic trajectories of a few degrees of freedom that describe the time evolution of the system under study. Since its foundations about 30 years ago, it has led to a series of fundamental results, including, among others, fluctuation theorems, theory of stochastic response, trajectory-level formulations of entropy production, and universal trade-off relations between accuracy, dissipation, and speed. These advances have been extensively tested in experiments on micro- and nanoscale systems, establishing stochastic thermodynamics as a central framework for non-equilibrium physics.

At the same time, the scope of stochastic thermodynamics has been rapidly expanding. Many complex systems of current interest --- ranging from active matter and biological systems to computation and social dynamics --- can be described within the framework of stochastic processes, yet operate under conditions that differ substantially from those originally assumed in foundational studies of stochastic thermodynamics. These include coarse-graining, hidden degrees of freedom, memory effects, and unidirectional transitions. Extending stochastic thermodynamics to such settings raises fundamental questions about the applicability and interpretation of the field's central concepts.

Recent work has approached these challenges along several complementary directions. These include the development of stochastic thermodynamics under partial information and non-Markovian dynamics, the study of irreversibility in interacting and active many-body systems, and geometric approaches that characterize nonequilibrium processes through optimal transport and thermodynamic constraints. At the same time, attempts to scale up the framework to larger systems have revealed limitations of the traditional connection between statistical irreversibility and thermodynamic dissipation, motivating more general formulations that extend beyond energy-based interpretations.
\rev{Across these directions, \textit{coarse-graining} provides a recurring cross-cutting challenge: being able to access only a partial set of the relevant non-equilibrium degrees of freedom may generate memory, obscure dissipation, and impact the thermodynamic interpretation of the variables retrieved upon coarse-graining.}

In this Perspective, we review some recent developments around a set of research directions that are reshaping stochastic thermodynamics. We first briefly summarize the conceptual foundations of the field, and then discuss their extension beyond their original domains. These developments raise broader questions about the nature of irreversibility, the roles of information and interactions, and the extent to which thermodynamic concepts can be applied to complex systems.

\section{Foundations  and  evolution --- an emerging, evolving field}

\begin{figure*}[t]
    \centering
    \includegraphics[width=0.75\linewidth]{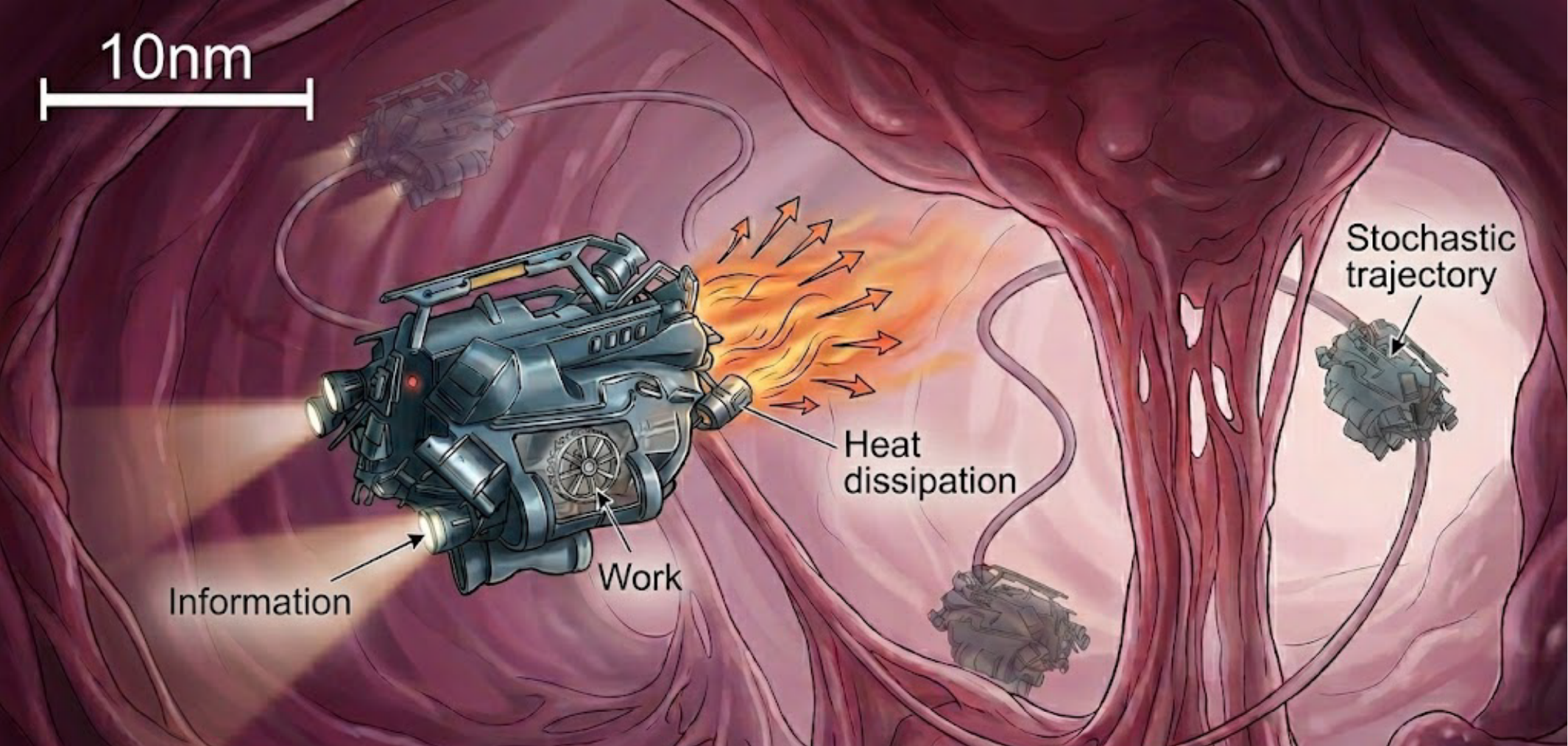}
    \caption{Illustration of the `dream' of early developments of stochastic thermodynamics.  A tiny nano-sized motor navigates through a biological tissue, tracing a stochastic trajectory. Such motion, powered by fuel and controlled through feedback upon information retrieval, is converted partially into useful work while the rest is dissipated to the surrounding media as heat. The first decades of stochastic thermodynamics research aimed to understand the statistical constraints for various dynamical and thermodynamic properties of such small-scale non-equilibrium processes. AI-generated image using Gemini, inspired by a presentation by Ignacio~A.~Mart\'inez and by the movie  {\em Innerspace} (1987).}
    \label{fig:placeholder}
\end{figure*}

\rev{Stochastic thermodynamics extends the laws of thermodynamics to individual stochastic trajectories of systems evolving arbitrarily far from equilibrium. 
Its emergence was driven by the need to reconcile macroscopic irreversibility with the time-reversal symmetry of microscopic dynamics in regimes where thermal fluctuations are significant~\cite{lebowitz1993boltzmann}, a problem already anticipated in the network thermodynamics of Schnakenberg~\cite{OSTER1971,Schnakenberg1976}. This challenge became particularly acute with the realization of  high-precision experiments on non-equilibrium mesoscopic systems, accompanied by molecular-dynamics and Monte Carlo simulations, which revealed nonnegligible, quantifiable fluctuations of work, heat, and entropy production.}
Since its foundation, the field has sought so-called `universal' principles that constrain the statistical properties of thermodynamic quantities and apply to a wide range of systems driven away from equilibrium. See Figure~\ref{fig:placeholder} for an illustration of the original aims of early developments in stochastic thermodynamics.

\rev{The first major breakthrough took place in the 1990s with the discovery and formulation of fluctuation theorems for nonequilibrium steady states~\cite{EvansCohenMorriss1993,EvansSearles1994,GallavottiCohen1995}}, 
which revealed a fundamental symmetry that governs the fluctuations of entropy production. Another critical development was the formulation of work, heat, and entropy production\footnote{We use the term ``entropy production'' and ``dissipation'' interchangeably in the rest of the Perspective.} at the level of individual fluctuating trajectories~\cite{Jarzynski1997,sekimoto1997complementarity,Sekimoto1998,Kurchan1998}. Closely related results, such as the Jarzynski equality \cite{Jarzynski1997} and Crooks fluctuation theorem \cite{Crooks1999}, further established exact connections between non-equilibrium fluctuations and equilibrium free-energy differences. Unification was provided by a general definition of stochastic entropy production associated with stochastic trajectories and analysis of its fluctuation properties~\cite{maes1999,maes2000,qian2001nonequilibrium}.

In the early 2000s, these insights were consolidated into a coherent theoretical framework, marking the emergence of stochastic thermodynamics as a trajectory-level theory. Building on the stochastic energetics approach pioneered by Sekimoto~\cite{Sekimoto1998,Sekimoto2010}, this period established consistent definitions of heat, work, and entropy production along individual stochastic trajectories, culminating in fluctuation theorems for entropy production~\cite{Seifert2005}. The framework was extended from transformations between equilibrium states to non-equilibrium steady states, drawing on earlier ideas in steady-state thermodynamics~\cite{OonoPaniconi1998} and yielding a rich development of novel theoretical results in response theory. They include extensions of fluctuation response relations to processes arbitrarily far from equilibrium~\cite{prost2009generalized,baiesi2009fluctuations}, relations between entropy production and the degree of violation of Kubo's fluctuation-dissipation theorem~\cite{harada2005equality}, and definitions of \emph{housekeeping} (adiabatic) and \emph{excess} (non-adiabatic) contributions of heat (entropy production)~\cite{HatanoSasa2001,SpeckSeifert2005,SpeckSeifert2006,Esposito10}. In parallel with these theoretical developments, major advances in experimental techniques enabled direct measurements of work, heat, and entropy production fluctuations and experimental verification of fluctuation relations~\cite{HummerSzabo2001,Liphardt2002,Collin2005}, establishing stochastic thermodynamics as an experimentally accessible framework. Ref.~\cite{Ciliberto2017} contains an extensive review on experiments in stochastic thermodynamics.

In the late 2000s and early 2010s, notable progress was achieved in information thermodynamics. The tools of stochastic thermodynamics shed new light on long-standing problems in the thermodynamics of information, such as reconciling Maxwell's demon and Landauer's principle with the laws of thermodynamics.
The correct interpretation of the thermodynamics of these milestone problems remained elusive until the development of fluctuation theorems and second laws of thermodynamics in the presence of feedback control,  pioneered by Sagawa and Ueda~\cite{sagawa2008second,SagawaUeda2010}, and followed up by other major endeavors of the community, both theoretical~\cite{parrondo2015thermodynamics,barato2014unifying,rosinberg2016continuous,HorowitzEsposito2014} and experimental~\cite{berut2012experimental,jun2014high,roldan2014universal,alemany2012experimental,Koski15,toyabe2010experimental}.

In parallel to thermodynamics of information, key studies in the first decade of the 21st century disentangled the structure of entropy production and its relation to information. A central contribution was the unifying formulation of non-equilibrium entropy production, applicable to Markovian dynamics described by either   Master equations and Markov-jump processes, or by  Fokker--Planck equations and Langevin dynamics~\cite{EspositoVanDenBroeck2010a,EspositoVanDenBroeck2010b}. 
Furthermore, a series of works~\cite{kawai2007dissipation,gomez2008footprints,Parrondo2009} showed that the arrow of time can be quantified using fluctuation relations, thereby linking dissipation to the degree of time-reversal asymmetry. 
\rev{Other work derived and experimentally tested fluctuation theorems for currents and response coefficients in nonequilibrium steady states~\cite{AndrieuxGaspard2007JSP,AndrieuxGaspard2007JSTAT,AndrieuxEtAl2007PRL}.}
These works show that entropy production can be estimated directly from data without detailed knowledge of the underlying mechanism by computing the Kullback–Leibler divergence between forward and time-reversed trajectories~\cite{roldan2010estimating}. These developments were also summarized in early tutorial reviews of stochastic thermodynamics \cite{Seifert2012,EspositoVanDenBroeck2015}.

A further conceptual shift occurred in the mid-2010s with the emergence of universal bounds and trade-off relations constraining non-equilibrium processes.  While fluctuation theorems and generalized second laws provide a statistical rationale for various thermodynamic fluctuating quantities, it remained unclear whether fundamental trade-off limits constrain them. One of the fundamental questions was whether there is a minimal cost (e.g., rate of heat dissipation) needed to sustain a prescribed accuracy of a mesoscopic (e.g., particle, energy) current. A central result of this period was the development of so-called \emph{thermodynamic uncertainty relations}, which establish trade-offs between the signal-to-noise ratio of time-integrated currents and the associated cost to sustain them (dissipation or total entropy production)~\cite{BaratoSeifert2015,Gingrich2016}. This perspective identified entropy production as a resource that controls fluctuations in non-equilibrium processes, leading to a broader class of uncertainty relations for currents, activities, and observables. Closely related developments extended these ideas to finite-time thermodynamics and dynamical constraints. These developments include \emph{thermodynamic speed limits}, bounds on the minimal time required for state transformations in terms of entropy production and dynamical activity \cite{Shiraishi2018,DechantSasa2018,Proesmans2020}, and finite-time trade-offs on power, efficiency, and dissipation in small engines and driven systems \cite{Pietzonka2016,Pietzonka2018,falasco2020dissipation,neri2022universal}. In parallel, the identification of martingale structures underlying entropy production~\cite{Neri2017,Manzano21} provided a swift route to derive fluctuation relations of thermodynamic quantities at stopping (e.g., first-passage) times as well as their extreme-value statistics, see the treatise~\cite{roldan2023martingales} for a review.

These developments laid the foundations of stochastic thermodynamics as a systematic framework for describing fluctuations in nonequilibrium systems.
Since then, it has been successfully applied
to a wide range of experimental platforms at the micro- and nanoscale~\cite{Ciliberto2017}:  electric circuits~\cite{Ciliberto13}, single electron transistors~\cite{koski2013distribution,Koski15,josefsson2018quantum}, mechanical resonators~\cite{Serra-Garcia16,yang2020phonon}, levitated nanoparticles~\cite{debiossac2020thermodynamics,Militaru21}, and colloidal particles in optical traps~\cite{gieseler2021optical} 
that enabled precise tests of key theoretical predictions~\cite{Trepagnier04,toyabe2010experimental,blickle2012realization,berut2012experimental,RoldanETAL:2015,Proesmans16,Proesmans2020,ibanez2024heating}. Applications have also included biophysics and chemistry, including single-molecule experiments on DNA and RNA~\cite{Liphardt2002,Collin2005,gupta2011experimental}, molecular motors~\cite{astumian2010thermodynamics,Hayashi10,toyabe2011thermodynamic,ariga2020experimental,leighton2025flow}, and chemical reaction networks~\cite{Gaspard04,Schmiedl07,Rao16}. 

\rev{
In addition to biochemical applications, stochastic thermodynamics is now being applied to the problem of reducing the energy consumption of computers, a key challenge in the current era of high-performance computing and artificial intelligence. 
By extending thermodynamics to engineered information-processing technologies (e.g., CMOS and electronic circuits), thermodynamics of computing is gaining considerable momentum as a field on its own, going far beyond a mere extension of information thermodynamics~\cite{FreitasDelvenneEsposito2021Circuits,FreitasDelvenneEsposito2020RLC,FreitasEtAl2026CMOS,MelansonEtAl2025ThermodynamicComputing,Lipka24,manzano2024thermodynamics,Wolpert24}.}

More recently, stochastic thermodynamics has been extended beyond some of its original idealizations. This includes developments addressing strong system--bath coupling through the \emph{Hamiltonian of mean force}~\cite{Seifert2016,Jarzynski2017} and related approaches \cite{Strasberg2017}, which show that thermodynamic quantities such as work, heat, and entropy production can remain well-defined even when standard reservoir assumptions break down. At the same time, increasing attention has been devoted to inference, the role of constraints, and partial information. A central challenge has been to quantify entropy production and irreversibility from limited or coarse-grained data~\cite{roldan2014irreversibility}, leading to inference-based approaches that extract thermodynamic bounds and dissipation estimates from time series \cite{SkinnerDunkel2021,seifert2019stochastic} and short experiments \cite{Manikandan2020}. 

Finally, several recent textbooks summarize the progress of stochastic thermodynamics, including key recent monographs \cite{PigolottiPeliti2021,roldan2023martingales,Shiraishi2023,Seifert2025}, marking its consolidation as a mature, self-contained yet still evolving field. 
The following sections explore several directions in which stochastic thermodynamics is being extended beyond the original scenarios considered in the field's early development. 
We do not cover developments of stochastic thermodynamics in quantum systems here, but we refer the interested reader to the recent perspective~\cite{Campbell2026} and textbooks~\cite {Deffner2019,Strasberg2022}.

\section{Dealing with the hidden ---  thermodynamics of non-Markovian dynamics}

A key recent development concerns the thermodynamics of partially observed systems, where some relevant non-equilibrium degrees of freedom are hidden; this situation naturally arises, for example, in single-molecule biophysics, where experiments routinely track the fluctuating motion of molecular machines while other internal variables remain often inaccessible. For instance, optical tweezers can resolve the position of a molecular motor with nanometer precision, while its chemical driving (e.g., ATP consumption) remains hidden. In such cases, the observed dynamics acquires memory, reflecting coupling to hidden degrees of freedom~\cite{Parrondo2009,roldan2010estimating,hartich2014stochastic,arias2021microscopically}. 

To tackle thermodynamics with hidden variables, a key theoretical task is to extend stochastic thermodynamics beyond the Markov regime. In experimentally relevant situations, memory arises, e.g., from strong system–environment coupling, time-delayed feedback, or coarse-graining over hidden internal degrees of freedom. In such cases, the dynamics of the observed variable becomes non-local in time and history-dependent, substantially complicating analytical descriptions and rendering standard results—such as state-based expressions for entropy production and free-energy dissipation—no longer directly applicable.

Building on earlier formulations of coarse-grained entropy production, recent work has sharpened trajectory-level and inference-based approaches. In particular, bounds based on variational or information-theoretic estimators relate observable irreversibility to hidden entropy production using time-series data from partial observations~\cite{roldan2010estimating,lacasa2012time,hartich2014stochastic,seifert2019stochastic,SkinnerDunkel2021,harunari2022learn,van2022thermodynamic,ferri2025conditional,nogare2025identifying,aguilera2026inferring}.

Another route embeds the observed non-Markovian dynamics into an enlarged Markovian description, often by introducing auxiliary degrees of freedom~\cite{ehrich2021tightest,kawaguchi2013fluctuation,dechant2021improving,degunther2024fluctuating}. This restores a consistent thermodynamic structure and clarifies that apparent violations of Markovian bounds reflect unaccounted contributions to entropy production and free energy stored in hidden variables and correlations. Despite their success, these approaches can suffer from non-uniqueness and ambiguities in thermodynamic interpretation. This embedding perspective builds on general results from coarse-graining and strong coupling \cite{esposito2012stochastic,Strasberg2017}, and has been applied to active or colored-noise systems~\cite{FodorETAL:2016,shankar2018hidden,mandal2017entropy,speck2016stochastic,Dabelow19,garcia2025optimal}, delayed-feedback controlled systems \cite{loos2021medium}, and optimal control problems for systems with memory~\cite{loos2024universal,alvarado2026optimal}.

A complementary line of theoretical work formulates thermodynamic relations directly for non-Markovian dynamics. This approach has been extensively developed for time-delayed systems~\cite{munakata2014entropy,rosinberg2015stochastic}, leading to generalized second-law-like inequalities in which memory enters through additional entropy-pumping contributions, thereby tightening thermodynamic bounds compared to Markovian descriptions. While this route avoids introducing auxiliary variables with (ambiguous) thermodynamic interpretations, it is, however, technically challenging, as it requires direct treatment of time-nonlocal dynamics and their associated path integrals, where time-reversal symmetry and causality become subtle. More recently, stochastic thermodynamic frameworks have been developed for general non-Markov jump processes, including consistent definitions of entropy production and time-reversal symmetry \cite{kanazawa2025stochastic}.
Notably, semi-Markov processes provide a transparent intermediate framework that incorporates memory through non-exponential waiting times while preserving a jump-process~structure. These approaches offer complementary perspectives: embedding restores thermodynamic consistency by enlarging the state space, while direct formulations and inference-based methods quantify irreversibility more directly, retaining the hidden character of the inaccessible degrees of freedom. 

Regarding experiments, colloidal particles in e.g., viscoelastic fluids or under time-delayed feedback control have emerged as experimentally accessible platforms for directly probing non-Markovian thermodynamics~\cite{debiossac2020thermodynamics,loos2024universal}.
\rev{In these systems, 
it was recently shown via feedback protocols that work can be extracted from (viscoelastic) hidden degrees of freedom, experimentally demonstrating the idea that environmental memory can be harvested as a thermodynamic resource through nonequilibrium driving~\cite{muruga2026extracting}.}
\rev{In addition, experiments tracking the thermal fluctuations of mechanical cantilevers (e.g., AFM tips) reported high-quality underdamped oscillations with a distinctive non-Markovian structure, which are amenable to memory-assisted thermodynamic protocols and information engines that exploit first-passage-time statistics as a resource~\cite{ArchambaultEtAl2024EPL,ArchambaultEtAl2025PRL}.}

Overall, these theoretical and experimental developments are reshaping stochastic thermodynamics into a framework that systematically incorporates memory, hidden variables, and time-delayed feedback control, thereby extending its applicability to realistic non-equilibrium scenarios and the constraints faced by complex systems.

\section{Active and interacting matter --- a microscopic thermodynamic perspective}

Active matter is a class of systems out of equilibrium, characterized by continuous input energy consumption and its efficient conversion to execute output tasks~\cite{FilyMarchetti:2012,FodorETAL:2016,thipmaungprom2026thermodynamic}. Typical examples involve agents that transform fuel into directed motion, such as motile bacteria and flocking birds. Quantifying irreversibility and energy dissipation in active-matter systems has attracted significant attention over the last few decades \cite{Cates:2012,roldan2021quantifying,gladrow2016broken,Pietzonka19,ghosal2026identification,NardiniETAL:2017,FodorETAL:2016,FodorETAL:2022,di2024brownian}, as a step towards understanding the thermodynamic properties of living matter.

A hallmark of active dynamics is the emergence of local currents that break time-reversal symmetry, thereby driving global circulation and fluxes fueled by the constant energy supply at the microscopic scale. Self-propulsion provides a clear example of such microscopic currents, which, in conjunction with anisotropy, can generate emergent global behavior. For instance, self-propelled particles subject to ratcheted landscapes exhibit particle rectification \cite{DiLeonardoETAL:2010,DiLeonardoETAL:2024,ReinETAL:2023,GrodzinskiJackCates:2025,ZhenPruessner:2022,RobertsZhen:2023}. However, emergent currents are not limited to physical transport in space. Currents also arise in phase space, for example, due to non-reciprocal interactions in chase-and-run dynamics, where a chaser tends to run behind a prey \cite{loos2020irreversibility,ZhangGarcia-Millan:2023}, while the reverse process is rarely observed. These examples highlight that irreversibility in active matter is inherently a many-body and dynamical phenomenon that often eludes standard coarse-grained descriptions.
Characterizing such microscopic and emergent macroscopic currents requires theoretical frameworks that retain the particle-level description \cite{BotheETAL:2023}. In particular, one must account for the multiplicative noise inherent to interacting particle systems, which is often lost in coarse-grained mesoscopic approaches~\cite{NardiniETAL:2017}. Microscopic theories such as the Doi--Peliti path integral formalism and Dean's equation provide exact representations of stochastic particle dynamics governed by Langevin or Fokker--Planck equations, and offer a systematic route to compute observables through perturbative expansions in non-linear couplings \cite{Doi:1976,Peliti:1985,Dean:1996}.

Within this framework, a central goal has been to derive closed-form expressions for thermodynamic quantities, in particular the entropy production rate. Building on its expression for Markov jump processes \cite{seifert2025universal,Gaspard:2004,Esposito10} and its extension to exactly solvable stochastic systems \cite{CocconiGarcia-MillanETAL:2020}, recent work has begun to address general interacting active particle systems. Microscopic approaches enable the calculation of entropy production for particles with persistent motion, correlated noise, $n$-point interactions, and confining potentials \cite{PruessnerGarcia-Millan:2025}. 

A key insight from analytical treatments based on Doi–Peliti field theory is that entropy production is dominated by short-time contributions, so only the lowest-order terms in the perturbative expansion are relevant. Moreover, the entropy production associated with $n$-point interactions involves the $(2n-1)$-point equal-time correlation function, independently of the total number of particles in the system. In particular, pairwise interactions require three-point correlation functions, whereas for short-ranged interactions the description effectively reduces to two-point correlations \cite{PruessnerGarcia-Millan:2025,BrossolletBiroli:2026}. These results illustrate how microscopic theories provide access to thermodynamic quantities that remain hidden at the coarse-grained level.
\rev{Another promising direction is the construction of thermodynamically consistent hydrodynamic descriptions of active matter via exact coarse-graining or related approaches. Exact coarse-graining approaches in active lattice models can preserve entropy production from the particle-based model to hydrodynamic field theories, which, in turn, can be formulated as non-ideal reaction–diffusion systems with an interpretable dissipation rate~\cite{agranov2024thermodynamically,aslyamov2023nonideal}.}

Despite these advances, several conceptual and technical open challenges remain. A central open problem concerns the systematic treatment of hidden internal non-equilibrium variables in active systems~\cite{roldan2021quantifying,harunari2022learn,hartich2021emergent,martinez2019inferring,CocconiKnightRoberts:2023}. For instance, the direction of propulsion in run-and-tumble particles \cite{TailleurCates:2008} and active Brownian particles \cite{SolonCatesTailleur:2015} and the instantaneous speed of active Ornstein-Uhlenbeck particles \cite{MartinETAL:2021} are both governed by evolving internal degrees of freedom that are often not directly observable. When such internal states are hidden, the observed trajectories of free, unbiased self-propelled particles can appear time-reversible, provided the dynamics is $\mathcal{PT}$-symmetric \cite{KnightKavehPruessner:2025}. However, time-reversal symmetry is typically broken once interactions or environmental couplings are introduced, even if the internal states remain unobserved \cite{Dabelow19,DabelowBoEichhorn:2021}. A systematic framework for quantifying irreversibility in such partially observed active systems remains an active area of research in the field. 

Microscopic theories offer promising avenues, for instance, by analyzing first-passage time statistics and splitting probabilities, which can be combined with thermodynamic bounds to estimate dissipation~\cite{Neri:2022,RoldanETAL:2015}. Another open challenge concerns systems with non-conserved particle number, such as branching or birth--death processes used to model cell division and population dynamics \cite{Harris:1963,CameronTjhung:2025}. In their simplest form, such processes are intrinsically irreversible due to absorbing states corresponding to extinction, which explicitly breaks the local detailed balance condition.

These challenges highlight that extending stochastic thermodynamics to active and interacting systems requires not only new computational tools, but also a rethinking of how irreversibility is defined and quantified in systems with hidden variables, interactions, and changing particle numbers.

\section{Geometry and optimal transport in stochastic thermodynamics}

Thermodynamics has long been connected to geometry, dating back to the pioneering work of Gibbs~\cite{gibbs1873graphical,gibbs1873method}. Geometric perspectives have since deepened our understanding of thermodynamic structure and enabled the derivation of new results. In equilibrium, thermodynamics can be formulated in terms of a contact geometry on the manifold of thermodynamic variables \cite{mrugala1978geometrical}, while Riemannian metrics~\cite{weinhold1975metric,ruppeiner1979thermodynamics} capture stability and fluctuations, thereby encoding aspects of the second law. This geometric picture extends to the linear-response regime, where \emph{thermodynamic length} quantifies dissipation during slow transformations, and geodesics prescribe optimal protocols~\cite{SalamonBerry1983,Crooks2007,SivakCrooks2012}. 

\newcommand\ft{1}
\newcommand\rhoinit{\rho_0}
\newcommand\rhofinal{\rho_f}
Recent developments have pushed this perspective far from equilibrium, where the state of a system is described by a probability distribution and stochastic thermodynamics becomes the natural framework~\cite{aurell2011optimal}. In this regime, optimal transport theory has emerged as a powerful tool to characterize the geometry of nonequilibrium transformations. Originally formulated by Monge in 1781 as a problem of minimizing transport cost~\cite{villani2009optimal}, optimal transport provides a natural Riemannian geometry on the space of probability distributions. Its associated geodesic distance, known as the Wasserstein-2 distance, quantifies the cost of redistribution.

A central result of stochastic thermodynamics is that this geometric structure has a direct thermodynamic interpretation. For overdamped Langevin systems with constant diffusivity, the entropy produced along a stochastic trajectory coincides with the optimal transport cost of redistribution~\cite{aurell2011optimal}. As a consequence, the Wasserstein-2 distance provides a tight lower bound on the entropy production required to transition from one thermodynamic state to another in finite time.
This result refines the second law by adding a finite-time correction that depends on the distance traversed, demonstrating that dissipation induces a geometry on the space of probability distributions.

Moreover, the solution to the optimal transport problem yields optimal control protocols that minimize dissipation. These control protocols turn out to be of conservative form, corresponding to dynamics that can be implemented through time-dependent potentials. Further, the overdamped ensemble dynamics can be expressed as gradient flows of a free energy functional with respect to the Wasserstein-2 metric~\cite{JKO1998}, thereby connecting stochastic thermodynamics to broader formalisms such as GENERIC~\cite{grmela1997dynamics,espanol2023statistical} and Steepest Entropy Ascent~\cite{beretta2014steepest,reina2015entropy}. This unifies thermodynamic dissipation, optimal control, and geometric structure within a single framework.

The optimal transport perspective has found numerous applications, including decomposing entropy production for systems that do not satisfy detailed balance~\cite{dechant2022geometric2}, deriving thermodynamic uncertainty relations~\cite{otsubo2020estimating}, and designing heat and information engines~\cite{movilla2021energy}. 
While optimal transport results are well understood for overdamped systems, extending them beyond this setting remains an active area of research. Generalizations have been proposed for underdamped dynamics~\cite{dechant2019thermodynamic,sabbagh2024wasserstein}, discrete stochastic systems~\cite{van2021geometrical,dechant2022minimum,van2023thermodynamic}, chemical reaction networks~\cite{yoshimura2023housekeeping,kolchinsky2026generalized}, reaction-diffusion systems~\cite{nagayama2025geometric}, and quantum systems~\cite{van2021geometrical,van2023thermodynamic,yoshimura2025force}. 
These developments demonstrate the broad applicability of geometric ideas, but also reveal that the extension of the Wasserstein framework is not unique.

Two main approaches have emerged. The first generalizes the Wasserstein-2 structure by introducing Onsager-type operators that relate thermodynamic forces to fluxes~\cite{van2021geometrical,yoshimura2023housekeeping,nagayama2025geometric,yoshimura2025force}. This preserves the connection to Riemannian geometry and gradient flows, and recovers known results in appropriate limits. However, it typically assumes fixed linear-response structures that may not be realistic far from equilibrium. 

The second approach abandons the strict Riemannian structure and instead defines generalized Wasserstein-1 geometries based on constraints on dynamical activity or related quantities~\cite{dechant2022minimum,van2023thermodynamic,nagayama2025infinite,kolchinsky2026generalized}. These formulations are often more flexible and physically interpretable, and can lead to tighter thermodynamic bounds, but at the cost of losing the geometric structure that underpins gradient-flow formulations and thermodynamic metrics.

These advances raise several open questions. In particular, it remains unclear how to construct geometric frameworks that faithfully capture realistic optimal control problems beyond idealized settings. Connections to large deviation theory and classical problems such as the Schr\"{o}dinger bridge are expected to play an important role in this direction. Extensions to the aforementioned non-Markovian dynamics and mixed conservative--dissipative systems also remain largely unexplored. 

More broadly, these efforts point toward a deeper question: to what extent is dissipation fundamentally geometric? While a unified geometric description of stochastic thermodynamics has not yet emerged, the search for such a framework continues to provide powerful insights and tools for understanding fundamental bounds on non-equilibrium transformations.

\section{Scaling up stochastic thermodynamics --- challenges and opportunities}

\newcommand\deltaS{\Delta S_\text{env}}
Stochastic thermodynamics was originally developed to describe the energetics of small systems where thermal fluctuations are comparable in scale to internal energy changes. 
A central assumption underlying this framework is the principle of \emph{local detailed balance} (LDB)~\cite{maes2021local}. 
For a Markov jump process with transition rates $k({x\to y})$, LDB takes the form $
 \deltaS/k_B= \ln[k({x\to y})/k({y \to x})]$.
Here, $\deltaS$ is the change of the environment's entropy associated with the system's transition  $x\to y$ ($k_B = 1.38 \times 10^{-23} \text{J/K}$ is Boltzmann's constant). 
This relation connects thermodynamic dissipation to statistical irreversibility, and it underpins many central results of stochastic thermodynamics, including fluctuation theorems~\cite{Seifert2012}, thermodynamic tradeoffs~\cite{mehta2016landauer,cao2025stochastic}, and inference schemes based on trajectory data~\cite{seifert2019stochastic,seifert2025universal}. 

Despite its success, the applicability of LDB becomes limited when scaling to larger or more complex systems. A first limitation arises from the microscopic energy scale involved. 
Once a transition dissipates significantly more than $\deltaS/k_B\sim 10$, the reverse transition is effectively never observed. As a consequence, the ratio $\ln[{ k(x\to y)}/k(y \to x)]$ becomes empirically inaccessible, and transition statistics decouple from thermodynamic dissipation~\rev{\cite{CilibertoJoubaudPetrosyan2010}}. This is particularly challenging since it corresponds to very small energy scales:   the \emph{in vivo} hydrolysis of a single ATP molecule dissipates $\deltaS/k_B \sim 10{-}20$~\cite[p.~184]{milo2015cell}, while the combustion of a single glucose molecule dissipates $\deltaS/k_B \sim 1.2\times 10^3$~\cite[p.~188]{milo2015cell}.

A second limitation concerns coarse-graining. Many systems of interest --- from intracellular structures~\cite{foster2023dissipation} to living cells~\cite{england2013statistical} and brains~\cite{lynn2021broken,nartallo2026nonequilibrium} --- are described at a macroscopic level where internal degrees of freedom remain out of equilibrium. In such cases, LDB holds only as an inequality, $\deltaS/k_B \ge \ln[{ P(x\to y)}/{P(y \to x)}]$~\cite{england2013statistical}, and the bound is typically very loose. Empirically, macroscopic irreversibility can underestimate thermodynamic dissipation by several orders of magnitude, ranging from $10^6$ in the cytoskeleton~\cite{foster2023dissipation} to more than $10^{20}$ in brain data~\cite{lynn2021broken}. Thus, at large scales, statistical irreversibility and thermodynamic dissipation become effectively decoupled.

Several approaches have been proposed to address these limitations. 
One direction extends standard methods by appropriately discounting the contribution from irreversible transitions~\cite{murashita2014nonequilibrium,pal2021thermodynamic,busiello2020entropy,kolchinsky2021dependence}. A second direction exploits decompositions of entropy production into non-negative components, such as excess and housekeeping contributions, to derive coarse-grained bounds~\cite{ge2016mesoscopic,Datta22,freitas2022emergent,manzano2024thermodynamics,kolchinsky2026generalized}. These methods are helpful in situations where either a set of a few transitions or a specific component of the entropy production is responsible for almost all the dissipation, allowing one to focus on the stochastic thermodynamics of the complementary part. 
\rev{A third direction seeks to extend fluctuation theorems to macroscopic driven systems by introducing ``effective'' temperatures or other phenomenological parameters. This approach has led to interesting experimental results in granular media~\cite{FeitosaMenon2004,JoubaudLohseVanDerMeer2012,LagoinCrausteThibiergeNaert2022}, although its interpretation and thermodynamic validity remain debated~\cite{PuglisiEtAl2005,CilibertoJoubaudPetrosyan2010}.} 
Finally, a more radical approach shifts focus away from thermodynamic dissipation and toward tradeoffs formulated in terms of alternative, non-energetic resources, such as dynamical activity~\cite{maes2020frenesy,di2019kinetic} or topological constraints~\cite{van2023topological}. 
 
These limitations suggest that the direct link between statistical irreversibility and thermodynamic dissipation becomes increasingly fragile at larger scales, raising the question of whether stochastic thermodynamic concepts can be meaningfully extended beyond their traditional domain of microscopic physics. Indeed, there is growing interest in applying ideas from stochastic thermodynamics to systems where fluctuations are not thermal, ranging from the aforementioned active matter~\cite{krishnamurthy2016micrometre,Dabelow19,Pietzonka19,Bebon25} to computational systems~\cite{Wolpert24}, brain dynamics~\cite{lynn2021broken,nartallo2026nonequilibrium}, evolutionary dynamics~\cite{Andrae2010,Qian2014,Rao22}, and social systems~\cite{Tome23,Hawthorne23,OLIVEIRA2024114694,irisarri2025stochasticthermodynamicssocialimitation}. 

In such contexts, statistical irreversibility is often treated as an observable in its own right, independent of thermodynamic dissipation. This perspective has proven useful in data-driven settings, for instance, for identifying brain states~\cite{lynn2021broken,nartallo2026nonequilibrium,aguilera2026inferring}, diagnosing diseases~\cite{Zanin2020,maldonado2024irreversibility}, and characterizing intracellular regulation~\cite{holehouse2026quantifying}. In this role, irreversibility is conceptually similar to information-theoretic measures of complexity~\cite{tononi1994measure}. 

Another promising direction is to identify generalized forms of LDB tailored to specific classes of systems, where entropy changes 
are related to relevant currents in the system. This approach allows the formal structure of stochastic thermodynamics to be extended beyond its traditional domain while preserving its predictive power. An example is provided in Ref.~\cite{irisarri2025stochasticthermodynamicssocialimitation}, where such a framework is developed for social imitation dynamics, with applications to cultural dissemination~\cite{Castello2006,Nowak22}, opinion formation~\cite{siedlecki2016interplay}, and financial behavior~\cite{Kirman93}. 

Nonetheless, it is important to emphasize that, in general, statistical irreversibility at macroscopic scales does not have a direct quantitative relationship to physical energy dissipation. Despite this,  this statistical irreversibility is typically referred to as ``entropy production'' and ``dissipation'' in the literature, due to its formal and conceptual similarity to definitions motivated by thermodynamics. 

\section{Summary and outlook}
The developments discussed in this Perspective highlight a field undergoing a rapid and ongoing transformation. Extensions to systems with memory and hidden degrees of freedom, the emergence of microscopic approaches to interacting and active matter, and the geometric reformulation of dissipation in terms of optimal transport all point toward a broadening of stochastic thermodynamics beyond its original domain. At the same time, attempts to scale the framework to larger and more complex systems reveal fundamental limitations of its traditional foundations. 

\rev{This Perspective reveals that several apparently independent directions share common conceptual challenges, including the thermodynamics of coarse-grained and partially observed systems and inference, and more broadly, the question of which thermodynamic structures survive the passage from microscopic dynamics to effective macroscopic descriptions. Some of these developments challenge the traditional identification between statistical irreversibility and thermodynamic dissipation, raising the broader question of which aspects of stochastic thermodynamics remain universal beyond its original setting.}

\rev{Beyond these} fundamental aspects, 
\rev{the Perspective} highlighted the ample room for opportunities to apply theoretical ideas from stochastic thermodynamics --- first developed to optimize small machines (see Figure~\ref{fig:placeholder}) --- to tackle complex systems spanning over a wide range of scales and impacting various disciplines.  Some promising directions are: optimal control of soft and biological matter towards maximal thermodynamic performance~\cite{alvarado2026optimal};   characterization of non-trivial response properties of active materials~\cite{liu2026cyclic,khodabandehlou2026bringing}; elucidation of the relevance of time-symmetric (frenetic) aspects of nonequilibria~\cite{maes2020frenesy}; and design of optimal protocols that enable efficient energy extraction from macroscopic  fluctuations~\cite{Falasco25}. 

These directions suggest that stochastic thermodynamics is evolving from a theory of small fluctuating systems into a more general framework for quantifying irreversibility, constraints, and tradeoffs in complex systems. Whether a unified formulation encompassing these diverse extensions exists remains an open question.

\begin{acknowledgments}
We would like to thank the organizers of the Joint European Thermodynamics Conference 2025 (JETC 2025), especially Velimir Ilic, for their help with preparing the manuscript. We also thank the JETC 2025 speakers Ken Sekimoto, Juan Parrondo, and Christian Maes for helpful conversations.
\end{acknowledgments}
\section*{Author Contributions}
All authors contributed equally to this perspective.

\section*{Use of Large Language Models, AI and Machine Learning Tools}
We used Gemini to create the illustration in Fig.~\ref{fig:placeholder}. This figure does not reflect any results of the paper; it is solely a high-level illustration of some original aspirations of stochastic thermodynamics.

\section*{Funding}
J.K. was supported by the Austrian Science Fund (FWF) under Grants No. 10.55776/P34994
and EFP5 ReMASS, funding from the Austrian Federal Ministry for Climate Action, Environment, Energy,
Mobility, Innovation, and Technology under GZ 2023-0.841.266, through the Postdoc Program for Complexity Science and Data Competence.
A.K. was supported by the European Union’s Horizon 2020 research and innovation programme under the Marie Skłodowska-Curie Grant Agreement No.~101068029 as well as the John Templeton Foundation (grant 62828). 
S.L. was supported by UK Research and Innovation (UKRI) under the UK government’s Horizon Europe funding Guarantee (grant number EP/X031926/1).
G.M. acknowledges support from CoQuSy project (No. PID2022-140506NB-C21 and C22) funded by MCIU/AEI/10.13039/501100011033, and Mar\'ia de Maeztu Program for units of Excellence in R\&D, grant CEX2021-001164-M.
O.M.M. was supported by the European Union's Horizon 2020 research and innovation
programme under the Marie Skłodowska-Curie Grant Agreement No.~101151140.

\section*{Conflict of Interest}
Authors declare no conflict of interest.

\section*{Data Availability}
Not applicable.

\bibliographystyle{apsrev4-2}
\bibliography{bibliography_arxiv}

\begin{thebibliography}{244}%
\makeatletter
\providecommand \@ifxundefined [1]{%
 \@ifx{#1\undefined}
}%
\providecommand \@ifnum [1]{%
 \ifnum #1\expandafter \@firstoftwo
 \else \expandafter \@secondoftwo
 \fi
}%
\providecommand \@ifx [1]{%
 \ifx #1\expandafter \@firstoftwo
 \else \expandafter \@secondoftwo
 \fi
}%
\providecommand \natexlab [1]{#1}%
\providecommand \enquote  [1]{``#1''}%
\providecommand \bibnamefont  [1]{#1}%
\providecommand \bibfnamefont [1]{#1}%
\providecommand \citenamefont [1]{#1}%
\providecommand \href@noop [0]{\@secondoftwo}%
\providecommand \href [0]{\begingroup \@sanitize@url \@href}%
\providecommand \@href[1]{\@@startlink{#1}\@@href}%
\providecommand \@@href[1]{\endgroup#1\@@endlink}%
\providecommand \@sanitize@url [0]{\catcode `\\12\catcode `\$12\catcode `\&12\catcode `\#12\catcode `\^12\catcode `\_12\catcode `\%12\relax}%
\providecommand \@@startlink[1]{}%
\providecommand \@@endlink[0]{}%
\providecommand \url  [0]{\begingroup\@sanitize@url \@url }%
\providecommand \@url [1]{\endgroup\@href {#1}{\urlprefix }}%
\providecommand \urlprefix  [0]{URL }%
\providecommand \Eprint [0]{\href }%
\providecommand \doibase [0]{https://doi.org/}%
\providecommand \selectlanguage [0]{\@gobble}%
\providecommand \bibinfo  [0]{\@secondoftwo}%
\providecommand \bibfield  [0]{\@secondoftwo}%
\providecommand \translation [1]{[#1]}%
\providecommand \BibitemOpen [0]{}%
\providecommand \bibitemStop [0]{}%
\providecommand \bibitemNoStop [0]{.\EOS\space}%
\providecommand \EOS [0]{\spacefactor3000\relax}%
\providecommand \BibitemShut  [1]{\csname bibitem#1\endcsname}%
\let\auto@bib@innerbib\@empty
\bibitem [{\citenamefont {Lebowitz}(1993)}]{lebowitz1993boltzmann}%
  \BibitemOpen
  \bibfield  {author} {\bibinfo {author} {\bibfnamefont {J.~L.}\ \bibnamefont {Lebowitz}},\ }\href@noop {} {\bibfield  {journal} {\bibinfo  {journal} {Physics today}\ }\textbf {\bibinfo {volume} {46}},\ \bibinfo {pages} {32} (\bibinfo {year} {1993})}\BibitemShut {NoStop}%
\bibitem [{\citenamefont {OSTER}\ \emph {et~al.}(1971)\citenamefont {OSTER}, \citenamefont {PERELSON},\ and\ \citenamefont {KATCHALSKY}}]{OSTER1971}%
  \BibitemOpen
  \bibfield  {author} {\bibinfo {author} {\bibfnamefont {G.}~\bibnamefont {OSTER}}, \bibinfo {author} {\bibfnamefont {A.}~\bibnamefont {PERELSON}},\ and\ \bibinfo {author} {\bibfnamefont {A.}~\bibnamefont {KATCHALSKY}},\ }\href {https://doi.org/10.1038/234393a0} {\bibfield  {journal} {\bibinfo  {journal} {Nature}\ }\textbf {\bibinfo {volume} {234}},\ \bibinfo {pages} {393} (\bibinfo {year} {1971})}\BibitemShut {NoStop}%
\bibitem [{\citenamefont {Schnakenberg}(1976)}]{Schnakenberg1976}%
  \BibitemOpen
  \bibfield  {author} {\bibinfo {author} {\bibfnamefont {J.}~\bibnamefont {Schnakenberg}},\ }\href {https://doi.org/10.1103/RevModPhys.48.571} {\bibfield  {journal} {\bibinfo  {journal} {Rev. Mod. Phys.}\ }\textbf {\bibinfo {volume} {48}},\ \bibinfo {pages} {571} (\bibinfo {year} {1976})}\BibitemShut {NoStop}%
\bibitem [{\citenamefont {Evans}\ \emph {et~al.}(1993)\citenamefont {Evans}, \citenamefont {Cohen},\ and\ \citenamefont {Morriss}}]{EvansCohenMorriss1993}%
  \BibitemOpen
  \bibfield  {author} {\bibinfo {author} {\bibfnamefont {D.~J.}\ \bibnamefont {Evans}}, \bibinfo {author} {\bibfnamefont {E.~G.~D.}\ \bibnamefont {Cohen}},\ and\ \bibinfo {author} {\bibfnamefont {G.~P.}\ \bibnamefont {Morriss}},\ }\href {https://doi.org/10.1103/PhysRevLett.71.2401} {\bibfield  {journal} {\bibinfo  {journal} {Physical Review Letters}\ }\textbf {\bibinfo {volume} {71}},\ \bibinfo {pages} {2401} (\bibinfo {year} {1993})}\BibitemShut {NoStop}%
\bibitem [{\citenamefont {Evans}\ and\ \citenamefont {Searles}(1994)}]{EvansSearles1994}%
  \BibitemOpen
  \bibfield  {author} {\bibinfo {author} {\bibfnamefont {D.~J.}\ \bibnamefont {Evans}}\ and\ \bibinfo {author} {\bibfnamefont {D.~J.}\ \bibnamefont {Searles}},\ }\href {https://doi.org/10.1103/PhysRevE.50.1645} {\bibfield  {journal} {\bibinfo  {journal} {Physical Review E}\ }\textbf {\bibinfo {volume} {50}},\ \bibinfo {pages} {1645} (\bibinfo {year} {1994})}\BibitemShut {NoStop}%
\bibitem [{\citenamefont {Gallavotti}\ and\ \citenamefont {Cohen}(1995)}]{GallavottiCohen1995}%
  \BibitemOpen
  \bibfield  {author} {\bibinfo {author} {\bibfnamefont {G.}~\bibnamefont {Gallavotti}}\ and\ \bibinfo {author} {\bibfnamefont {E.~G.~D.}\ \bibnamefont {Cohen}},\ }\href {https://doi.org/10.1103/PhysRevLett.74.2694} {\bibfield  {journal} {\bibinfo  {journal} {Physical Review Letters}\ }\textbf {\bibinfo {volume} {74}},\ \bibinfo {pages} {2694} (\bibinfo {year} {1995})}\BibitemShut {NoStop}%
\bibitem [{\citenamefont {Jarzynski}(1997)}]{Jarzynski1997}%
  \BibitemOpen
  \bibfield  {author} {\bibinfo {author} {\bibfnamefont {C.}~\bibnamefont {Jarzynski}},\ }\href {https://doi.org/10.1103/PhysRevLett.78.2690} {\bibfield  {journal} {\bibinfo  {journal} {Physical Review Letters}\ }\textbf {\bibinfo {volume} {78}},\ \bibinfo {pages} {2690} (\bibinfo {year} {1997})}\BibitemShut {NoStop}%
\bibitem [{\citenamefont {Sekimoto}\ and\ \citenamefont {Sasa}(1997)}]{sekimoto1997complementarity}%
  \BibitemOpen
  \bibfield  {author} {\bibinfo {author} {\bibfnamefont {K.}~\bibnamefont {Sekimoto}}\ and\ \bibinfo {author} {\bibfnamefont {S.-i.}\ \bibnamefont {Sasa}},\ }\href@noop {} {\bibfield  {journal} {\bibinfo  {journal} {Journal of the Physical Society of Japan}\ }\textbf {\bibinfo {volume} {66}},\ \bibinfo {pages} {3326} (\bibinfo {year} {1997})}\BibitemShut {NoStop}%
\bibitem [{\citenamefont {Sekimoto}(1998)}]{Sekimoto1998}%
  \BibitemOpen
  \bibfield  {author} {\bibinfo {author} {\bibfnamefont {K.}~\bibnamefont {Sekimoto}},\ }\href {https://doi.org/10.1143/PTPS.130.17} {\bibfield  {journal} {\bibinfo  {journal} {Progress of Theoretical Physics Supplement}\ }\textbf {\bibinfo {volume} {130}},\ \bibinfo {pages} {17} (\bibinfo {year} {1998})}\BibitemShut {NoStop}%
\bibitem [{\citenamefont {Kurchan}(1998)}]{Kurchan1998}%
  \BibitemOpen
  \bibfield  {author} {\bibinfo {author} {\bibfnamefont {J.}~\bibnamefont {Kurchan}},\ }\href {https://doi.org/10.1088/0305-4470/31/16/003} {\bibfield  {journal} {\bibinfo  {journal} {Journal of Physics A: Mathematical and General}\ }\textbf {\bibinfo {volume} {31}},\ \bibinfo {pages} {3719} (\bibinfo {year} {1998})}\BibitemShut {NoStop}%
\bibitem [{\citenamefont {Crooks}(1999)}]{Crooks1999}%
  \BibitemOpen
  \bibfield  {author} {\bibinfo {author} {\bibfnamefont {G.~E.}\ \bibnamefont {Crooks}},\ }\href {https://doi.org/10.1103/PhysRevE.60.2721} {\bibfield  {journal} {\bibinfo  {journal} {Physical Review E}\ }\textbf {\bibinfo {volume} {60}},\ \bibinfo {pages} {2721} (\bibinfo {year} {1999})}\BibitemShut {NoStop}%
\bibitem [{\citenamefont {Maes}(1999)}]{maes1999}%
  \BibitemOpen
  \bibfield  {author} {\bibinfo {author} {\bibfnamefont {C.}~\bibnamefont {Maes}},\ }\href {https://doi.org/10.1023/A:1004541830999} {\bibfield  {journal} {\bibinfo  {journal} {Journal of Statistical Physics}\ }\textbf {\bibinfo {volume} {95}},\ \bibinfo {pages} {367} (\bibinfo {year} {1999})}\BibitemShut {NoStop}%
\bibitem [{\citenamefont {Maes}\ \emph {et~al.}(2000)\citenamefont {Maes}, \citenamefont {Redig},\ and\ \citenamefont {Moffaert}}]{maes2000}%
  \BibitemOpen
  \bibfield  {author} {\bibinfo {author} {\bibfnamefont {C.}~\bibnamefont {Maes}}, \bibinfo {author} {\bibfnamefont {F.}~\bibnamefont {Redig}},\ and\ \bibinfo {author} {\bibfnamefont {A.~V.}\ \bibnamefont {Moffaert}},\ }\href {https://doi.org/10.1063/1.533195} {\bibfield  {journal} {\bibinfo  {journal} {Journal of Mathematical Physics}\ }\textbf {\bibinfo {volume} {41}},\ \bibinfo {pages} {1528} (\bibinfo {year} {2000})}\BibitemShut {NoStop}%
\bibitem [{\citenamefont {Qian}(2001)}]{qian2001nonequilibrium}%
  \BibitemOpen
  \bibfield  {author} {\bibinfo {author} {\bibfnamefont {H.}~\bibnamefont {Qian}},\ }\href@noop {} {\bibfield  {journal} {\bibinfo  {journal} {Physical Review E}\ }\textbf {\bibinfo {volume} {64}},\ \bibinfo {pages} {022101} (\bibinfo {year} {2001})}\BibitemShut {NoStop}%
\bibitem [{\citenamefont {Sekimoto}(2010)}]{Sekimoto2010}%
  \BibitemOpen
  \bibfield  {author} {\bibinfo {author} {\bibfnamefont {K.}~\bibnamefont {Sekimoto}},\ }\href {https://doi.org/10.1007/978-3-642-05411-2} {\emph {\bibinfo {title} {Stochastic Energetics}}}\ (\bibinfo  {publisher} {Springer},\ \bibinfo {year} {2010})\BibitemShut {NoStop}%
\bibitem [{\citenamefont {Seifert}(2005)}]{Seifert2005}%
  \BibitemOpen
  \bibfield  {author} {\bibinfo {author} {\bibfnamefont {U.}~\bibnamefont {Seifert}},\ }\href {https://doi.org/10.1103/PhysRevLett.95.040602} {\bibfield  {journal} {\bibinfo  {journal} {Physical Review Letters}\ }\textbf {\bibinfo {volume} {95}},\ \bibinfo {pages} {040602} (\bibinfo {year} {2005})}\BibitemShut {NoStop}%
\bibitem [{\citenamefont {Oono}\ and\ \citenamefont {Paniconi}(1998)}]{OonoPaniconi1998}%
  \BibitemOpen
  \bibfield  {author} {\bibinfo {author} {\bibfnamefont {Y.}~\bibnamefont {Oono}}\ and\ \bibinfo {author} {\bibfnamefont {M.}~\bibnamefont {Paniconi}},\ }\href {https://doi.org/10.1143/PTPS.130.29} {\bibfield  {journal} {\bibinfo  {journal} {Progress of Theoretical Physics Supplement}\ }\textbf {\bibinfo {volume} {130}},\ \bibinfo {pages} {29} (\bibinfo {year} {1998})}\BibitemShut {NoStop}%
\bibitem [{\citenamefont {Prost}\ \emph {et~al.}(2009)\citenamefont {Prost}, \citenamefont {Joanny},\ and\ \citenamefont {Parrondo}}]{prost2009generalized}%
  \BibitemOpen
  \bibfield  {author} {\bibinfo {author} {\bibfnamefont {J.}~\bibnamefont {Prost}}, \bibinfo {author} {\bibfnamefont {J.-F.}\ \bibnamefont {Joanny}},\ and\ \bibinfo {author} {\bibfnamefont {J.~M.}\ \bibnamefont {Parrondo}},\ }\href@noop {} {\bibfield  {journal} {\bibinfo  {journal} {Physical review letters}\ }\textbf {\bibinfo {volume} {103}},\ \bibinfo {pages} {090601} (\bibinfo {year} {2009})}\BibitemShut {NoStop}%
\bibitem [{\citenamefont {Baiesi}\ \emph {et~al.}(2009)\citenamefont {Baiesi}, \citenamefont {Maes},\ and\ \citenamefont {Wynants}}]{baiesi2009fluctuations}%
  \BibitemOpen
  \bibfield  {author} {\bibinfo {author} {\bibfnamefont {M.}~\bibnamefont {Baiesi}}, \bibinfo {author} {\bibfnamefont {C.}~\bibnamefont {Maes}},\ and\ \bibinfo {author} {\bibfnamefont {B.}~\bibnamefont {Wynants}},\ }\href@noop {} {\bibfield  {journal} {\bibinfo  {journal} {Physical review letters}\ }\textbf {\bibinfo {volume} {103}},\ \bibinfo {pages} {010602} (\bibinfo {year} {2009})}\BibitemShut {NoStop}%
\bibitem [{\citenamefont {Harada}\ and\ \citenamefont {Sasa}(2005)}]{harada2005equality}%
  \BibitemOpen
  \bibfield  {author} {\bibinfo {author} {\bibfnamefont {T.}~\bibnamefont {Harada}}\ and\ \bibinfo {author} {\bibfnamefont {S.-i.}\ \bibnamefont {Sasa}},\ }\href@noop {} {\bibfield  {journal} {\bibinfo  {journal} {Physical review letters}\ }\textbf {\bibinfo {volume} {95}},\ \bibinfo {pages} {130602} (\bibinfo {year} {2005})}\BibitemShut {NoStop}%
\bibitem [{\citenamefont {Hatano}\ and\ \citenamefont {Sasa}(2001)}]{HatanoSasa2001}%
  \BibitemOpen
  \bibfield  {author} {\bibinfo {author} {\bibfnamefont {T.}~\bibnamefont {Hatano}}\ and\ \bibinfo {author} {\bibfnamefont {S.-i.}\ \bibnamefont {Sasa}},\ }\href {https://doi.org/10.1103/PhysRevLett.86.3463} {\bibfield  {journal} {\bibinfo  {journal} {Physical Review Letters}\ }\textbf {\bibinfo {volume} {86}},\ \bibinfo {pages} {3463} (\bibinfo {year} {2001})}\BibitemShut {NoStop}%
\bibitem [{\citenamefont {Speck}\ and\ \citenamefont {Seifert}(2005)}]{SpeckSeifert2005}%
  \BibitemOpen
  \bibfield  {author} {\bibinfo {author} {\bibfnamefont {T.}~\bibnamefont {Speck}}\ and\ \bibinfo {author} {\bibfnamefont {U.}~\bibnamefont {Seifert}},\ }\href {https://doi.org/10.1088/0305-4470/38/34/L03} {\bibfield  {journal} {\bibinfo  {journal} {Journal of Physics A: Mathematical and General}\ }\textbf {\bibinfo {volume} {38}},\ \bibinfo {pages} {L581} (\bibinfo {year} {2005})}\BibitemShut {NoStop}%
\bibitem [{\citenamefont {Speck}\ and\ \citenamefont {Seifert}(2006)}]{SpeckSeifert2006}%
  \BibitemOpen
  \bibfield  {author} {\bibinfo {author} {\bibfnamefont {T.}~\bibnamefont {Speck}}\ and\ \bibinfo {author} {\bibfnamefont {U.}~\bibnamefont {Seifert}},\ }\href {https://doi.org/10.1209/epl/i2005-10562-8} {\bibfield  {journal} {\bibinfo  {journal} {Europhysics Letters}\ }\textbf {\bibinfo {volume} {74}},\ \bibinfo {pages} {391} (\bibinfo {year} {2006})}\BibitemShut {NoStop}%
\bibitem [{\citenamefont {Esposito}\ and\ \citenamefont {Van~den Broeck}(2010{\natexlab{a}})}]{Esposito10}%
  \BibitemOpen
  \bibfield  {author} {\bibinfo {author} {\bibfnamefont {M.}~\bibnamefont {Esposito}}\ and\ \bibinfo {author} {\bibfnamefont {C.}~\bibnamefont {Van~den Broeck}},\ }\href {https://doi.org/10.1103/PhysRevLett.104.090601} {\bibfield  {journal} {\bibinfo  {journal} {Phys. Rev. Lett.}\ }\textbf {\bibinfo {volume} {104}},\ \bibinfo {pages} {090601} (\bibinfo {year} {2010}{\natexlab{a}})}\BibitemShut {NoStop}%
\bibitem [{\citenamefont {Hummer}\ and\ \citenamefont {Szabo}(2001)}]{HummerSzabo2001}%
  \BibitemOpen
  \bibfield  {author} {\bibinfo {author} {\bibfnamefont {G.}~\bibnamefont {Hummer}}\ and\ \bibinfo {author} {\bibfnamefont {A.}~\bibnamefont {Szabo}},\ }\href {https://doi.org/10.1073/pnas.071034098} {\bibfield  {journal} {\bibinfo  {journal} {Proceedings of the National Academy of Sciences}\ }\textbf {\bibinfo {volume} {98}},\ \bibinfo {pages} {3658} (\bibinfo {year} {2001})}\BibitemShut {NoStop}%
\bibitem [{\citenamefont {Liphardt}\ \emph {et~al.}(2002)\citenamefont {Liphardt}, \citenamefont {Dumont}, \citenamefont {Smith}, \citenamefont {Tinoco},\ and\ \citenamefont {Bustamante}}]{Liphardt2002}%
  \BibitemOpen
  \bibfield  {author} {\bibinfo {author} {\bibfnamefont {J.}~\bibnamefont {Liphardt}}, \bibinfo {author} {\bibfnamefont {S.}~\bibnamefont {Dumont}}, \bibinfo {author} {\bibfnamefont {S.~B.}\ \bibnamefont {Smith}}, \bibinfo {author} {\bibfnamefont {I.}~\bibnamefont {Tinoco}},\ and\ \bibinfo {author} {\bibfnamefont {C.}~\bibnamefont {Bustamante}},\ }\href {https://doi.org/10.1126/science.1071152} {\bibfield  {journal} {\bibinfo  {journal} {Science}\ }\textbf {\bibinfo {volume} {296}},\ \bibinfo {pages} {1832} (\bibinfo {year} {2002})}\BibitemShut {NoStop}%
\bibitem [{\citenamefont {Collin}\ \emph {et~al.}(2005)\citenamefont {Collin}, \citenamefont {Ritort}, \citenamefont {Jarzynski}, \citenamefont {Smith}, \citenamefont {Tinoco},\ and\ \citenamefont {Bustamante}}]{Collin2005}%
  \BibitemOpen
  \bibfield  {author} {\bibinfo {author} {\bibfnamefont {D.}~\bibnamefont {Collin}}, \bibinfo {author} {\bibfnamefont {F.}~\bibnamefont {Ritort}}, \bibinfo {author} {\bibfnamefont {C.}~\bibnamefont {Jarzynski}}, \bibinfo {author} {\bibfnamefont {S.~B.}\ \bibnamefont {Smith}}, \bibinfo {author} {\bibfnamefont {I.}~\bibnamefont {Tinoco}},\ and\ \bibinfo {author} {\bibfnamefont {C.}~\bibnamefont {Bustamante}},\ }\href {https://doi.org/10.1038/nature04061} {\bibfield  {journal} {\bibinfo  {journal} {Nature}\ }\textbf {\bibinfo {volume} {437}},\ \bibinfo {pages} {231} (\bibinfo {year} {2005})}\BibitemShut {NoStop}%
\bibitem [{\citenamefont {Ciliberto}(2017)}]{Ciliberto2017}%
  \BibitemOpen
  \bibfield  {author} {\bibinfo {author} {\bibfnamefont {S.}~\bibnamefont {Ciliberto}},\ }\href {https://doi.org/10.1103/PhysRevX.7.021051} {\bibfield  {journal} {\bibinfo  {journal} {Phys. Rev. X}\ }\textbf {\bibinfo {volume} {7}},\ \bibinfo {pages} {021051} (\bibinfo {year} {2017})}\BibitemShut {NoStop}%
\bibitem [{\citenamefont {Sagawa}\ and\ \citenamefont {Ueda}(2008)}]{sagawa2008second}%
  \BibitemOpen
  \bibfield  {author} {\bibinfo {author} {\bibfnamefont {T.}~\bibnamefont {Sagawa}}\ and\ \bibinfo {author} {\bibfnamefont {M.}~\bibnamefont {Ueda}},\ }\href@noop {} {\bibfield  {journal} {\bibinfo  {journal} {Physical review letters}\ }\textbf {\bibinfo {volume} {100}},\ \bibinfo {pages} {080403} (\bibinfo {year} {2008})}\BibitemShut {NoStop}%
\bibitem [{\citenamefont {Sagawa}\ and\ \citenamefont {Ueda}(2010)}]{SagawaUeda2010}%
  \BibitemOpen
  \bibfield  {author} {\bibinfo {author} {\bibfnamefont {T.}~\bibnamefont {Sagawa}}\ and\ \bibinfo {author} {\bibfnamefont {M.}~\bibnamefont {Ueda}},\ }\href {https://doi.org/10.1103/PhysRevLett.104.090602} {\bibfield  {journal} {\bibinfo  {journal} {Physical Review Letters}\ }\textbf {\bibinfo {volume} {104}},\ \bibinfo {pages} {090602} (\bibinfo {year} {2010})}\BibitemShut {NoStop}%
\bibitem [{\citenamefont {Parrondo}\ \emph {et~al.}(2015)\citenamefont {Parrondo}, \citenamefont {Horowitz},\ and\ \citenamefont {Sagawa}}]{parrondo2015thermodynamics}%
  \BibitemOpen
  \bibfield  {author} {\bibinfo {author} {\bibfnamefont {J.~M.}\ \bibnamefont {Parrondo}}, \bibinfo {author} {\bibfnamefont {J.~M.}\ \bibnamefont {Horowitz}},\ and\ \bibinfo {author} {\bibfnamefont {T.}~\bibnamefont {Sagawa}},\ }\href@noop {} {\bibfield  {journal} {\bibinfo  {journal} {Nature physics}\ }\textbf {\bibinfo {volume} {11}},\ \bibinfo {pages} {131} (\bibinfo {year} {2015})}\BibitemShut {NoStop}%
\bibitem [{\citenamefont {Barato}\ and\ \citenamefont {Seifert}(2014)}]{barato2014unifying}%
  \BibitemOpen
  \bibfield  {author} {\bibinfo {author} {\bibfnamefont {A.}~\bibnamefont {Barato}}\ and\ \bibinfo {author} {\bibfnamefont {U.}~\bibnamefont {Seifert}},\ }\href@noop {} {\bibfield  {journal} {\bibinfo  {journal} {Physical review letters}\ }\textbf {\bibinfo {volume} {112}},\ \bibinfo {pages} {090601} (\bibinfo {year} {2014})}\BibitemShut {NoStop}%
\bibitem [{\citenamefont {Rosinberg}\ and\ \citenamefont {Horowitz}(2016)}]{rosinberg2016continuous}%
  \BibitemOpen
  \bibfield  {author} {\bibinfo {author} {\bibfnamefont {M.~L.}\ \bibnamefont {Rosinberg}}\ and\ \bibinfo {author} {\bibfnamefont {J.~M.}\ \bibnamefont {Horowitz}},\ }\href@noop {} {\bibfield  {journal} {\bibinfo  {journal} {Europhysics Letters}\ }\textbf {\bibinfo {volume} {116}},\ \bibinfo {pages} {10007} (\bibinfo {year} {2016})}\BibitemShut {NoStop}%
\bibitem [{\citenamefont {Horowitz}\ and\ \citenamefont {Esposito}(2014)}]{HorowitzEsposito2014}%
  \BibitemOpen
  \bibfield  {author} {\bibinfo {author} {\bibfnamefont {J.~M.}\ \bibnamefont {Horowitz}}\ and\ \bibinfo {author} {\bibfnamefont {M.}~\bibnamefont {Esposito}},\ }\href {https://doi.org/10.1103/PhysRevX.4.031015} {\bibfield  {journal} {\bibinfo  {journal} {Physical Review X}\ }\textbf {\bibinfo {volume} {4}},\ \bibinfo {pages} {031015} (\bibinfo {year} {2014})}\BibitemShut {NoStop}%
\bibitem [{\citenamefont {B{\'e}rut}\ \emph {et~al.}(2012)\citenamefont {B{\'e}rut}, \citenamefont {Arakelyan}, \citenamefont {Petrosyan}, \citenamefont {Ciliberto}, \citenamefont {Dillenschneider},\ and\ \citenamefont {Lutz}}]{berut2012experimental}%
  \BibitemOpen
  \bibfield  {author} {\bibinfo {author} {\bibfnamefont {A.}~\bibnamefont {B{\'e}rut}}, \bibinfo {author} {\bibfnamefont {A.}~\bibnamefont {Arakelyan}}, \bibinfo {author} {\bibfnamefont {A.}~\bibnamefont {Petrosyan}}, \bibinfo {author} {\bibfnamefont {S.}~\bibnamefont {Ciliberto}}, \bibinfo {author} {\bibfnamefont {R.}~\bibnamefont {Dillenschneider}},\ and\ \bibinfo {author} {\bibfnamefont {E.}~\bibnamefont {Lutz}},\ }\href@noop {} {\bibfield  {journal} {\bibinfo  {journal} {Nature}\ }\textbf {\bibinfo {volume} {483}},\ \bibinfo {pages} {187} (\bibinfo {year} {2012})}\BibitemShut {NoStop}%
\bibitem [{\citenamefont {Jun}\ \emph {et~al.}(2014)\citenamefont {Jun}, \citenamefont {Gavrilov},\ and\ \citenamefont {Bechhoefer}}]{jun2014high}%
  \BibitemOpen
  \bibfield  {author} {\bibinfo {author} {\bibfnamefont {Y.}~\bibnamefont {Jun}}, \bibinfo {author} {\bibfnamefont {M.}~\bibnamefont {Gavrilov}},\ and\ \bibinfo {author} {\bibfnamefont {J.}~\bibnamefont {Bechhoefer}},\ }\href@noop {} {\bibfield  {journal} {\bibinfo  {journal} {Physical review letters}\ }\textbf {\bibinfo {volume} {113}},\ \bibinfo {pages} {190601} (\bibinfo {year} {2014})}\BibitemShut {NoStop}%
\bibitem [{\citenamefont {Rold{\'a}n}\ \emph {et~al.}(2014)\citenamefont {Rold{\'a}n}, \citenamefont {Martinez}, \citenamefont {Parrondo},\ and\ \citenamefont {Petrov}}]{roldan2014universal}%
  \BibitemOpen
  \bibfield  {author} {\bibinfo {author} {\bibfnamefont {{\'E}.}~\bibnamefont {Rold{\'a}n}}, \bibinfo {author} {\bibfnamefont {I.~A.}\ \bibnamefont {Martinez}}, \bibinfo {author} {\bibfnamefont {J.~M.}\ \bibnamefont {Parrondo}},\ and\ \bibinfo {author} {\bibfnamefont {D.}~\bibnamefont {Petrov}},\ }\href@noop {} {\bibfield  {journal} {\bibinfo  {journal} {Nature Physics}\ }\textbf {\bibinfo {volume} {10}},\ \bibinfo {pages} {457} (\bibinfo {year} {2014})}\BibitemShut {NoStop}%
\bibitem [{\citenamefont {Alemany}\ \emph {et~al.}(2012)\citenamefont {Alemany}, \citenamefont {Mossa}, \citenamefont {Junier},\ and\ \citenamefont {Ritort}}]{alemany2012experimental}%
  \BibitemOpen
  \bibfield  {author} {\bibinfo {author} {\bibfnamefont {A.}~\bibnamefont {Alemany}}, \bibinfo {author} {\bibfnamefont {A.}~\bibnamefont {Mossa}}, \bibinfo {author} {\bibfnamefont {I.}~\bibnamefont {Junier}},\ and\ \bibinfo {author} {\bibfnamefont {F.}~\bibnamefont {Ritort}},\ }\href@noop {} {\bibfield  {journal} {\bibinfo  {journal} {Nature Physics}\ }\textbf {\bibinfo {volume} {8}},\ \bibinfo {pages} {688} (\bibinfo {year} {2012})}\BibitemShut {NoStop}%
\bibitem [{\citenamefont {Koski}\ \emph {et~al.}(2015)\citenamefont {Koski}, \citenamefont {Kutvonen}, \citenamefont {Khaymovich}, \citenamefont {Ala-Nissila},\ and\ \citenamefont {Pekola}}]{Koski15}%
  \BibitemOpen
  \bibfield  {author} {\bibinfo {author} {\bibfnamefont {J.~V.}\ \bibnamefont {Koski}}, \bibinfo {author} {\bibfnamefont {A.}~\bibnamefont {Kutvonen}}, \bibinfo {author} {\bibfnamefont {I.~M.}\ \bibnamefont {Khaymovich}}, \bibinfo {author} {\bibfnamefont {T.}~\bibnamefont {Ala-Nissila}},\ and\ \bibinfo {author} {\bibfnamefont {J.~P.}\ \bibnamefont {Pekola}},\ }\href {https://doi.org/10.1103/PhysRevLett.115.260602} {\bibfield  {journal} {\bibinfo  {journal} {Phys. Rev. Lett.}\ }\textbf {\bibinfo {volume} {115}},\ \bibinfo {pages} {260602} (\bibinfo {year} {2015})}\BibitemShut {NoStop}%
\bibitem [{\citenamefont {Toyabe}\ \emph {et~al.}(2010)\citenamefont {Toyabe}, \citenamefont {Sagawa}, \citenamefont {Ueda}, \citenamefont {Muneyuki},\ and\ \citenamefont {Sano}}]{toyabe2010experimental}%
  \BibitemOpen
  \bibfield  {author} {\bibinfo {author} {\bibfnamefont {S.}~\bibnamefont {Toyabe}}, \bibinfo {author} {\bibfnamefont {T.}~\bibnamefont {Sagawa}}, \bibinfo {author} {\bibfnamefont {M.}~\bibnamefont {Ueda}}, \bibinfo {author} {\bibfnamefont {E.}~\bibnamefont {Muneyuki}},\ and\ \bibinfo {author} {\bibfnamefont {M.}~\bibnamefont {Sano}},\ }\href@noop {} {\bibfield  {journal} {\bibinfo  {journal} {Nature physics}\ }\textbf {\bibinfo {volume} {6}},\ \bibinfo {pages} {988} (\bibinfo {year} {2010})}\BibitemShut {NoStop}%
\bibitem [{\citenamefont {Esposito}\ and\ \citenamefont {Van~den Broeck}(2010{\natexlab{b}})}]{EspositoVanDenBroeck2010a}%
  \BibitemOpen
  \bibfield  {author} {\bibinfo {author} {\bibfnamefont {M.}~\bibnamefont {Esposito}}\ and\ \bibinfo {author} {\bibfnamefont {C.}~\bibnamefont {Van~den Broeck}},\ }\href {https://doi.org/10.1103/PhysRevE.82.011143} {\bibfield  {journal} {\bibinfo  {journal} {Physical Review E}\ }\textbf {\bibinfo {volume} {82}},\ \bibinfo {pages} {011143} (\bibinfo {year} {2010}{\natexlab{b}})}\BibitemShut {NoStop}%
\bibitem [{\citenamefont {Esposito}\ and\ \citenamefont {Van~den Broeck}(2010{\natexlab{c}})}]{EspositoVanDenBroeck2010b}%
  \BibitemOpen
  \bibfield  {author} {\bibinfo {author} {\bibfnamefont {M.}~\bibnamefont {Esposito}}\ and\ \bibinfo {author} {\bibfnamefont {C.}~\bibnamefont {Van~den Broeck}},\ }\href {https://doi.org/10.1103/PhysRevE.82.011144} {\bibfield  {journal} {\bibinfo  {journal} {Physical Review E}\ }\textbf {\bibinfo {volume} {82}},\ \bibinfo {pages} {011144} (\bibinfo {year} {2010}{\natexlab{c}})}\BibitemShut {NoStop}%
\bibitem [{\citenamefont {Kawai}\ \emph {et~al.}(2007)\citenamefont {Kawai}, \citenamefont {Parrondo},\ and\ \citenamefont {den Broeck}}]{kawai2007dissipation}%
  \BibitemOpen
  \bibfield  {author} {\bibinfo {author} {\bibfnamefont {R.}~\bibnamefont {Kawai}}, \bibinfo {author} {\bibfnamefont {J.~M.}\ \bibnamefont {Parrondo}},\ and\ \bibinfo {author} {\bibfnamefont {C.~V.}\ \bibnamefont {den Broeck}},\ }\href@noop {} {\bibfield  {journal} {\bibinfo  {journal} {Physical review letters}\ }\textbf {\bibinfo {volume} {98}},\ \bibinfo {pages} {080602} (\bibinfo {year} {2007})}\BibitemShut {NoStop}%
\bibitem [{\citenamefont {Gomez-Marin}\ \emph {et~al.}(2008)\citenamefont {Gomez-Marin}, \citenamefont {Parrondo},\ and\ \citenamefont {Van~den Broeck}}]{gomez2008footprints}%
  \BibitemOpen
  \bibfield  {author} {\bibinfo {author} {\bibfnamefont {A.}~\bibnamefont {Gomez-Marin}}, \bibinfo {author} {\bibfnamefont {J.~M.}\ \bibnamefont {Parrondo}},\ and\ \bibinfo {author} {\bibfnamefont {C.}~\bibnamefont {Van~den Broeck}},\ }\href@noop {} {\bibfield  {journal} {\bibinfo  {journal} {EPL (Europhysics Letters)}\ }\textbf {\bibinfo {volume} {82}},\ \bibinfo {pages} {50002} (\bibinfo {year} {2008})}\BibitemShut {NoStop}%
\bibitem [{\citenamefont {Parrondo}\ \emph {et~al.}(2009)\citenamefont {Parrondo}, \citenamefont {Van~den Broeck},\ and\ \citenamefont {Kawai}}]{Parrondo2009}%
  \BibitemOpen
  \bibfield  {author} {\bibinfo {author} {\bibfnamefont {J.~M.~R.}\ \bibnamefont {Parrondo}}, \bibinfo {author} {\bibfnamefont {C.}~\bibnamefont {Van~den Broeck}},\ and\ \bibinfo {author} {\bibfnamefont {R.}~\bibnamefont {Kawai}},\ }\href {https://doi.org/10.1088/1367-2630/11/7/073008} {\bibfield  {journal} {\bibinfo  {journal} {New Journal of Physics}\ }\textbf {\bibinfo {volume} {11}},\ \bibinfo {pages} {073008} (\bibinfo {year} {2009})}\BibitemShut {NoStop}%
\bibitem [{\citenamefont {Andrieux}\ and\ \citenamefont {Gaspard}(2007{\natexlab{a}})}]{AndrieuxGaspard2007JSP}%
  \BibitemOpen
  \bibfield  {author} {\bibinfo {author} {\bibfnamefont {D.}~\bibnamefont {Andrieux}}\ and\ \bibinfo {author} {\bibfnamefont {P.}~\bibnamefont {Gaspard}},\ }\href {https://doi.org/10.1007/s10955-006-9233-5} {\bibfield  {journal} {\bibinfo  {journal} {Journal of Statistical Physics}\ }\textbf {\bibinfo {volume} {127}},\ \bibinfo {pages} {107} (\bibinfo {year} {2007}{\natexlab{a}})}\BibitemShut {NoStop}%
\bibitem [{\citenamefont {Andrieux}\ and\ \citenamefont {Gaspard}(2007{\natexlab{b}})}]{AndrieuxGaspard2007JSTAT}%
  \BibitemOpen
  \bibfield  {author} {\bibinfo {author} {\bibfnamefont {D.}~\bibnamefont {Andrieux}}\ and\ \bibinfo {author} {\bibfnamefont {P.}~\bibnamefont {Gaspard}},\ }\href {https://doi.org/10.1088/1742-5468/2007/02/p02006} {\bibfield  {journal} {\bibinfo  {journal} {Journal of Statistical Mechanics: Theory and Experiment}\ }\textbf {\bibinfo {volume} {2007}},\ \bibinfo {pages} {P02006} (\bibinfo {year} {2007}{\natexlab{b}})}\BibitemShut {NoStop}%
\bibitem [{\citenamefont {Andrieux}\ \emph {et~al.}(2007)\citenamefont {Andrieux}, \citenamefont {Gaspard}, \citenamefont {Ciliberto}, \citenamefont {Garnier}, \citenamefont {Joubaud},\ and\ \citenamefont {Petrosyan}}]{AndrieuxEtAl2007PRL}%
  \BibitemOpen
  \bibfield  {author} {\bibinfo {author} {\bibfnamefont {D.}~\bibnamefont {Andrieux}}, \bibinfo {author} {\bibfnamefont {P.}~\bibnamefont {Gaspard}}, \bibinfo {author} {\bibfnamefont {S.}~\bibnamefont {Ciliberto}}, \bibinfo {author} {\bibfnamefont {N.}~\bibnamefont {Garnier}}, \bibinfo {author} {\bibfnamefont {S.}~\bibnamefont {Joubaud}},\ and\ \bibinfo {author} {\bibfnamefont {A.}~\bibnamefont {Petrosyan}},\ }\href {https://doi.org/10.1103/physrevlett.98.150601} {\bibfield  {journal} {\bibinfo  {journal} {Physical Review Letters}\ }\textbf {\bibinfo {volume} {98}},\ \bibinfo {pages} {150601} (\bibinfo {year} {2007})}\BibitemShut {NoStop}%
\bibitem [{\citenamefont {Rold{\'a}n}\ and\ \citenamefont {Parrondo}(2010)}]{roldan2010estimating}%
  \BibitemOpen
  \bibfield  {author} {\bibinfo {author} {\bibfnamefont {{\'E}.}~\bibnamefont {Rold{\'a}n}}\ and\ \bibinfo {author} {\bibfnamefont {J.~M.}\ \bibnamefont {Parrondo}},\ }\href@noop {} {\bibfield  {journal} {\bibinfo  {journal} {Physical review letters}\ }\textbf {\bibinfo {volume} {105}},\ \bibinfo {pages} {150607} (\bibinfo {year} {2010})}\BibitemShut {NoStop}%
\bibitem [{\citenamefont {Seifert}(2012)}]{Seifert2012}%
  \BibitemOpen
  \bibfield  {author} {\bibinfo {author} {\bibfnamefont {U.}~\bibnamefont {Seifert}},\ }\href {https://doi.org/10.1088/0034-4885/75/12/126001} {\bibfield  {journal} {\bibinfo  {journal} {Reports on Progress in Physics}\ }\textbf {\bibinfo {volume} {75}},\ \bibinfo {pages} {126001} (\bibinfo {year} {2012})}\BibitemShut {NoStop}%
\bibitem [{\citenamefont {Esposito}\ and\ \citenamefont {Van~den Broeck}(2015)}]{EspositoVanDenBroeck2015}%
  \BibitemOpen
  \bibfield  {author} {\bibinfo {author} {\bibfnamefont {M.}~\bibnamefont {Esposito}}\ and\ \bibinfo {author} {\bibfnamefont {C.}~\bibnamefont {Van~den Broeck}},\ }\href {https://doi.org/10.1016/j.physa.2014.04.019} {\bibfield  {journal} {\bibinfo  {journal} {Physica A}\ }\textbf {\bibinfo {volume} {418}},\ \bibinfo {pages} {6} (\bibinfo {year} {2015})}\BibitemShut {NoStop}%
\bibitem [{\citenamefont {Barato}\ and\ \citenamefont {Seifert}(2015)}]{BaratoSeifert2015}%
  \BibitemOpen
  \bibfield  {author} {\bibinfo {author} {\bibfnamefont {A.~C.}\ \bibnamefont {Barato}}\ and\ \bibinfo {author} {\bibfnamefont {U.}~\bibnamefont {Seifert}},\ }\href {https://doi.org/10.1103/PhysRevLett.114.158101} {\bibfield  {journal} {\bibinfo  {journal} {Physical Review Letters}\ }\textbf {\bibinfo {volume} {114}},\ \bibinfo {pages} {158101} (\bibinfo {year} {2015})}\BibitemShut {NoStop}%
\bibitem [{\citenamefont {Gingrich}\ \emph {et~al.}(2016)\citenamefont {Gingrich}, \citenamefont {Horowitz}, \citenamefont {Perunov},\ and\ \citenamefont {England}}]{Gingrich2016}%
  \BibitemOpen
  \bibfield  {author} {\bibinfo {author} {\bibfnamefont {T.~R.}\ \bibnamefont {Gingrich}}, \bibinfo {author} {\bibfnamefont {J.~M.}\ \bibnamefont {Horowitz}}, \bibinfo {author} {\bibfnamefont {N.}~\bibnamefont {Perunov}},\ and\ \bibinfo {author} {\bibfnamefont {J.~L.}\ \bibnamefont {England}},\ }\href {https://doi.org/10.1103/PhysRevLett.116.120601} {\bibfield  {journal} {\bibinfo  {journal} {Physical Review Letters}\ }\textbf {\bibinfo {volume} {116}},\ \bibinfo {pages} {120601} (\bibinfo {year} {2016})}\BibitemShut {NoStop}%
\bibitem [{\citenamefont {Shiraishi}\ \emph {et~al.}(2018)\citenamefont {Shiraishi}, \citenamefont {Funo},\ and\ \citenamefont {Saito}}]{Shiraishi2018}%
  \BibitemOpen
  \bibfield  {author} {\bibinfo {author} {\bibfnamefont {N.}~\bibnamefont {Shiraishi}}, \bibinfo {author} {\bibfnamefont {K.}~\bibnamefont {Funo}},\ and\ \bibinfo {author} {\bibfnamefont {K.}~\bibnamefont {Saito}},\ }\href {https://doi.org/10.1103/PhysRevLett.121.070601} {\bibfield  {journal} {\bibinfo  {journal} {Physical Review Letters}\ }\textbf {\bibinfo {volume} {121}},\ \bibinfo {pages} {070601} (\bibinfo {year} {2018})}\BibitemShut {NoStop}%
\bibitem [{\citenamefont {Dechant}\ and\ \citenamefont {Sasa}(2018)}]{DechantSasa2018}%
  \BibitemOpen
  \bibfield  {author} {\bibinfo {author} {\bibfnamefont {A.}~\bibnamefont {Dechant}}\ and\ \bibinfo {author} {\bibfnamefont {S.-i.}\ \bibnamefont {Sasa}},\ }\href {https://doi.org/10.1103/PhysRevE.97.062101} {\bibfield  {journal} {\bibinfo  {journal} {Physical Review E}\ }\textbf {\bibinfo {volume} {97}},\ \bibinfo {pages} {062101} (\bibinfo {year} {2018})}\BibitemShut {NoStop}%
\bibitem [{\citenamefont {Proesmans}\ \emph {et~al.}(2020)\citenamefont {Proesmans}, \citenamefont {Ehrich},\ and\ \citenamefont {Bechhoefer}}]{Proesmans2020}%
  \BibitemOpen
  \bibfield  {author} {\bibinfo {author} {\bibfnamefont {K.}~\bibnamefont {Proesmans}}, \bibinfo {author} {\bibfnamefont {J.}~\bibnamefont {Ehrich}},\ and\ \bibinfo {author} {\bibfnamefont {J.}~\bibnamefont {Bechhoefer}},\ }\href {https://doi.org/10.1103/PhysRevLett.125.100602} {\bibfield  {journal} {\bibinfo  {journal} {Physical Review Letters}\ }\textbf {\bibinfo {volume} {125}},\ \bibinfo {pages} {100602} (\bibinfo {year} {2020})}\BibitemShut {NoStop}%
\bibitem [{\citenamefont {Pietzonka}\ and\ \citenamefont {Seifert}(2018)}]{Pietzonka2016}%
  \BibitemOpen
  \bibfield  {author} {\bibinfo {author} {\bibfnamefont {P.}~\bibnamefont {Pietzonka}}\ and\ \bibinfo {author} {\bibfnamefont {U.}~\bibnamefont {Seifert}},\ }\href {https://doi.org/10.1103/PhysRevLett.120.190602} {\bibfield  {journal} {\bibinfo  {journal} {Physical Review Letters}\ }\textbf {\bibinfo {volume} {120}},\ \bibinfo {pages} {190602} (\bibinfo {year} {2018})}\BibitemShut {NoStop}%
\bibitem [{\citenamefont {Pietzonka}\ \emph {et~al.}(2017)\citenamefont {Pietzonka}, \citenamefont {Ritort},\ and\ \citenamefont {Seifert}}]{Pietzonka2018}%
  \BibitemOpen
  \bibfield  {author} {\bibinfo {author} {\bibfnamefont {P.}~\bibnamefont {Pietzonka}}, \bibinfo {author} {\bibfnamefont {F.}~\bibnamefont {Ritort}},\ and\ \bibinfo {author} {\bibfnamefont {U.}~\bibnamefont {Seifert}},\ }\href {https://doi.org/10.1103/PhysRevE.96.012101} {\bibfield  {journal} {\bibinfo  {journal} {Physical Review E}\ }\textbf {\bibinfo {volume} {96}},\ \bibinfo {pages} {012101} (\bibinfo {year} {2017})}\BibitemShut {NoStop}%
\bibitem [{\citenamefont {Falasco}\ and\ \citenamefont {Esposito}(2020)}]{falasco2020dissipation}%
  \BibitemOpen
  \bibfield  {author} {\bibinfo {author} {\bibfnamefont {G.}~\bibnamefont {Falasco}}\ and\ \bibinfo {author} {\bibfnamefont {M.}~\bibnamefont {Esposito}},\ }\href@noop {} {\bibfield  {journal} {\bibinfo  {journal} {Physical Review Letters}\ }\textbf {\bibinfo {volume} {125}},\ \bibinfo {pages} {120604} (\bibinfo {year} {2020})}\BibitemShut {NoStop}%
\bibitem [{\citenamefont {Neri}(2022{\natexlab{a}})}]{neri2022universal}%
  \BibitemOpen
  \bibfield  {author} {\bibinfo {author} {\bibfnamefont {I.}~\bibnamefont {Neri}},\ }\href@noop {} {\bibfield  {journal} {\bibinfo  {journal} {SciPost Physics}\ }\textbf {\bibinfo {volume} {12}},\ \bibinfo {pages} {139} (\bibinfo {year} {2022}{\natexlab{a}})}\BibitemShut {NoStop}%
\bibitem [{\citenamefont {Neri}\ \emph {et~al.}(2017)\citenamefont {Neri}, \citenamefont {Rold{\'a}n},\ and\ \citenamefont {J{\"u}licher}}]{Neri2017}%
  \BibitemOpen
  \bibfield  {author} {\bibinfo {author} {\bibfnamefont {I.}~\bibnamefont {Neri}}, \bibinfo {author} {\bibfnamefont {{\'E}.}~\bibnamefont {Rold{\'a}n}},\ and\ \bibinfo {author} {\bibfnamefont {F.}~\bibnamefont {J{\"u}licher}},\ }\href {https://doi.org/10.1103/PhysRevX.7.011019} {\bibfield  {journal} {\bibinfo  {journal} {Physical Review X}\ }\textbf {\bibinfo {volume} {7}},\ \bibinfo {pages} {011019} (\bibinfo {year} {2017})}\BibitemShut {NoStop}%
\bibitem [{\citenamefont {Manzano}\ \emph {et~al.}(2021)\citenamefont {Manzano}, \citenamefont {Subero}, \citenamefont {Maillet}, \citenamefont {Fazio}, \citenamefont {Pekola},\ and\ \citenamefont {Rold\'an}}]{Manzano21}%
  \BibitemOpen
  \bibfield  {author} {\bibinfo {author} {\bibfnamefont {G.}~\bibnamefont {Manzano}}, \bibinfo {author} {\bibfnamefont {D.}~\bibnamefont {Subero}}, \bibinfo {author} {\bibfnamefont {O.}~\bibnamefont {Maillet}}, \bibinfo {author} {\bibfnamefont {R.}~\bibnamefont {Fazio}}, \bibinfo {author} {\bibfnamefont {J.~P.}\ \bibnamefont {Pekola}},\ and\ \bibinfo {author} {\bibfnamefont {E.}~\bibnamefont {Rold\'an}},\ }\href {https://doi.org/10.1103/PhysRevLett.126.080603} {\bibfield  {journal} {\bibinfo  {journal} {Phys. Rev. Lett.}\ }\textbf {\bibinfo {volume} {126}},\ \bibinfo {pages} {080603} (\bibinfo {year} {2021})}\BibitemShut {NoStop}%
\bibitem [{\citenamefont {Rold{\'a}n}\ \emph {et~al.}(2023)\citenamefont {Rold{\'a}n}, \citenamefont {Neri}, \citenamefont {Chetrite}, \citenamefont {Gupta}, \citenamefont {Pigolotti}, \citenamefont {J{\"u}licher},\ and\ \citenamefont {Sekimoto}}]{roldan2023martingales}%
  \BibitemOpen
  \bibfield  {author} {\bibinfo {author} {\bibfnamefont {{\'E}.}~\bibnamefont {Rold{\'a}n}}, \bibinfo {author} {\bibfnamefont {I.}~\bibnamefont {Neri}}, \bibinfo {author} {\bibfnamefont {R.}~\bibnamefont {Chetrite}}, \bibinfo {author} {\bibfnamefont {S.}~\bibnamefont {Gupta}}, \bibinfo {author} {\bibfnamefont {S.}~\bibnamefont {Pigolotti}}, \bibinfo {author} {\bibfnamefont {F.}~\bibnamefont {J{\"u}licher}},\ and\ \bibinfo {author} {\bibfnamefont {K.}~\bibnamefont {Sekimoto}},\ }\href@noop {} {\bibfield  {journal} {\bibinfo  {journal} {Advances in Physics}\ }\textbf {\bibinfo {volume} {72}},\ \bibinfo {pages} {1} (\bibinfo {year} {2023})}\BibitemShut {NoStop}%
\bibitem [{\citenamefont {Ciliberto}\ \emph {et~al.}(2013)\citenamefont {Ciliberto}, \citenamefont {Imparato}, \citenamefont {Naert},\ and\ \citenamefont {Tanase}}]{Ciliberto13}%
  \BibitemOpen
  \bibfield  {author} {\bibinfo {author} {\bibfnamefont {S.}~\bibnamefont {Ciliberto}}, \bibinfo {author} {\bibfnamefont {A.}~\bibnamefont {Imparato}}, \bibinfo {author} {\bibfnamefont {A.}~\bibnamefont {Naert}},\ and\ \bibinfo {author} {\bibfnamefont {M.}~\bibnamefont {Tanase}},\ }\href {https://doi.org/10.1103/PhysRevLett.110.180601} {\bibfield  {journal} {\bibinfo  {journal} {Phys. Rev. Lett.}\ }\textbf {\bibinfo {volume} {110}},\ \bibinfo {pages} {180601} (\bibinfo {year} {2013})}\BibitemShut {NoStop}%
\bibitem [{\citenamefont {Koski}\ \emph {et~al.}(2013)\citenamefont {Koski}, \citenamefont {Sagawa}, \citenamefont {Saira}, \citenamefont {Yoon}, \citenamefont {Kutvonen}, \citenamefont {Solinas}, \citenamefont {M{\"o}tt{\"o}nen}, \citenamefont {Ala-Nissila},\ and\ \citenamefont {Pekola}}]{koski2013distribution}%
  \BibitemOpen
  \bibfield  {author} {\bibinfo {author} {\bibfnamefont {J.}~\bibnamefont {Koski}}, \bibinfo {author} {\bibfnamefont {T.}~\bibnamefont {Sagawa}}, \bibinfo {author} {\bibfnamefont {O.}~\bibnamefont {Saira}}, \bibinfo {author} {\bibfnamefont {Y.}~\bibnamefont {Yoon}}, \bibinfo {author} {\bibfnamefont {A.}~\bibnamefont {Kutvonen}}, \bibinfo {author} {\bibfnamefont {P.}~\bibnamefont {Solinas}}, \bibinfo {author} {\bibfnamefont {M.}~\bibnamefont {M{\"o}tt{\"o}nen}}, \bibinfo {author} {\bibfnamefont {T.}~\bibnamefont {Ala-Nissila}},\ and\ \bibinfo {author} {\bibfnamefont {J.}~\bibnamefont {Pekola}},\ }\href@noop {} {\bibfield  {journal} {\bibinfo  {journal} {Nature Physics}\ }\textbf {\bibinfo {volume} {9}},\ \bibinfo {pages} {644} (\bibinfo {year} {2013})}\BibitemShut {NoStop}%
\bibitem [{\citenamefont {Josefsson}\ \emph {et~al.}(2018)\citenamefont {Josefsson}, \citenamefont {Svilans}, \citenamefont {Burke}, \citenamefont {Hoffmann}, \citenamefont {Fahlvik}, \citenamefont {Thelander}, \citenamefont {Leijnse},\ and\ \citenamefont {Linke}}]{josefsson2018quantum}%
  \BibitemOpen
  \bibfield  {author} {\bibinfo {author} {\bibfnamefont {M.}~\bibnamefont {Josefsson}}, \bibinfo {author} {\bibfnamefont {A.}~\bibnamefont {Svilans}}, \bibinfo {author} {\bibfnamefont {A.~M.}\ \bibnamefont {Burke}}, \bibinfo {author} {\bibfnamefont {E.~A.}\ \bibnamefont {Hoffmann}}, \bibinfo {author} {\bibfnamefont {S.}~\bibnamefont {Fahlvik}}, \bibinfo {author} {\bibfnamefont {C.}~\bibnamefont {Thelander}}, \bibinfo {author} {\bibfnamefont {M.}~\bibnamefont {Leijnse}},\ and\ \bibinfo {author} {\bibfnamefont {H.}~\bibnamefont {Linke}},\ }\href@noop {} {\bibfield  {journal} {\bibinfo  {journal} {Nature nanotechnology}\ }\textbf {\bibinfo {volume} {13}},\ \bibinfo {pages} {920} (\bibinfo {year} {2018})}\BibitemShut {NoStop}%
\bibitem [{\citenamefont {Serra-Garcia}\ \emph {et~al.}(2016)\citenamefont {Serra-Garcia}, \citenamefont {Foehr}, \citenamefont {Moler\'on}, \citenamefont {Lydon}, \citenamefont {Chong},\ and\ \citenamefont {Daraio}}]{Serra-Garcia16}%
  \BibitemOpen
  \bibfield  {author} {\bibinfo {author} {\bibfnamefont {M.}~\bibnamefont {Serra-Garcia}}, \bibinfo {author} {\bibfnamefont {A.}~\bibnamefont {Foehr}}, \bibinfo {author} {\bibfnamefont {M.}~\bibnamefont {Moler\'on}}, \bibinfo {author} {\bibfnamefont {J.}~\bibnamefont {Lydon}}, \bibinfo {author} {\bibfnamefont {C.}~\bibnamefont {Chong}},\ and\ \bibinfo {author} {\bibfnamefont {C.}~\bibnamefont {Daraio}},\ }\href {https://doi.org/10.1103/PhysRevLett.117.010602} {\bibfield  {journal} {\bibinfo  {journal} {Phys. Rev. Lett.}\ }\textbf {\bibinfo {volume} {117}},\ \bibinfo {pages} {010602} (\bibinfo {year} {2016})}\BibitemShut {NoStop}%
\bibitem [{\citenamefont {Yang}\ \emph {et~al.}(2020)\citenamefont {Yang}, \citenamefont {Wei}, \citenamefont {Sheng},\ and\ \citenamefont {Wu}}]{yang2020phonon}%
  \BibitemOpen
  \bibfield  {author} {\bibinfo {author} {\bibfnamefont {C.}~\bibnamefont {Yang}}, \bibinfo {author} {\bibfnamefont {X.}~\bibnamefont {Wei}}, \bibinfo {author} {\bibfnamefont {J.}~\bibnamefont {Sheng}},\ and\ \bibinfo {author} {\bibfnamefont {H.}~\bibnamefont {Wu}},\ }\href@noop {} {\bibfield  {journal} {\bibinfo  {journal} {Nature communications}\ }\textbf {\bibinfo {volume} {11}},\ \bibinfo {pages} {4656} (\bibinfo {year} {2020})}\BibitemShut {NoStop}%
\bibitem [{\citenamefont {Debiossac}\ \emph {et~al.}(2020)\citenamefont {Debiossac}, \citenamefont {Grass}, \citenamefont {Alonso}, \citenamefont {Lutz},\ and\ \citenamefont {Kiesel}}]{debiossac2020thermodynamics}%
  \BibitemOpen
  \bibfield  {author} {\bibinfo {author} {\bibfnamefont {M.}~\bibnamefont {Debiossac}}, \bibinfo {author} {\bibfnamefont {D.}~\bibnamefont {Grass}}, \bibinfo {author} {\bibfnamefont {J.~J.}\ \bibnamefont {Alonso}}, \bibinfo {author} {\bibfnamefont {E.}~\bibnamefont {Lutz}},\ and\ \bibinfo {author} {\bibfnamefont {N.}~\bibnamefont {Kiesel}},\ }\href@noop {} {\bibfield  {journal} {\bibinfo  {journal} {Nature communications}\ }\textbf {\bibinfo {volume} {11}},\ \bibinfo {pages} {1360} (\bibinfo {year} {2020})}\BibitemShut {NoStop}%
\bibitem [{\citenamefont {Militaru}\ \emph {et~al.}(2021)\citenamefont {Militaru}, \citenamefont {Lasanta}, \citenamefont {Frimmer}, \citenamefont {Bonilla}, \citenamefont {Novotny},\ and\ \citenamefont {Rica}}]{Militaru21}%
  \BibitemOpen
  \bibfield  {author} {\bibinfo {author} {\bibfnamefont {A.}~\bibnamefont {Militaru}}, \bibinfo {author} {\bibfnamefont {A.}~\bibnamefont {Lasanta}}, \bibinfo {author} {\bibfnamefont {M.}~\bibnamefont {Frimmer}}, \bibinfo {author} {\bibfnamefont {L.~L.}\ \bibnamefont {Bonilla}}, \bibinfo {author} {\bibfnamefont {L.}~\bibnamefont {Novotny}},\ and\ \bibinfo {author} {\bibfnamefont {R.~A.}\ \bibnamefont {Rica}},\ }\href {https://doi.org/10.1103/PhysRevLett.127.130603} {\bibfield  {journal} {\bibinfo  {journal} {Phys. Rev. Lett.}\ }\textbf {\bibinfo {volume} {127}},\ \bibinfo {pages} {130603} (\bibinfo {year} {2021})}\BibitemShut {NoStop}%
\bibitem [{\citenamefont {Gieseler}\ \emph {et~al.}(2021)\citenamefont {Gieseler}, \citenamefont {Gomez-Solano}, \citenamefont {Magazz{\`u}}, \citenamefont {P{\'e}rez~Castillo}, \citenamefont {P{\'e}rez~Garc{\'\i}a} \emph {et~al.}}]{gieseler2021optical}%
  \BibitemOpen
  \bibfield  {author} {\bibinfo {author} {\bibfnamefont {J.}~\bibnamefont {Gieseler}}, \bibinfo {author} {\bibfnamefont {J.~R.}\ \bibnamefont {Gomez-Solano}}, \bibinfo {author} {\bibfnamefont {A.}~\bibnamefont {Magazz{\`u}}}, \bibinfo {author} {\bibfnamefont {I.}~\bibnamefont {P{\'e}rez~Castillo}}, \bibinfo {author} {\bibfnamefont {L.}~\bibnamefont {P{\'e}rez~Garc{\'\i}a}}, \emph {et~al.},\ }\href@noop {} {\bibfield  {journal} {\bibinfo  {journal} {Advances in Optics and Photonics}\ }\textbf {\bibinfo {volume} {13}},\ \bibinfo {pages} {74} (\bibinfo {year} {2021})}\BibitemShut {NoStop}%
\bibitem [{\citenamefont {Trepagnier}\ \emph {et~al.}(2004)\citenamefont {Trepagnier}, \citenamefont {Jarzynski}, \citenamefont {Ritort}, \citenamefont {Crooks}, \citenamefont {Bustamante},\ and\ \citenamefont {Liphardt}}]{Trepagnier04}%
  \BibitemOpen
  \bibfield  {author} {\bibinfo {author} {\bibfnamefont {E.~H.}\ \bibnamefont {Trepagnier}}, \bibinfo {author} {\bibfnamefont {C.}~\bibnamefont {Jarzynski}}, \bibinfo {author} {\bibfnamefont {F.}~\bibnamefont {Ritort}}, \bibinfo {author} {\bibfnamefont {G.~E.}\ \bibnamefont {Crooks}}, \bibinfo {author} {\bibfnamefont {C.~J.}\ \bibnamefont {Bustamante}},\ and\ \bibinfo {author} {\bibfnamefont {J.}~\bibnamefont {Liphardt}},\ }\href {https://doi.org/10.1073/pnas.0406405101} {\bibfield  {journal} {\bibinfo  {journal} {Proceedings of the National Academy of Sciences}\ }\textbf {\bibinfo {volume} {101}},\ \bibinfo {pages} {15038} (\bibinfo {year} {2004})},\ \Eprint {https://arxiv.org/abs/https://www.pnas.org/doi/pdf/10.1073/pnas.0406405101} {https://www.pnas.org/doi/pdf/10.1073/pnas.0406405101} \BibitemShut {NoStop}%
\bibitem [{\citenamefont {Blickle}\ and\ \citenamefont {Bechinger}(2012)}]{blickle2012realization}%
  \BibitemOpen
  \bibfield  {author} {\bibinfo {author} {\bibfnamefont {V.}~\bibnamefont {Blickle}}\ and\ \bibinfo {author} {\bibfnamefont {C.}~\bibnamefont {Bechinger}},\ }\href@noop {} {\bibfield  {journal} {\bibinfo  {journal} {Nature Physics}\ }\textbf {\bibinfo {volume} {8}},\ \bibinfo {pages} {143} (\bibinfo {year} {2012})}\BibitemShut {NoStop}%
\bibitem [{\citenamefont {Rold{\'a}n}\ \emph {et~al.}(2015)\citenamefont {Rold{\'a}n}, \citenamefont {Neri}, \citenamefont {D{\"o}rpinghaus}, \citenamefont {Meyr},\ and\ \citenamefont {J{\"u}licher}}]{RoldanETAL:2015}%
  \BibitemOpen
  \bibfield  {author} {\bibinfo {author} {\bibfnamefont {{\'E}.}~\bibnamefont {Rold{\'a}n}}, \bibinfo {author} {\bibfnamefont {I.}~\bibnamefont {Neri}}, \bibinfo {author} {\bibfnamefont {M.}~\bibnamefont {D{\"o}rpinghaus}}, \bibinfo {author} {\bibfnamefont {H.}~\bibnamefont {Meyr}},\ and\ \bibinfo {author} {\bibfnamefont {F.}~\bibnamefont {J{\"u}licher}},\ }\href@noop {} {\bibfield  {journal} {\bibinfo  {journal} {Phys. Rev. Lett.}\ }\textbf {\bibinfo {volume} {115}},\ \bibinfo {pages} {250602} (\bibinfo {year} {2015})}\BibitemShut {NoStop}%
\bibitem [{\citenamefont {Proesmans}\ \emph {et~al.}(2016)\citenamefont {Proesmans}, \citenamefont {Dreher}, \citenamefont {Gavrilov}, \citenamefont {Bechhoefer},\ and\ \citenamefont {Van~den Broeck}}]{Proesmans16}%
  \BibitemOpen
  \bibfield  {author} {\bibinfo {author} {\bibfnamefont {K.}~\bibnamefont {Proesmans}}, \bibinfo {author} {\bibfnamefont {Y.}~\bibnamefont {Dreher}}, \bibinfo {author} {\bibfnamefont {M.}~\bibnamefont {Gavrilov}}, \bibinfo {author} {\bibfnamefont {J.}~\bibnamefont {Bechhoefer}},\ and\ \bibinfo {author} {\bibfnamefont {C.}~\bibnamefont {Van~den Broeck}},\ }\href {https://doi.org/10.1103/PhysRevX.6.041010} {\bibfield  {journal} {\bibinfo  {journal} {Phys. Rev. X}\ }\textbf {\bibinfo {volume} {6}},\ \bibinfo {pages} {041010} (\bibinfo {year} {2016})}\BibitemShut {NoStop}%
\bibitem [{\citenamefont {Ib{\'a}{\~n}ez}\ \emph {et~al.}(2024)\citenamefont {Ib{\'a}{\~n}ez}, \citenamefont {Dieball}, \citenamefont {Lasanta}, \citenamefont {Godec},\ and\ \citenamefont {Rica}}]{ibanez2024heating}%
  \BibitemOpen
  \bibfield  {author} {\bibinfo {author} {\bibfnamefont {M.}~\bibnamefont {Ib{\'a}{\~n}ez}}, \bibinfo {author} {\bibfnamefont {C.}~\bibnamefont {Dieball}}, \bibinfo {author} {\bibfnamefont {A.}~\bibnamefont {Lasanta}}, \bibinfo {author} {\bibfnamefont {A.}~\bibnamefont {Godec}},\ and\ \bibinfo {author} {\bibfnamefont {R.~A.}\ \bibnamefont {Rica}},\ }\href@noop {} {\bibfield  {journal} {\bibinfo  {journal} {Nature Physics}\ }\textbf {\bibinfo {volume} {20}},\ \bibinfo {pages} {135} (\bibinfo {year} {2024})}\BibitemShut {NoStop}%
\bibitem [{\citenamefont {Gupta}\ \emph {et~al.}(2011)\citenamefont {Gupta}, \citenamefont {Vincent}, \citenamefont {Neupane}, \citenamefont {Yu}, \citenamefont {Wang},\ and\ \citenamefont {Woodside}}]{gupta2011experimental}%
  \BibitemOpen
  \bibfield  {author} {\bibinfo {author} {\bibfnamefont {A.~N.}\ \bibnamefont {Gupta}}, \bibinfo {author} {\bibfnamefont {A.}~\bibnamefont {Vincent}}, \bibinfo {author} {\bibfnamefont {K.}~\bibnamefont {Neupane}}, \bibinfo {author} {\bibfnamefont {H.}~\bibnamefont {Yu}}, \bibinfo {author} {\bibfnamefont {F.}~\bibnamefont {Wang}},\ and\ \bibinfo {author} {\bibfnamefont {M.~T.}\ \bibnamefont {Woodside}},\ }\href@noop {} {\bibfield  {journal} {\bibinfo  {journal} {Nature Physics}\ }\textbf {\bibinfo {volume} {7}},\ \bibinfo {pages} {631} (\bibinfo {year} {2011})}\BibitemShut {NoStop}%
\bibitem [{\citenamefont {Astumian}(2010)}]{astumian2010thermodynamics}%
  \BibitemOpen
  \bibfield  {author} {\bibinfo {author} {\bibfnamefont {R.~D.}\ \bibnamefont {Astumian}},\ }\href@noop {} {\bibfield  {journal} {\bibinfo  {journal} {Biophysical journal}\ }\textbf {\bibinfo {volume} {98}},\ \bibinfo {pages} {2401} (\bibinfo {year} {2010})}\BibitemShut {NoStop}%
\bibitem [{\citenamefont {Hayashi}\ \emph {et~al.}(2010)\citenamefont {Hayashi}, \citenamefont {Ueno}, \citenamefont {Iino},\ and\ \citenamefont {Noji}}]{Hayashi10}%
  \BibitemOpen
  \bibfield  {author} {\bibinfo {author} {\bibfnamefont {K.}~\bibnamefont {Hayashi}}, \bibinfo {author} {\bibfnamefont {H.}~\bibnamefont {Ueno}}, \bibinfo {author} {\bibfnamefont {R.}~\bibnamefont {Iino}},\ and\ \bibinfo {author} {\bibfnamefont {H.}~\bibnamefont {Noji}},\ }\href {https://doi.org/10.1103/PhysRevLett.104.218103} {\bibfield  {journal} {\bibinfo  {journal} {Phys. Rev. Lett.}\ }\textbf {\bibinfo {volume} {104}},\ \bibinfo {pages} {218103} (\bibinfo {year} {2010})}\BibitemShut {NoStop}%
\bibitem [{\citenamefont {Toyabe}\ \emph {et~al.}(2011)\citenamefont {Toyabe}, \citenamefont {Watanabe-Nakayama}, \citenamefont {Okamoto}, \citenamefont {Kudo},\ and\ \citenamefont {Muneyuki}}]{toyabe2011thermodynamic}%
  \BibitemOpen
  \bibfield  {author} {\bibinfo {author} {\bibfnamefont {S.}~\bibnamefont {Toyabe}}, \bibinfo {author} {\bibfnamefont {T.}~\bibnamefont {Watanabe-Nakayama}}, \bibinfo {author} {\bibfnamefont {T.}~\bibnamefont {Okamoto}}, \bibinfo {author} {\bibfnamefont {S.}~\bibnamefont {Kudo}},\ and\ \bibinfo {author} {\bibfnamefont {E.}~\bibnamefont {Muneyuki}},\ }\href@noop {} {\bibfield  {journal} {\bibinfo  {journal} {Proceedings of the National Academy of Sciences}\ }\textbf {\bibinfo {volume} {108}},\ \bibinfo {pages} {17951} (\bibinfo {year} {2011})}\BibitemShut {NoStop}%
\bibitem [{\citenamefont {Ariga}\ \emph {et~al.}(2020)\citenamefont {Ariga}, \citenamefont {Tomishige},\ and\ \citenamefont {Mizuno}}]{ariga2020experimental}%
  \BibitemOpen
  \bibfield  {author} {\bibinfo {author} {\bibfnamefont {T.}~\bibnamefont {Ariga}}, \bibinfo {author} {\bibfnamefont {M.}~\bibnamefont {Tomishige}},\ and\ \bibinfo {author} {\bibfnamefont {D.}~\bibnamefont {Mizuno}},\ }\href@noop {} {\bibfield  {journal} {\bibinfo  {journal} {Biophysical reviews}\ }\textbf {\bibinfo {volume} {12}},\ \bibinfo {pages} {503} (\bibinfo {year} {2020})}\BibitemShut {NoStop}%
\bibitem [{\citenamefont {Leighton}\ and\ \citenamefont {Sivak}(2025)}]{leighton2025flow}%
  \BibitemOpen
  \bibfield  {author} {\bibinfo {author} {\bibfnamefont {M.~P.}\ \bibnamefont {Leighton}}\ and\ \bibinfo {author} {\bibfnamefont {D.~A.}\ \bibnamefont {Sivak}},\ }\href@noop {} {\bibfield  {journal} {\bibinfo  {journal} {Annual Review of Physical Chemistry}\ }\textbf {\bibinfo {volume} {76}},\ \bibinfo {pages} {379} (\bibinfo {year} {2025})}\BibitemShut {NoStop}%
\bibitem [{\citenamefont {Gaspard}(2004{\natexlab{a}})}]{Gaspard04}%
  \BibitemOpen
  \bibfield  {author} {\bibinfo {author} {\bibfnamefont {P.}~\bibnamefont {Gaspard}},\ }\href {https://doi.org/10.1063/1.1688758} {\bibfield  {journal} {\bibinfo  {journal} {The Journal of Chemical Physics}\ }\textbf {\bibinfo {volume} {120}},\ \bibinfo {pages} {8898} (\bibinfo {year} {2004}{\natexlab{a}})}\BibitemShut {NoStop}%
\bibitem [{\citenamefont {Schmiedl}\ and\ \citenamefont {Seifert}(2007)}]{Schmiedl07}%
  \BibitemOpen
  \bibfield  {author} {\bibinfo {author} {\bibfnamefont {T.}~\bibnamefont {Schmiedl}}\ and\ \bibinfo {author} {\bibfnamefont {U.}~\bibnamefont {Seifert}},\ }\href {https://doi.org/10.1063/1.2428297} {\bibfield  {journal} {\bibinfo  {journal} {The Journal of Chemical Physics}\ }\textbf {\bibinfo {volume} {126}},\ \bibinfo {pages} {044101} (\bibinfo {year} {2007})}\BibitemShut {NoStop}%
\bibitem [{\citenamefont {Rao}\ and\ \citenamefont {Esposito}(2016)}]{Rao16}%
  \BibitemOpen
  \bibfield  {author} {\bibinfo {author} {\bibfnamefont {R.}~\bibnamefont {Rao}}\ and\ \bibinfo {author} {\bibfnamefont {M.}~\bibnamefont {Esposito}},\ }\href {https://doi.org/10.1103/PhysRevX.6.041064} {\bibfield  {journal} {\bibinfo  {journal} {Phys. Rev. X}\ }\textbf {\bibinfo {volume} {6}},\ \bibinfo {pages} {041064} (\bibinfo {year} {2016})}\BibitemShut {NoStop}%
\bibitem [{\citenamefont {Freitas}\ \emph {et~al.}(2021)\citenamefont {Freitas}, \citenamefont {Delvenne},\ and\ \citenamefont {Esposito}}]{FreitasDelvenneEsposito2021Circuits}%
  \BibitemOpen
  \bibfield  {author} {\bibinfo {author} {\bibfnamefont {N.}~\bibnamefont {Freitas}}, \bibinfo {author} {\bibfnamefont {J.-C.}\ \bibnamefont {Delvenne}},\ and\ \bibinfo {author} {\bibfnamefont {M.}~\bibnamefont {Esposito}},\ }\href {https://doi.org/10.1103/physrevx.11.031064} {\bibfield  {journal} {\bibinfo  {journal} {Physical Review X}\ }\textbf {\bibinfo {volume} {11}},\ \bibinfo {pages} {031064} (\bibinfo {year} {2021})}\BibitemShut {NoStop}%
\bibitem [{\citenamefont {Freitas}\ \emph {et~al.}(2020)\citenamefont {Freitas}, \citenamefont {Delvenne},\ and\ \citenamefont {Esposito}}]{FreitasDelvenneEsposito2020RLC}%
  \BibitemOpen
  \bibfield  {author} {\bibinfo {author} {\bibfnamefont {N.}~\bibnamefont {Freitas}}, \bibinfo {author} {\bibfnamefont {J.-C.}\ \bibnamefont {Delvenne}},\ and\ \bibinfo {author} {\bibfnamefont {M.}~\bibnamefont {Esposito}},\ }\href {https://doi.org/10.1103/physrevx.10.031005} {\bibfield  {journal} {\bibinfo  {journal} {Physical Review X}\ }\textbf {\bibinfo {volume} {10}},\ \bibinfo {pages} {031005} (\bibinfo {year} {2020})}\BibitemShut {NoStop}%
\bibitem [{\citenamefont {Freitas}\ \emph {et~al.}(2026)\citenamefont {Freitas}, \citenamefont {Massarelli}, \citenamefont {Rothschild}, \citenamefont {Keane}, \citenamefont {Dawe}, \citenamefont {Hwang}, \citenamefont {Garlapati},\ and\ \citenamefont {McCourt}}]{FreitasEtAl2026CMOS}%
  \BibitemOpen
  \bibfield  {author} {\bibinfo {author} {\bibfnamefont {N.}~\bibnamefont {Freitas}}, \bibinfo {author} {\bibfnamefont {G.}~\bibnamefont {Massarelli}}, \bibinfo {author} {\bibfnamefont {J.}~\bibnamefont {Rothschild}}, \bibinfo {author} {\bibfnamefont {D.}~\bibnamefont {Keane}}, \bibinfo {author} {\bibfnamefont {E.}~\bibnamefont {Dawe}}, \bibinfo {author} {\bibfnamefont {S.}~\bibnamefont {Hwang}}, \bibinfo {author} {\bibfnamefont {A.}~\bibnamefont {Garlapati}},\ and\ \bibinfo {author} {\bibfnamefont {T.}~\bibnamefont {McCourt}},\ }\bibfield  {journal} {\bibinfo  {journal} {Physical Review Applied}\ }\textbf {\bibinfo {volume} {25}},\ \href {https://doi.org/10.1103/5l2z-3f1h} {10.1103/5l2z-3f1h} (\bibinfo {year} {2026})\BibitemShut {NoStop}%
\bibitem [{\citenamefont {Melanson}\ \emph {et~al.}(2025)\citenamefont {Melanson}, \citenamefont {Abu~Khater}, \citenamefont {Aifer}, \citenamefont {Donatella}, \citenamefont {Hunter~Gordon}, \citenamefont {Ahle}, \citenamefont {Crooks}, \citenamefont {Martinez}, \citenamefont {Sbahi},\ and\ \citenamefont {Coles}}]{MelansonEtAl2025ThermodynamicComputing}%
  \BibitemOpen
  \bibfield  {author} {\bibinfo {author} {\bibfnamefont {D.}~\bibnamefont {Melanson}}, \bibinfo {author} {\bibfnamefont {M.}~\bibnamefont {Abu~Khater}}, \bibinfo {author} {\bibfnamefont {M.}~\bibnamefont {Aifer}}, \bibinfo {author} {\bibfnamefont {K.}~\bibnamefont {Donatella}}, \bibinfo {author} {\bibfnamefont {M.}~\bibnamefont {Hunter~Gordon}}, \bibinfo {author} {\bibfnamefont {T.}~\bibnamefont {Ahle}}, \bibinfo {author} {\bibfnamefont {G.}~\bibnamefont {Crooks}}, \bibinfo {author} {\bibfnamefont {A.~J.}\ \bibnamefont {Martinez}}, \bibinfo {author} {\bibfnamefont {F.}~\bibnamefont {Sbahi}},\ and\ \bibinfo {author} {\bibfnamefont {P.~J.}\ \bibnamefont {Coles}},\ }\href {https://doi.org/10.1038/s41467-025-59011-x} {\bibfield  {journal} {\bibinfo  {journal} {Nature Communications}\ }\textbf {\bibinfo {volume} {16}},\ \bibinfo {pages} {3757} (\bibinfo {year} {2025})}\BibitemShut {NoStop}%
\bibitem [{\citenamefont {Lipka-Bartosik}\ \emph {et~al.}(2024)\citenamefont {Lipka-Bartosik}, \citenamefont {Perarnau-Llobet},\ and\ \citenamefont {Brunner}}]{Lipka24}%
  \BibitemOpen
  \bibfield  {author} {\bibinfo {author} {\bibfnamefont {P.}~\bibnamefont {Lipka-Bartosik}}, \bibinfo {author} {\bibfnamefont {M.}~\bibnamefont {Perarnau-Llobet}},\ and\ \bibinfo {author} {\bibfnamefont {N.}~\bibnamefont {Brunner}},\ }\href {https://doi.org/10.1126/sciadv.adm8792} {\bibfield  {journal} {\bibinfo  {journal} {Science Advances}\ }\textbf {\bibinfo {volume} {10}},\ \bibinfo {pages} {eadm8792} (\bibinfo {year} {2024})},\ \Eprint {https://arxiv.org/abs/https://www.science.org/doi/pdf/10.1126/sciadv.adm8792} {https://www.science.org/doi/pdf/10.1126/sciadv.adm8792} \BibitemShut {NoStop}%
\bibitem [{\citenamefont {Manzano}\ \emph {et~al.}(2024)\citenamefont {Manzano}, \citenamefont {Karde{\c{s}}}, \citenamefont {Rold{\'a}n},\ and\ \citenamefont {Wolpert}}]{manzano2024thermodynamics}%
  \BibitemOpen
  \bibfield  {author} {\bibinfo {author} {\bibfnamefont {G.}~\bibnamefont {Manzano}}, \bibinfo {author} {\bibfnamefont {G.}~\bibnamefont {Karde{\c{s}}}}, \bibinfo {author} {\bibfnamefont {{\'E}.}~\bibnamefont {Rold{\'a}n}},\ and\ \bibinfo {author} {\bibfnamefont {D.~H.}\ \bibnamefont {Wolpert}},\ }\href@noop {} {\bibfield  {journal} {\bibinfo  {journal} {Physical Review X}\ }\textbf {\bibinfo {volume} {14}},\ \bibinfo {pages} {021026} (\bibinfo {year} {2024})}\BibitemShut {NoStop}%
\bibitem [{\citenamefont {Wolpert}\ \emph {et~al.}(2024)\citenamefont {Wolpert}, \citenamefont {Korbel}, \citenamefont {Lynn}, \citenamefont {Tasnim}, \citenamefont {Grochow} \emph {et~al.}}]{Wolpert24}%
  \BibitemOpen
  \bibfield  {author} {\bibinfo {author} {\bibfnamefont {D.~H.}\ \bibnamefont {Wolpert}}, \bibinfo {author} {\bibfnamefont {J.}~\bibnamefont {Korbel}}, \bibinfo {author} {\bibfnamefont {C.~W.}\ \bibnamefont {Lynn}}, \bibinfo {author} {\bibfnamefont {F.}~\bibnamefont {Tasnim}}, \bibinfo {author} {\bibfnamefont {J.~A.}\ \bibnamefont {Grochow}}, \emph {et~al.},\ }\href {https://doi.org/10.1073/pnas.2321112121} {\bibfield  {journal} {\bibinfo  {journal} {Proceedings of the National Academy of Sciences}\ }\textbf {\bibinfo {volume} {121}},\ \bibinfo {pages} {e2321112121} (\bibinfo {year} {2024})},\ \Eprint {https://arxiv.org/abs/https://www.pnas.org/doi/pdf/10.1073/pnas.2321112121} {https://www.pnas.org/doi/pdf/10.1073/pnas.2321112121} \BibitemShut {NoStop}%
\bibitem [{\citenamefont {Seifert}(2016)}]{Seifert2016}%
  \BibitemOpen
  \bibfield  {author} {\bibinfo {author} {\bibfnamefont {U.}~\bibnamefont {Seifert}},\ }\href {https://doi.org/10.1103/PhysRevLett.116.020601} {\bibfield  {journal} {\bibinfo  {journal} {Physical Review Letters}\ }\textbf {\bibinfo {volume} {116}},\ \bibinfo {pages} {020601} (\bibinfo {year} {2016})}\BibitemShut {NoStop}%
\bibitem [{\citenamefont {Jarzynski}(2017)}]{Jarzynski2017}%
  \BibitemOpen
  \bibfield  {author} {\bibinfo {author} {\bibfnamefont {C.}~\bibnamefont {Jarzynski}},\ }\href {https://doi.org/10.1103/PhysRevX.7.011008} {\bibfield  {journal} {\bibinfo  {journal} {Physical Review X}\ }\textbf {\bibinfo {volume} {7}},\ \bibinfo {pages} {011008} (\bibinfo {year} {2017})}\BibitemShut {NoStop}%
\bibitem [{\citenamefont {Strasberg}\ and\ \citenamefont {Esposito}(2017)}]{Strasberg2017}%
  \BibitemOpen
  \bibfield  {author} {\bibinfo {author} {\bibfnamefont {P.}~\bibnamefont {Strasberg}}\ and\ \bibinfo {author} {\bibfnamefont {M.}~\bibnamefont {Esposito}},\ }\href {https://doi.org/10.1103/PhysRevE.95.062101} {\bibfield  {journal} {\bibinfo  {journal} {Physical Review E}\ }\textbf {\bibinfo {volume} {95}},\ \bibinfo {pages} {062101} (\bibinfo {year} {2017})}\BibitemShut {NoStop}%
\bibitem [{\citenamefont {Rold{\'a}n}(2014)}]{roldan2014irreversibility}%
  \BibitemOpen
  \bibfield  {author} {\bibinfo {author} {\bibfnamefont {{\'E}.}~\bibnamefont {Rold{\'a}n}},\ }\href@noop {} {\emph {\bibinfo {title} {Irreversibility and dissipation in microscopic systems}}}\ (\bibinfo  {publisher} {Springer},\ \bibinfo {year} {2014})\BibitemShut {NoStop}%
\bibitem [{\citenamefont {Skinner}\ and\ \citenamefont {Dunkel}(2021)}]{SkinnerDunkel2021}%
  \BibitemOpen
  \bibfield  {author} {\bibinfo {author} {\bibfnamefont {D.~J.}\ \bibnamefont {Skinner}}\ and\ \bibinfo {author} {\bibfnamefont {J.}~\bibnamefont {Dunkel}},\ }\href {https://doi.org/10.1103/PhysRevLett.127.198101} {\bibfield  {journal} {\bibinfo  {journal} {Physical Review Letters}\ }\textbf {\bibinfo {volume} {127}},\ \bibinfo {pages} {198101} (\bibinfo {year} {2021})}\BibitemShut {NoStop}%
\bibitem [{\citenamefont {Seifert}(2019)}]{seifert2019stochastic}%
  \BibitemOpen
  \bibfield  {author} {\bibinfo {author} {\bibfnamefont {U.}~\bibnamefont {Seifert}},\ }\href@noop {} {\bibfield  {journal} {\bibinfo  {journal} {Annual Review of Condensed Matter Physics}\ }\textbf {\bibinfo {volume} {10}},\ \bibinfo {pages} {171} (\bibinfo {year} {2019})}\BibitemShut {NoStop}%
\bibitem [{\citenamefont {Manikandan}\ \emph {et~al.}(2020)\citenamefont {Manikandan}, \citenamefont {Gupta},\ and\ \citenamefont {Krishnamurthy}}]{Manikandan2020}%
  \BibitemOpen
  \bibfield  {author} {\bibinfo {author} {\bibfnamefont {S.~K.}\ \bibnamefont {Manikandan}}, \bibinfo {author} {\bibfnamefont {D.}~\bibnamefont {Gupta}},\ and\ \bibinfo {author} {\bibfnamefont {S.}~\bibnamefont {Krishnamurthy}},\ }\href {https://doi.org/10.1103/PhysRevLett.124.120603} {\bibfield  {journal} {\bibinfo  {journal} {Phys. Rev. Lett.}\ }\textbf {\bibinfo {volume} {124}},\ \bibinfo {pages} {120603} (\bibinfo {year} {2020})}\BibitemShut {NoStop}%
\bibitem [{\citenamefont {Pigolotti}\ and\ \citenamefont {Peliti}(2021)}]{PigolottiPeliti2021}%
  \BibitemOpen
  \bibfield  {author} {\bibinfo {author} {\bibfnamefont {S.}~\bibnamefont {Pigolotti}}\ and\ \bibinfo {author} {\bibfnamefont {L.}~\bibnamefont {Peliti}},\ }\href@noop {} {\emph {\bibinfo {title} {Stochastic Thermodynamics: An Introduction}}}\ (\bibinfo  {publisher} {Princeton University Press},\ \bibinfo {year} {2021})\BibitemShut {NoStop}%
\bibitem [{\citenamefont {Shiraishi}(2023)}]{Shiraishi2023}%
  \BibitemOpen
  \bibfield  {author} {\bibinfo {author} {\bibfnamefont {N.}~\bibnamefont {Shiraishi}},\ }\href@noop {} {\emph {\bibinfo {title} {An Introduction to Stochastic Thermodynamics}}}\ (\bibinfo  {publisher} {Springer},\ \bibinfo {year} {2023})\BibitemShut {NoStop}%
\bibitem [{\citenamefont {Seifert}(2025{\natexlab{a}})}]{Seifert2025}%
  \BibitemOpen
  \bibfield  {author} {\bibinfo {author} {\bibfnamefont {U.}~\bibnamefont {Seifert}},\ }\href@noop {} {\emph {\bibinfo {title} {Stochastic Thermodynamics}}}\ (\bibinfo  {publisher} {Cambridge University Press},\ \bibinfo {year} {2025})\BibitemShut {NoStop}%
\bibitem [{\citenamefont {Campbell}\ \emph {et~al.}(2026)\citenamefont {Campbell}, \citenamefont {D’Amico}, \citenamefont {Ciampini}, \citenamefont {Anders}, \citenamefont {Ares}, \citenamefont {Artini} \emph {et~al.}}]{Campbell2026}%
  \BibitemOpen
  \bibfield  {author} {\bibinfo {author} {\bibfnamefont {S.}~\bibnamefont {Campbell}}, \bibinfo {author} {\bibfnamefont {I.}~\bibnamefont {D’Amico}}, \bibinfo {author} {\bibfnamefont {M.~A.}\ \bibnamefont {Ciampini}}, \bibinfo {author} {\bibfnamefont {J.}~\bibnamefont {Anders}}, \bibinfo {author} {\bibfnamefont {N.}~\bibnamefont {Ares}}, \bibinfo {author} {\bibnamefont {Artini}}, \emph {et~al.},\ }\href {https://doi.org/10.1088/2058-9565/ae1e27} {\bibfield  {journal} {\bibinfo  {journal} {Quantum Science and Technology}\ }\textbf {\bibinfo {volume} {11}},\ \bibinfo {pages} {012501} (\bibinfo {year} {2026})}\BibitemShut {NoStop}%
\bibitem [{\citenamefont {Deffner}\ and\ \citenamefont {Campbell}(2019)}]{Deffner2019}%
  \BibitemOpen
  \bibfield  {author} {\bibinfo {author} {\bibfnamefont {S.}~\bibnamefont {Deffner}}\ and\ \bibinfo {author} {\bibfnamefont {S.}~\bibnamefont {Campbell}},\ }\href {https://doi.org/10.1088/2053-2571/ab21c6} {\emph {\bibinfo {title} {Quantum Thermodynamics}}},\ 2053-2571\ (\bibinfo  {publisher} {Morgan \& Claypool Publishers},\ \bibinfo {year} {2019})\BibitemShut {NoStop}%
\bibitem [{\citenamefont {Strasberg}(2022)}]{Strasberg2022}%
  \BibitemOpen
  \bibfield  {author} {\bibinfo {author} {\bibfnamefont {P.}~\bibnamefont {Strasberg}},\ }\href@noop {} {\emph {\bibinfo {title} {Quantum Stochastic Thermodynamics}}}\ (\bibinfo  {publisher} {Oxford University Press},\ \bibinfo {year} {2022})\BibitemShut {NoStop}%
\bibitem [{\citenamefont {Hartich}\ \emph {et~al.}(2014)\citenamefont {Hartich}, \citenamefont {Barato},\ and\ \citenamefont {Seifert}}]{hartich2014stochastic}%
  \BibitemOpen
  \bibfield  {author} {\bibinfo {author} {\bibfnamefont {D.}~\bibnamefont {Hartich}}, \bibinfo {author} {\bibfnamefont {A.~C.}\ \bibnamefont {Barato}},\ and\ \bibinfo {author} {\bibfnamefont {U.}~\bibnamefont {Seifert}},\ }\href@noop {} {\bibfield  {journal} {\bibinfo  {journal} {Journal of Statistical Mechanics: Theory and Experiment}\ }\textbf {\bibinfo {volume} {2014}},\ \bibinfo {pages} {P02016} (\bibinfo {year} {2014})}\BibitemShut {NoStop}%
\bibitem [{\citenamefont {Arias-Gonzalez}(2021)}]{arias2021microscopically}%
  \BibitemOpen
  \bibfield  {author} {\bibinfo {author} {\bibfnamefont {J.~R.}\ \bibnamefont {Arias-Gonzalez}},\ }\href@noop {} {\bibfield  {journal} {\bibinfo  {journal} {Mathematics}\ }\textbf {\bibinfo {volume} {9}},\ \bibinfo {pages} {127} (\bibinfo {year} {2021})}\BibitemShut {NoStop}%
\bibitem [{\citenamefont {Lacasa}\ \emph {et~al.}(2012)\citenamefont {Lacasa}, \citenamefont {Nunez}, \citenamefont {Rold{\'a}n}, \citenamefont {Parrondo},\ and\ \citenamefont {Luque}}]{lacasa2012time}%
  \BibitemOpen
  \bibfield  {author} {\bibinfo {author} {\bibfnamefont {L.}~\bibnamefont {Lacasa}}, \bibinfo {author} {\bibfnamefont {A.}~\bibnamefont {Nunez}}, \bibinfo {author} {\bibfnamefont {{\'E}.}~\bibnamefont {Rold{\'a}n}}, \bibinfo {author} {\bibfnamefont {J.~M.}\ \bibnamefont {Parrondo}},\ and\ \bibinfo {author} {\bibfnamefont {B.}~\bibnamefont {Luque}},\ }\href@noop {} {\bibfield  {journal} {\bibinfo  {journal} {The European Physical Journal B}\ }\textbf {\bibinfo {volume} {85}},\ \bibinfo {pages} {217} (\bibinfo {year} {2012})}\BibitemShut {NoStop}%
\bibitem [{\citenamefont {Harunari}\ \emph {et~al.}(2022)\citenamefont {Harunari}, \citenamefont {Dutta}, \citenamefont {Polettini},\ and\ \citenamefont {Rold{\'a}n}}]{harunari2022learn}%
  \BibitemOpen
  \bibfield  {author} {\bibinfo {author} {\bibfnamefont {P.~E.}\ \bibnamefont {Harunari}}, \bibinfo {author} {\bibfnamefont {A.}~\bibnamefont {Dutta}}, \bibinfo {author} {\bibfnamefont {M.}~\bibnamefont {Polettini}},\ and\ \bibinfo {author} {\bibfnamefont {{\'E}.}~\bibnamefont {Rold{\'a}n}},\ }\href@noop {} {\bibfield  {journal} {\bibinfo  {journal} {Physical Review X}\ }\textbf {\bibinfo {volume} {12}},\ \bibinfo {pages} {041026} (\bibinfo {year} {2022})}\BibitemShut {NoStop}%
\bibitem [{\citenamefont {Van~der Meer}\ \emph {et~al.}(2022)\citenamefont {Van~der Meer}, \citenamefont {Ertel},\ and\ \citenamefont {Seifert}}]{van2022thermodynamic}%
  \BibitemOpen
  \bibfield  {author} {\bibinfo {author} {\bibfnamefont {J.}~\bibnamefont {Van~der Meer}}, \bibinfo {author} {\bibfnamefont {B.}~\bibnamefont {Ertel}},\ and\ \bibinfo {author} {\bibfnamefont {U.}~\bibnamefont {Seifert}},\ }\href@noop {} {\bibfield  {journal} {\bibinfo  {journal} {Physical Review X}\ }\textbf {\bibinfo {volume} {12}},\ \bibinfo {pages} {031025} (\bibinfo {year} {2022})}\BibitemShut {NoStop}%
\bibitem [{\citenamefont {Ferri-Cort\'es}\ \emph {et~al.}(2025)\citenamefont {Ferri-Cort\'es}, \citenamefont {Almanza-Marrero}, \citenamefont {L\'opez}, \citenamefont {Zambrini},\ and\ \citenamefont {Manzano}}]{ferri2025conditional}%
  \BibitemOpen
  \bibfield  {author} {\bibinfo {author} {\bibfnamefont {M.}~\bibnamefont {Ferri-Cort\'es}}, \bibinfo {author} {\bibfnamefont {J.~A.}\ \bibnamefont {Almanza-Marrero}}, \bibinfo {author} {\bibfnamefont {R.}~\bibnamefont {L\'opez}}, \bibinfo {author} {\bibfnamefont {R.}~\bibnamefont {Zambrini}},\ and\ \bibinfo {author} {\bibfnamefont {G.}~\bibnamefont {Manzano}},\ }\href {https://doi.org/10.1103/PhysRevResearch.7.013077} {\bibfield  {journal} {\bibinfo  {journal} {Phys. Rev. Res.}\ }\textbf {\bibinfo {volume} {7}},\ \bibinfo {pages} {013077} (\bibinfo {year} {2025})}\BibitemShut {NoStop}%
\bibitem [{\citenamefont {Nogare}\ and\ \citenamefont {Fulcher}(2025)}]{nogare2025identifying}%
  \BibitemOpen
  \bibfield  {author} {\bibinfo {author} {\bibfnamefont {T.~D.}\ \bibnamefont {Nogare}}\ and\ \bibinfo {author} {\bibfnamefont {B.~D.}\ \bibnamefont {Fulcher}},\ }\href@noop {} {\bibfield  {journal} {\bibinfo  {journal} {arXiv preprint arXiv:2511.15991}\ } (\bibinfo {year} {2025})}\BibitemShut {NoStop}%
\bibitem [{\citenamefont {Aguilera}\ \emph {et~al.}(2026)\citenamefont {Aguilera}, \citenamefont {Ito},\ and\ \citenamefont {Kolchinsky}}]{aguilera2026inferring}%
  \BibitemOpen
  \bibfield  {author} {\bibinfo {author} {\bibfnamefont {M.}~\bibnamefont {Aguilera}}, \bibinfo {author} {\bibfnamefont {S.}~\bibnamefont {Ito}},\ and\ \bibinfo {author} {\bibfnamefont {A.}~\bibnamefont {Kolchinsky}},\ }\href@noop {} {\bibfield  {journal} {\bibinfo  {journal} {Physical Review Letters}\ }\textbf {\bibinfo {volume} {136}},\ \bibinfo {pages} {077101} (\bibinfo {year} {2026})}\BibitemShut {NoStop}%
\bibitem [{\citenamefont {Ehrich}(2021)}]{ehrich2021tightest}%
  \BibitemOpen
  \bibfield  {author} {\bibinfo {author} {\bibfnamefont {J.}~\bibnamefont {Ehrich}},\ }\href@noop {} {\bibfield  {journal} {\bibinfo  {journal} {Journal of Statistical Mechanics: Theory and Experiment}\ }\textbf {\bibinfo {volume} {2021}},\ \bibinfo {pages} {083214} (\bibinfo {year} {2021})}\BibitemShut {NoStop}%
\bibitem [{\citenamefont {Kawaguchi}\ and\ \citenamefont {Nakayama}(2013)}]{kawaguchi2013fluctuation}%
  \BibitemOpen
  \bibfield  {author} {\bibinfo {author} {\bibfnamefont {K.}~\bibnamefont {Kawaguchi}}\ and\ \bibinfo {author} {\bibfnamefont {Y.}~\bibnamefont {Nakayama}},\ }\href@noop {} {\bibfield  {journal} {\bibinfo  {journal} {Physical Review E—Statistical, Nonlinear, and Soft Matter Physics}\ }\textbf {\bibinfo {volume} {88}},\ \bibinfo {pages} {022147} (\bibinfo {year} {2013})}\BibitemShut {NoStop}%
\bibitem [{\citenamefont {Dechant}\ and\ \citenamefont {Sasa}(2021)}]{dechant2021improving}%
  \BibitemOpen
  \bibfield  {author} {\bibinfo {author} {\bibfnamefont {A.}~\bibnamefont {Dechant}}\ and\ \bibinfo {author} {\bibfnamefont {S.-i.}\ \bibnamefont {Sasa}},\ }\href@noop {} {\bibfield  {journal} {\bibinfo  {journal} {Physical Review X}\ }\textbf {\bibinfo {volume} {11}},\ \bibinfo {pages} {041061} (\bibinfo {year} {2021})}\BibitemShut {NoStop}%
\bibitem [{\citenamefont {Deg{\"u}nther}\ \emph {et~al.}(2024)\citenamefont {Deg{\"u}nther}, \citenamefont {Van Der~Meer},\ and\ \citenamefont {Seifert}}]{degunther2024fluctuating}%
  \BibitemOpen
  \bibfield  {author} {\bibinfo {author} {\bibfnamefont {J.}~\bibnamefont {Deg{\"u}nther}}, \bibinfo {author} {\bibfnamefont {J.}~\bibnamefont {Van Der~Meer}},\ and\ \bibinfo {author} {\bibfnamefont {U.}~\bibnamefont {Seifert}},\ }\href@noop {} {\bibfield  {journal} {\bibinfo  {journal} {Physical Review Research}\ }\textbf {\bibinfo {volume} {6}},\ \bibinfo {pages} {023175} (\bibinfo {year} {2024})}\BibitemShut {NoStop}%
\bibitem [{\citenamefont {Esposito}(2012)}]{esposito2012stochastic}%
  \BibitemOpen
  \bibfield  {author} {\bibinfo {author} {\bibfnamefont {M.}~\bibnamefont {Esposito}},\ }\href@noop {} {\bibfield  {journal} {\bibinfo  {journal} {Physical Review E—Statistical, Nonlinear, and Soft Matter Physics}\ }\textbf {\bibinfo {volume} {85}},\ \bibinfo {pages} {041125} (\bibinfo {year} {2012})}\BibitemShut {NoStop}%
\bibitem [{\citenamefont {Fodor}\ \emph {et~al.}(2016)\citenamefont {Fodor}, \citenamefont {Nardini}, \citenamefont {Cates}, \citenamefont {Tailleur}, \citenamefont {Visco},\ and\ \citenamefont {Van~Wijland}}]{FodorETAL:2016}%
  \BibitemOpen
  \bibfield  {author} {\bibinfo {author} {\bibfnamefont {{\'E}.}~\bibnamefont {Fodor}}, \bibinfo {author} {\bibfnamefont {C.}~\bibnamefont {Nardini}}, \bibinfo {author} {\bibfnamefont {M.~E.}\ \bibnamefont {Cates}}, \bibinfo {author} {\bibfnamefont {J.}~\bibnamefont {Tailleur}}, \bibinfo {author} {\bibfnamefont {P.}~\bibnamefont {Visco}},\ and\ \bibinfo {author} {\bibfnamefont {F.}~\bibnamefont {Van~Wijland}},\ }\href@noop {} {\bibfield  {journal} {\bibinfo  {journal} {Phys.~Rev.~Lett.}\ }\textbf {\bibinfo {volume} {117}},\ \bibinfo {pages} {038103} (\bibinfo {year} {2016})}\BibitemShut {NoStop}%
\bibitem [{\citenamefont {Shankar}\ and\ \citenamefont {Marchetti}(2018)}]{shankar2018hidden}%
  \BibitemOpen
  \bibfield  {author} {\bibinfo {author} {\bibfnamefont {S.}~\bibnamefont {Shankar}}\ and\ \bibinfo {author} {\bibfnamefont {M.~C.}\ \bibnamefont {Marchetti}},\ }\href@noop {} {\bibfield  {journal} {\bibinfo  {journal} {Physical Review E}\ }\textbf {\bibinfo {volume} {98}},\ \bibinfo {pages} {020604} (\bibinfo {year} {2018})}\BibitemShut {NoStop}%
\bibitem [{\citenamefont {Mandal}\ \emph {et~al.}(2017)\citenamefont {Mandal}, \citenamefont {Klymko},\ and\ \citenamefont {DeWeese}}]{mandal2017entropy}%
  \BibitemOpen
  \bibfield  {author} {\bibinfo {author} {\bibfnamefont {D.}~\bibnamefont {Mandal}}, \bibinfo {author} {\bibfnamefont {K.}~\bibnamefont {Klymko}},\ and\ \bibinfo {author} {\bibfnamefont {M.~R.}\ \bibnamefont {DeWeese}},\ }\href@noop {} {\bibfield  {journal} {\bibinfo  {journal} {Physical review letters}\ }\textbf {\bibinfo {volume} {119}},\ \bibinfo {pages} {258001} (\bibinfo {year} {2017})}\BibitemShut {NoStop}%
\bibitem [{\citenamefont {Speck}(2016)}]{speck2016stochastic}%
  \BibitemOpen
  \bibfield  {author} {\bibinfo {author} {\bibfnamefont {T.}~\bibnamefont {Speck}},\ }\href@noop {} {\bibfield  {journal} {\bibinfo  {journal} {Europhysics Letters}\ }\textbf {\bibinfo {volume} {114}},\ \bibinfo {pages} {30006} (\bibinfo {year} {2016})}\BibitemShut {NoStop}%
\bibitem [{\citenamefont {Dabelow}\ \emph {et~al.}(2019)\citenamefont {Dabelow}, \citenamefont {Bo},\ and\ \citenamefont {Eichhorn}}]{Dabelow19}%
  \BibitemOpen
  \bibfield  {author} {\bibinfo {author} {\bibfnamefont {L.}~\bibnamefont {Dabelow}}, \bibinfo {author} {\bibfnamefont {S.}~\bibnamefont {Bo}},\ and\ \bibinfo {author} {\bibfnamefont {R.}~\bibnamefont {Eichhorn}},\ }\href {https://doi.org/10.1103/PhysRevX.9.021009} {\bibfield  {journal} {\bibinfo  {journal} {Phys. Rev. X}\ }\textbf {\bibinfo {volume} {9}},\ \bibinfo {pages} {021009} (\bibinfo {year} {2019})}\BibitemShut {NoStop}%
\bibitem [{\citenamefont {Garcia-Millan}\ \emph {et~al.}(2025)\citenamefont {Garcia-Millan}, \citenamefont {Sch{\"u}ttler}, \citenamefont {Cates},\ and\ \citenamefont {Loos}}]{garcia2025optimal}%
  \BibitemOpen
  \bibfield  {author} {\bibinfo {author} {\bibfnamefont {R.}~\bibnamefont {Garcia-Millan}}, \bibinfo {author} {\bibfnamefont {J.}~\bibnamefont {Sch{\"u}ttler}}, \bibinfo {author} {\bibfnamefont {M.~E.}\ \bibnamefont {Cates}},\ and\ \bibinfo {author} {\bibfnamefont {S.~A.}\ \bibnamefont {Loos}},\ }\href@noop {} {\bibfield  {journal} {\bibinfo  {journal} {Physical Review Letters}\ }\textbf {\bibinfo {volume} {135}},\ \bibinfo {pages} {088301} (\bibinfo {year} {2025})}\BibitemShut {NoStop}%
\bibitem [{\citenamefont {Loos}\ \emph {et~al.}(2021)\citenamefont {Loos}, \citenamefont {Hermann},\ and\ \citenamefont {Klapp}}]{loos2021medium}%
  \BibitemOpen
  \bibfield  {author} {\bibinfo {author} {\bibfnamefont {S.~A.}\ \bibnamefont {Loos}}, \bibinfo {author} {\bibfnamefont {S.}~\bibnamefont {Hermann}},\ and\ \bibinfo {author} {\bibfnamefont {S.~H.}\ \bibnamefont {Klapp}},\ }\href@noop {} {\bibfield  {journal} {\bibinfo  {journal} {Entropy}\ }\textbf {\bibinfo {volume} {23}},\ \bibinfo {pages} {696} (\bibinfo {year} {2021})}\BibitemShut {NoStop}%
\bibitem [{\citenamefont {Loos}\ \emph {et~al.}(2024)\citenamefont {Loos}, \citenamefont {Monter}, \citenamefont {Ginot},\ and\ \citenamefont {Bechinger}}]{loos2024universal}%
  \BibitemOpen
  \bibfield  {author} {\bibinfo {author} {\bibfnamefont {S.~A.}\ \bibnamefont {Loos}}, \bibinfo {author} {\bibfnamefont {S.}~\bibnamefont {Monter}}, \bibinfo {author} {\bibfnamefont {F.}~\bibnamefont {Ginot}},\ and\ \bibinfo {author} {\bibfnamefont {C.}~\bibnamefont {Bechinger}},\ }\href@noop {} {\bibfield  {journal} {\bibinfo  {journal} {Physical Review X}\ }\textbf {\bibinfo {volume} {14}},\ \bibinfo {pages} {021032} (\bibinfo {year} {2024})}\BibitemShut {NoStop}%
\bibitem [{\citenamefont {Alvarado}\ \emph {et~al.}(2026)\citenamefont {Alvarado}, \citenamefont {Teich}, \citenamefont {Sivak},\ and\ \citenamefont {Bechhoefer}}]{alvarado2026optimal}%
  \BibitemOpen
  \bibfield  {author} {\bibinfo {author} {\bibfnamefont {J.}~\bibnamefont {Alvarado}}, \bibinfo {author} {\bibfnamefont {E.~G.}\ \bibnamefont {Teich}}, \bibinfo {author} {\bibfnamefont {D.~A.}\ \bibnamefont {Sivak}},\ and\ \bibinfo {author} {\bibfnamefont {J.}~\bibnamefont {Bechhoefer}},\ }\href@noop {} {\bibfield  {journal} {\bibinfo  {journal} {Annual Review of Condensed Matter Physics}\ }\textbf {\bibinfo {volume} {17}} (\bibinfo {year} {2026})}\BibitemShut {NoStop}%
\bibitem [{\citenamefont {Munakata}\ and\ \citenamefont {Rosinberg}(2014)}]{munakata2014entropy}%
  \BibitemOpen
  \bibfield  {author} {\bibinfo {author} {\bibfnamefont {T.}~\bibnamefont {Munakata}}\ and\ \bibinfo {author} {\bibfnamefont {M.}~\bibnamefont {Rosinberg}},\ }\href@noop {} {\bibfield  {journal} {\bibinfo  {journal} {Physical review letters}\ }\textbf {\bibinfo {volume} {112}},\ \bibinfo {pages} {180601} (\bibinfo {year} {2014})}\BibitemShut {NoStop}%
\bibitem [{\citenamefont {Rosinberg}\ \emph {et~al.}(2015)\citenamefont {Rosinberg}, \citenamefont {Munakata},\ and\ \citenamefont {Tarjus}}]{rosinberg2015stochastic}%
  \BibitemOpen
  \bibfield  {author} {\bibinfo {author} {\bibfnamefont {M.}~\bibnamefont {Rosinberg}}, \bibinfo {author} {\bibfnamefont {T.}~\bibnamefont {Munakata}},\ and\ \bibinfo {author} {\bibfnamefont {G.}~\bibnamefont {Tarjus}},\ }\href@noop {} {\bibfield  {journal} {\bibinfo  {journal} {Physical Review E}\ }\textbf {\bibinfo {volume} {91}},\ \bibinfo {pages} {042114} (\bibinfo {year} {2015})}\BibitemShut {NoStop}%
\bibitem [{\citenamefont {Kanazawa}\ and\ \citenamefont {Dechant}(2025)}]{kanazawa2025stochastic}%
  \BibitemOpen
  \bibfield  {author} {\bibinfo {author} {\bibfnamefont {K.}~\bibnamefont {Kanazawa}}\ and\ \bibinfo {author} {\bibfnamefont {A.}~\bibnamefont {Dechant}},\ }\href@noop {} {\bibfield  {journal} {\bibinfo  {journal} {arXiv preprint arXiv:2506.04726}\ } (\bibinfo {year} {2025})}\BibitemShut {NoStop}%
\bibitem [{\citenamefont {Muruga}\ \emph {et~al.}(2026)\citenamefont {Muruga}, \citenamefont {Ginot}, \citenamefont {Loos},\ and\ \citenamefont {Bechinger}}]{muruga2026extracting}%
  \BibitemOpen
  \bibfield  {author} {\bibinfo {author} {\bibfnamefont {L.}~\bibnamefont {Muruga}}, \bibinfo {author} {\bibfnamefont {F.}~\bibnamefont {Ginot}}, \bibinfo {author} {\bibfnamefont {S.~A.}\ \bibnamefont {Loos}},\ and\ \bibinfo {author} {\bibfnamefont {C.}~\bibnamefont {Bechinger}},\ }\href@noop {} {\bibfield  {journal} {\bibinfo  {journal} {arXiv preprint arXiv:2603.06160}\ } (\bibinfo {year} {2026})}\BibitemShut {NoStop}%
\bibitem [{\citenamefont {Archambault}\ \emph {et~al.}(2024)\citenamefont {Archambault}, \citenamefont {Crauste-Thibierge}, \citenamefont {Ciliberto},\ and\ \citenamefont {Bellon}}]{ArchambaultEtAl2024EPL}%
  \BibitemOpen
  \bibfield  {author} {\bibinfo {author} {\bibfnamefont {A.}~\bibnamefont {Archambault}}, \bibinfo {author} {\bibfnamefont {C.}~\bibnamefont {Crauste-Thibierge}}, \bibinfo {author} {\bibfnamefont {S.}~\bibnamefont {Ciliberto}},\ and\ \bibinfo {author} {\bibfnamefont {L.}~\bibnamefont {Bellon}},\ }\href {https://doi.org/10.1209/0295-5075/ad8bf0} {\bibfield  {journal} {\bibinfo  {journal} {Europhysics Letters}\ }\textbf {\bibinfo {volume} {148}},\ \bibinfo {pages} {41002} (\bibinfo {year} {2024})}\BibitemShut {NoStop}%
\bibitem [{\citenamefont {Archambault}\ \emph {et~al.}(2025)\citenamefont {Archambault}, \citenamefont {Crauste-Thibierge}, \citenamefont {Imparato}, \citenamefont {Jarzynski},\ and\ \citenamefont {Ciliberto}}]{ArchambaultEtAl2025PRL}%
  \BibitemOpen
  \bibfield  {author} {\bibinfo {author} {\bibfnamefont {A.}~\bibnamefont {Archambault}}, \bibinfo {author} {\bibfnamefont {C.}~\bibnamefont {Crauste-Thibierge}}, \bibinfo {author} {\bibfnamefont {A.}~\bibnamefont {Imparato}}, \bibinfo {author} {\bibfnamefont {C.}~\bibnamefont {Jarzynski}},\ and\ \bibinfo {author} {\bibfnamefont {S.}~\bibnamefont {Ciliberto}},\ }\href {https://doi.org/10.1103/s9kj-lczm} {\bibfield  {journal} {\bibinfo  {journal} {Physical Review Letters}\ }\textbf {\bibinfo {volume} {135}},\ \bibinfo {pages} {147101} (\bibinfo {year} {2025})}\BibitemShut {NoStop}%
\bibitem [{\citenamefont {Fily}\ and\ \citenamefont {Marchetti}(2012)}]{FilyMarchetti:2012}%
  \BibitemOpen
  \bibfield  {author} {\bibinfo {author} {\bibfnamefont {Y.}~\bibnamefont {Fily}}\ and\ \bibinfo {author} {\bibfnamefont {M.~C.}\ \bibnamefont {Marchetti}},\ }\href@noop {} {\bibfield  {journal} {\bibinfo  {journal} {Phys. Rev. Lett.}\ }\textbf {\bibinfo {volume} {108}},\ \bibinfo {pages} {235702} (\bibinfo {year} {2012})}\BibitemShut {NoStop}%
\bibitem [{\citenamefont {Thipmaungprom}\ \emph {et~al.}(2026)\citenamefont {Thipmaungprom}, \citenamefont {Saliekh}, \citenamefont {Alonso}, \citenamefont {Rold{\'a}n}, \citenamefont {Berger},\ and\ \citenamefont {Belousov}}]{thipmaungprom2026thermodynamic}%
  \BibitemOpen
  \bibfield  {author} {\bibinfo {author} {\bibfnamefont {Y.}~\bibnamefont {Thipmaungprom}}, \bibinfo {author} {\bibfnamefont {L.}~\bibnamefont {Saliekh}}, \bibinfo {author} {\bibfnamefont {R.}~\bibnamefont {Alonso}}, \bibinfo {author} {\bibfnamefont {{\'E}.}~\bibnamefont {Rold{\'a}n}}, \bibinfo {author} {\bibfnamefont {F.}~\bibnamefont {Berger}},\ and\ \bibinfo {author} {\bibfnamefont {R.}~\bibnamefont {Belousov}},\ }\href@noop {} {\bibfield  {journal} {\bibinfo  {journal} {PRX Life}\ }\textbf {\bibinfo {volume} {4}},\ \bibinfo {pages} {013039} (\bibinfo {year} {2026})}\BibitemShut {NoStop}%
\bibitem [{\citenamefont {Cates}(2012)}]{Cates:2012}%
  \BibitemOpen
  \bibfield  {author} {\bibinfo {author} {\bibfnamefont {M.~E.}\ \bibnamefont {Cates}},\ }\href@noop {} {\bibfield  {journal} {\bibinfo  {journal} {Rep. Progr. Phys.}\ }\textbf {\bibinfo {volume} {75}},\ \bibinfo {pages} {042601} (\bibinfo {year} {2012})}\BibitemShut {NoStop}%
\bibitem [{\citenamefont {Rold{\'a}n}\ \emph {et~al.}(2021)\citenamefont {Rold{\'a}n}, \citenamefont {Barral}, \citenamefont {Martin}, \citenamefont {Parrondo},\ and\ \citenamefont {J{\"u}licher}}]{roldan2021quantifying}%
  \BibitemOpen
  \bibfield  {author} {\bibinfo {author} {\bibfnamefont {{\'E}.}~\bibnamefont {Rold{\'a}n}}, \bibinfo {author} {\bibfnamefont {J.}~\bibnamefont {Barral}}, \bibinfo {author} {\bibfnamefont {P.}~\bibnamefont {Martin}}, \bibinfo {author} {\bibfnamefont {J.~M.}\ \bibnamefont {Parrondo}},\ and\ \bibinfo {author} {\bibfnamefont {F.}~\bibnamefont {J{\"u}licher}},\ }\href@noop {} {\bibfield  {journal} {\bibinfo  {journal} {New Journal of Physics}\ }\textbf {\bibinfo {volume} {23}},\ \bibinfo {pages} {083013} (\bibinfo {year} {2021})}\BibitemShut {NoStop}%
\bibitem [{\citenamefont {Gladrow}\ \emph {et~al.}(2016)\citenamefont {Gladrow}, \citenamefont {Fakhri}, \citenamefont {MacKintosh}, \citenamefont {Schmidt},\ and\ \citenamefont {Broedersz}}]{gladrow2016broken}%
  \BibitemOpen
  \bibfield  {author} {\bibinfo {author} {\bibfnamefont {J.}~\bibnamefont {Gladrow}}, \bibinfo {author} {\bibfnamefont {N.}~\bibnamefont {Fakhri}}, \bibinfo {author} {\bibfnamefont {F.~C.}\ \bibnamefont {MacKintosh}}, \bibinfo {author} {\bibfnamefont {C.}~\bibnamefont {Schmidt}},\ and\ \bibinfo {author} {\bibfnamefont {C.~P.}\ \bibnamefont {Broedersz}},\ }\href@noop {} {\bibfield  {journal} {\bibinfo  {journal} {Physical review letters}\ }\textbf {\bibinfo {volume} {116}},\ \bibinfo {pages} {248301} (\bibinfo {year} {2016})}\BibitemShut {NoStop}%
\bibitem [{\citenamefont {Pietzonka}\ \emph {et~al.}(2019)\citenamefont {Pietzonka}, \citenamefont {Fodor}, \citenamefont {Lohrmann}, \citenamefont {Cates},\ and\ \citenamefont {Seifert}}]{Pietzonka19}%
  \BibitemOpen
  \bibfield  {author} {\bibinfo {author} {\bibfnamefont {P.}~\bibnamefont {Pietzonka}}, \bibinfo {author} {\bibfnamefont {E.}~\bibnamefont {Fodor}}, \bibinfo {author} {\bibfnamefont {C.}~\bibnamefont {Lohrmann}}, \bibinfo {author} {\bibfnamefont {M.~E.}\ \bibnamefont {Cates}},\ and\ \bibinfo {author} {\bibfnamefont {U.}~\bibnamefont {Seifert}},\ }\href {https://doi.org/10.1103/PhysRevX.9.041032} {\bibfield  {journal} {\bibinfo  {journal} {Phys. Rev. X}\ }\textbf {\bibinfo {volume} {9}},\ \bibinfo {pages} {041032} (\bibinfo {year} {2019})}\BibitemShut {NoStop}%
\bibitem [{\citenamefont {Ghosal}\ and\ \citenamefont {Bisker}(2026)}]{ghosal2026identification}%
  \BibitemOpen
  \bibfield  {author} {\bibinfo {author} {\bibfnamefont {A.}~\bibnamefont {Ghosal}}\ and\ \bibinfo {author} {\bibfnamefont {G.}~\bibnamefont {Bisker}},\ }\href@noop {} {\bibfield  {journal} {\bibinfo  {journal} {Physical Chemistry Chemical Physics}\ } (\bibinfo {year} {2026})}\BibitemShut {NoStop}%
\bibitem [{\citenamefont {Nardini}\ \emph {et~al.}(2017)\citenamefont {Nardini}, \citenamefont {Fodor}, \citenamefont {Tjhung}, \citenamefont {Van~Wijland}, \citenamefont {Tailleur},\ and\ \citenamefont {Cates}}]{NardiniETAL:2017}%
  \BibitemOpen
  \bibfield  {author} {\bibinfo {author} {\bibfnamefont {C.}~\bibnamefont {Nardini}}, \bibinfo {author} {\bibfnamefont {{\'E}.}~\bibnamefont {Fodor}}, \bibinfo {author} {\bibfnamefont {E.}~\bibnamefont {Tjhung}}, \bibinfo {author} {\bibfnamefont {F.}~\bibnamefont {Van~Wijland}}, \bibinfo {author} {\bibfnamefont {J.}~\bibnamefont {Tailleur}},\ and\ \bibinfo {author} {\bibfnamefont {M.~E.}\ \bibnamefont {Cates}},\ }\href@noop {} {\bibfield  {journal} {\bibinfo  {journal} {Phys. Rev. X}\ }\textbf {\bibinfo {volume} {7}},\ \bibinfo {pages} {021007} (\bibinfo {year} {2017})}\BibitemShut {NoStop}%
\bibitem [{\citenamefont {Fodor}\ \emph {et~al.}(2022)\citenamefont {Fodor}, \citenamefont {Jack},\ and\ \citenamefont {Cates}}]{FodorETAL:2022}%
  \BibitemOpen
  \bibfield  {author} {\bibinfo {author} {\bibfnamefont {{\'E}.}~\bibnamefont {Fodor}}, \bibinfo {author} {\bibfnamefont {R.~L.}\ \bibnamefont {Jack}},\ and\ \bibinfo {author} {\bibfnamefont {M.~E.}\ \bibnamefont {Cates}},\ }\href@noop {} {\bibfield  {journal} {\bibinfo  {journal} {Annual Review of Condensed Matter Physics}\ }\textbf {\bibinfo {volume} {13}},\ \bibinfo {pages} {215} (\bibinfo {year} {2022})}\BibitemShut {NoStop}%
\bibitem [{\citenamefont {Di~Bello}\ \emph {et~al.}(2024)\citenamefont {Di~Bello}, \citenamefont {Majumdar}, \citenamefont {Marathe}, \citenamefont {Metzler},\ and\ \citenamefont {Rold{\'a}n}}]{di2024brownian}%
  \BibitemOpen
  \bibfield  {author} {\bibinfo {author} {\bibfnamefont {C.}~\bibnamefont {Di~Bello}}, \bibinfo {author} {\bibfnamefont {R.}~\bibnamefont {Majumdar}}, \bibinfo {author} {\bibfnamefont {R.}~\bibnamefont {Marathe}}, \bibinfo {author} {\bibfnamefont {R.}~\bibnamefont {Metzler}},\ and\ \bibinfo {author} {\bibfnamefont {{\'E}.}~\bibnamefont {Rold{\'a}n}},\ }\href@noop {} {\bibfield  {journal} {\bibinfo  {journal} {Annalen der Physik}\ }\textbf {\bibinfo {volume} {536}},\ \bibinfo {pages} {2300427} (\bibinfo {year} {2024})}\BibitemShut {NoStop}%
\bibitem [{\citenamefont {Di~Leonardo}\ \emph {et~al.}(2010)\citenamefont {Di~Leonardo}, \citenamefont {Angelani}, \citenamefont {Dell'Arciprete}, \citenamefont {Ruocco}, \citenamefont {Iebba}, \citenamefont {Schippa}, \citenamefont {Conte}, \citenamefont {Mecarini}, \citenamefont {De~Angelis},\ and\ \citenamefont {Di~Fabrizio}}]{DiLeonardoETAL:2010}%
  \BibitemOpen
  \bibfield  {author} {\bibinfo {author} {\bibfnamefont {R.}~\bibnamefont {Di~Leonardo}}, \bibinfo {author} {\bibfnamefont {L.}~\bibnamefont {Angelani}}, \bibinfo {author} {\bibfnamefont {D.}~\bibnamefont {Dell'Arciprete}}, \bibinfo {author} {\bibfnamefont {G.}~\bibnamefont {Ruocco}}, \bibinfo {author} {\bibfnamefont {V.}~\bibnamefont {Iebba}}, \bibinfo {author} {\bibfnamefont {S.}~\bibnamefont {Schippa}}, \bibinfo {author} {\bibfnamefont {M.~P.}\ \bibnamefont {Conte}}, \bibinfo {author} {\bibfnamefont {F.}~\bibnamefont {Mecarini}}, \bibinfo {author} {\bibfnamefont {F.}~\bibnamefont {De~Angelis}},\ and\ \bibinfo {author} {\bibfnamefont {E.}~\bibnamefont {Di~Fabrizio}},\ }\href@noop {} {\bibfield  {journal} {\bibinfo  {journal} {Proc. Natl. Acad. Sci. USA}\ }\textbf {\bibinfo {volume} {107}},\ \bibinfo {pages} {9541} (\bibinfo {year} {2010})}\BibitemShut {NoStop}%
\bibitem [{\citenamefont {Di~Leonardo}\ \emph {et~al.}(2024)\citenamefont {Di~Leonardo}, \citenamefont {B{\'u}z{\'a}s}, \citenamefont {Kelemen}, \citenamefont {T{\'o}th}, \citenamefont {T{\'o}th}, \citenamefont {Ormos},\ and\ \citenamefont {Vizsnyiczai}}]{DiLeonardoETAL:2024}%
  \BibitemOpen
  \bibfield  {author} {\bibinfo {author} {\bibfnamefont {R.}~\bibnamefont {Di~Leonardo}}, \bibinfo {author} {\bibfnamefont {A.}~\bibnamefont {B{\'u}z{\'a}s}}, \bibinfo {author} {\bibfnamefont {L.}~\bibnamefont {Kelemen}}, \bibinfo {author} {\bibfnamefont {D.}~\bibnamefont {T{\'o}th}}, \bibinfo {author} {\bibfnamefont {S.}~\bibnamefont {T{\'o}th}}, \bibinfo {author} {\bibfnamefont {P.}~\bibnamefont {Ormos}},\ and\ \bibinfo {author} {\bibfnamefont {G.}~\bibnamefont {Vizsnyiczai}},\ }\href@noop {} {\bibfield  {journal} {\bibinfo  {journal} {arXiv:2410.01916}\ } (\bibinfo {year} {2024})}\BibitemShut {NoStop}%
\bibitem [{\citenamefont {Rein}\ \emph {et~al.}(2023)\citenamefont {Rein}, \citenamefont {Kol{\'a}{\v{r}}}, \citenamefont {Kroy},\ and\ \citenamefont {Holubec}}]{ReinETAL:2023}%
  \BibitemOpen
  \bibfield  {author} {\bibinfo {author} {\bibfnamefont {C.}~\bibnamefont {Rein}}, \bibinfo {author} {\bibfnamefont {M.}~\bibnamefont {Kol{\'a}{\v{r}}}}, \bibinfo {author} {\bibfnamefont {K.}~\bibnamefont {Kroy}},\ and\ \bibinfo {author} {\bibfnamefont {V.}~\bibnamefont {Holubec}},\ }\href@noop {} {\bibfield  {journal} {\bibinfo  {journal} {Europhys. Lett.}\ }\textbf {\bibinfo {volume} {142}},\ \bibinfo {pages} {31001} (\bibinfo {year} {2023})}\BibitemShut {NoStop}%
\bibitem [{\citenamefont {Grodzinski}\ \emph {et~al.}(2025)\citenamefont {Grodzinski}, \citenamefont {Jack},\ and\ \citenamefont {Cates}}]{GrodzinskiJackCates:2025}%
  \BibitemOpen
  \bibfield  {author} {\bibinfo {author} {\bibfnamefont {N.}~\bibnamefont {Grodzinski}}, \bibinfo {author} {\bibfnamefont {R.~L.}\ \bibnamefont {Jack}},\ and\ \bibinfo {author} {\bibfnamefont {M.~E.}\ \bibnamefont {Cates}},\ }\href@noop {} {\bibfield  {journal} {\bibinfo  {journal} {arXiv:2506.14559}\ } (\bibinfo {year} {2025})}\BibitemShut {NoStop}%
\bibitem [{\citenamefont {Zhen}\ and\ \citenamefont {Pruessner}(2022)}]{ZhenPruessner:2022}%
  \BibitemOpen
  \bibfield  {author} {\bibinfo {author} {\bibfnamefont {Z.}~\bibnamefont {Zhen}}\ and\ \bibinfo {author} {\bibfnamefont {G.}~\bibnamefont {Pruessner}},\ }\href@noop {} {\bibfield  {journal} {\bibinfo  {journal} {arXiv:2204.04070}\ } (\bibinfo {year} {2022})}\BibitemShut {NoStop}%
\bibitem [{\citenamefont {Roberts}\ and\ \citenamefont {Zhen}(2023)}]{RobertsZhen:2023}%
  \BibitemOpen
  \bibfield  {author} {\bibinfo {author} {\bibfnamefont {C.}~\bibnamefont {Roberts}}\ and\ \bibinfo {author} {\bibfnamefont {Z.}~\bibnamefont {Zhen}},\ }\href@noop {} {\bibfield  {journal} {\bibinfo  {journal} {Phys. Rev. E}\ }\textbf {\bibinfo {volume} {108}},\ \bibinfo {pages} {014139} (\bibinfo {year} {2023})}\BibitemShut {NoStop}%
\bibitem [{\citenamefont {Loos}\ and\ \citenamefont {Klapp}(2020)}]{loos2020irreversibility}%
  \BibitemOpen
  \bibfield  {author} {\bibinfo {author} {\bibfnamefont {S.~A.}\ \bibnamefont {Loos}}\ and\ \bibinfo {author} {\bibfnamefont {S.~H.}\ \bibnamefont {Klapp}},\ }\href@noop {} {\bibfield  {journal} {\bibinfo  {journal} {New Journal of Physics}\ }\textbf {\bibinfo {volume} {22}},\ \bibinfo {pages} {123051} (\bibinfo {year} {2020})}\BibitemShut {NoStop}%
\bibitem [{\citenamefont {Zhang}\ and\ \citenamefont {Garcia-Millan}(2023)}]{ZhangGarcia-Millan:2023}%
  \BibitemOpen
  \bibfield  {author} {\bibinfo {author} {\bibfnamefont {Z.}~\bibnamefont {Zhang}}\ and\ \bibinfo {author} {\bibfnamefont {R.}~\bibnamefont {Garcia-Millan}},\ }\href@noop {} {\bibfield  {journal} {\bibinfo  {journal} {Phys. Rev. Res.}\ }\textbf {\bibinfo {volume} {5}},\ \bibinfo {pages} {L022033} (\bibinfo {year} {2023})}\BibitemShut {NoStop}%
\bibitem [{\citenamefont {Bothe}\ \emph {et~al.}(2023)\citenamefont {Bothe}, \citenamefont {Cocconi}, \citenamefont {Zhen},\ and\ \citenamefont {Pruessner}}]{BotheETAL:2023}%
  \BibitemOpen
  \bibfield  {author} {\bibinfo {author} {\bibfnamefont {M.}~\bibnamefont {Bothe}}, \bibinfo {author} {\bibfnamefont {L.}~\bibnamefont {Cocconi}}, \bibinfo {author} {\bibfnamefont {Z.}~\bibnamefont {Zhen}},\ and\ \bibinfo {author} {\bibfnamefont {G.}~\bibnamefont {Pruessner}},\ }\href@noop {} {\bibfield  {journal} {\bibinfo  {journal} {J. Phys. A: Math. Theor.}\ }\textbf {\bibinfo {volume} {56}},\ \bibinfo {pages} {175002} (\bibinfo {year} {2023})}\BibitemShut {NoStop}%
\bibitem [{\citenamefont {Doi}(1976)}]{Doi:1976}%
  \BibitemOpen
  \bibfield  {author} {\bibinfo {author} {\bibfnamefont {M.}~\bibnamefont {Doi}},\ }\href@noop {} {\bibfield  {journal} {\bibinfo  {journal} {J. Phys. A: Math. Gen.}\ }\textbf {\bibinfo {volume} {9}},\ \bibinfo {pages} {1465} (\bibinfo {year} {1976})}\BibitemShut {NoStop}%
\bibitem [{\citenamefont {Peliti}(1985)}]{Peliti:1985}%
  \BibitemOpen
  \bibfield  {author} {\bibinfo {author} {\bibfnamefont {L.}~\bibnamefont {Peliti}},\ }\href@noop {} {\bibfield  {journal} {\bibinfo  {journal} {J. Phys. (Paris)}\ }\textbf {\bibinfo {volume} {46}},\ \bibinfo {pages} {1469} (\bibinfo {year} {1985})}\BibitemShut {NoStop}%
\bibitem [{\citenamefont {Dean}(1996)}]{Dean:1996}%
  \BibitemOpen
  \bibfield  {author} {\bibinfo {author} {\bibfnamefont {D.~S.}\ \bibnamefont {Dean}},\ }\href@noop {} {\bibfield  {journal} {\bibinfo  {journal} {J. Phys. A: Math. Gen.}\ }\textbf {\bibinfo {volume} {29}},\ \bibinfo {pages} {L613} (\bibinfo {year} {1996})}\BibitemShut {NoStop}%
\bibitem [{\citenamefont {Seifert}(2025{\natexlab{b}})}]{seifert2025universal}%
  \BibitemOpen
  \bibfield  {author} {\bibinfo {author} {\bibfnamefont {U.}~\bibnamefont {Seifert}},\ }\href@noop {} {\bibfield  {journal} {\bibinfo  {journal} {arXiv preprint arXiv:2512.07772}\ } (\bibinfo {year} {2025}{\natexlab{b}})}\BibitemShut {NoStop}%
\bibitem [{\citenamefont {Gaspard}(2004{\natexlab{b}})}]{Gaspard:2004}%
  \BibitemOpen
  \bibfield  {author} {\bibinfo {author} {\bibfnamefont {P.}~\bibnamefont {Gaspard}},\ }\href@noop {} {\bibfield  {journal} {\bibinfo  {journal} {J. Stat. Phys.}\ }\textbf {\bibinfo {volume} {117}},\ \bibinfo {pages} {599} (\bibinfo {year} {2004}{\natexlab{b}})}\BibitemShut {NoStop}%
\bibitem [{\citenamefont {Cocconi}\ \emph {et~al.}(2020)\citenamefont {Cocconi}, \citenamefont {Garcia-Millan}, \citenamefont {Zhen}, \citenamefont {Buturca},\ and\ \citenamefont {Pruessner}}]{CocconiGarcia-MillanETAL:2020}%
  \BibitemOpen
  \bibfield  {author} {\bibinfo {author} {\bibfnamefont {L.}~\bibnamefont {Cocconi}}, \bibinfo {author} {\bibfnamefont {R.}~\bibnamefont {Garcia-Millan}}, \bibinfo {author} {\bibfnamefont {Z.}~\bibnamefont {Zhen}}, \bibinfo {author} {\bibfnamefont {B.}~\bibnamefont {Buturca}},\ and\ \bibinfo {author} {\bibfnamefont {G.}~\bibnamefont {Pruessner}},\ }\href {https://doi.org/10.3390/e22111252} {\bibfield  {journal} {\bibinfo  {journal} {Entropy}\ }\textbf {\bibinfo {volume} {22}},\ \bibinfo {pages} {1252} (\bibinfo {year} {2020})}\BibitemShut {NoStop}%
\bibitem [{\citenamefont {Pruessner}\ and\ \citenamefont {Garcia-Millan}(2025)}]{PruessnerGarcia-Millan:2025}%
  \BibitemOpen
  \bibfield  {author} {\bibinfo {author} {\bibfnamefont {G.}~\bibnamefont {Pruessner}}\ and\ \bibinfo {author} {\bibfnamefont {R.}~\bibnamefont {Garcia-Millan}},\ }\href@noop {} {\bibfield  {journal} {\bibinfo  {journal} {Rep. Progr. Phys.}\ }\textbf {\bibinfo {volume} {88}},\ \bibinfo {pages} {097601} (\bibinfo {year} {2025})}\BibitemShut {NoStop}%
\bibitem [{\citenamefont {Brossollet}\ and\ \citenamefont {Biroli}(2026)}]{BrossolletBiroli:2026}%
  \BibitemOpen
  \bibfield  {author} {\bibinfo {author} {\bibfnamefont {A.}~\bibnamefont {Brossollet}}\ and\ \bibinfo {author} {\bibfnamefont {G.}~\bibnamefont {Biroli}},\ }\href@noop {} {\bibfield  {journal} {\bibinfo  {journal} {J. Phys. A: Math. Theor.}\ }\textbf {\bibinfo {volume} {59}},\ \bibinfo {pages} {045002} (\bibinfo {year} {2026})}\BibitemShut {NoStop}%
\bibitem [{\citenamefont {Agranov}\ \emph {et~al.}(2024)\citenamefont {Agranov}, \citenamefont {Jack}, \citenamefont {Cates},\ and\ \citenamefont {Fodor}}]{agranov2024thermodynamically}%
  \BibitemOpen
  \bibfield  {author} {\bibinfo {author} {\bibfnamefont {T.}~\bibnamefont {Agranov}}, \bibinfo {author} {\bibfnamefont {R.~L.}\ \bibnamefont {Jack}}, \bibinfo {author} {\bibfnamefont {M.~E.}\ \bibnamefont {Cates}},\ and\ \bibinfo {author} {\bibfnamefont {{\'E}.}~\bibnamefont {Fodor}},\ }\href@noop {} {\bibfield  {journal} {\bibinfo  {journal} {New Journal of Physics}\ }\textbf {\bibinfo {volume} {26}},\ \bibinfo {pages} {063006} (\bibinfo {year} {2024})}\BibitemShut {NoStop}%
\bibitem [{\citenamefont {Aslyamov}\ \emph {et~al.}(2023)\citenamefont {Aslyamov}, \citenamefont {Avanzini}, \citenamefont {Fodor},\ and\ \citenamefont {Esposito}}]{aslyamov2023nonideal}%
  \BibitemOpen
  \bibfield  {author} {\bibinfo {author} {\bibfnamefont {T.}~\bibnamefont {Aslyamov}}, \bibinfo {author} {\bibfnamefont {F.}~\bibnamefont {Avanzini}}, \bibinfo {author} {\bibfnamefont {{\'E}.}~\bibnamefont {Fodor}},\ and\ \bibinfo {author} {\bibfnamefont {M.}~\bibnamefont {Esposito}},\ }\href@noop {} {\bibfield  {journal} {\bibinfo  {journal} {Physical Review Letters}\ }\textbf {\bibinfo {volume} {131}},\ \bibinfo {pages} {138301} (\bibinfo {year} {2023})}\BibitemShut {NoStop}%
\bibitem [{\citenamefont {Hartich}\ and\ \citenamefont {Godec}(2021)}]{hartich2021emergent}%
  \BibitemOpen
  \bibfield  {author} {\bibinfo {author} {\bibfnamefont {D.}~\bibnamefont {Hartich}}\ and\ \bibinfo {author} {\bibfnamefont {A.}~\bibnamefont {Godec}},\ }\href@noop {} {\bibfield  {journal} {\bibinfo  {journal} {Physical Review X}\ }\textbf {\bibinfo {volume} {11}},\ \bibinfo {pages} {041047} (\bibinfo {year} {2021})}\BibitemShut {NoStop}%
\bibitem [{\citenamefont {Mart{\'\i}nez}\ \emph {et~al.}(2019)\citenamefont {Mart{\'\i}nez}, \citenamefont {Bisker}, \citenamefont {Horowitz},\ and\ \citenamefont {Parrondo}}]{martinez2019inferring}%
  \BibitemOpen
  \bibfield  {author} {\bibinfo {author} {\bibfnamefont {I.~A.}\ \bibnamefont {Mart{\'\i}nez}}, \bibinfo {author} {\bibfnamefont {G.}~\bibnamefont {Bisker}}, \bibinfo {author} {\bibfnamefont {J.~M.}\ \bibnamefont {Horowitz}},\ and\ \bibinfo {author} {\bibfnamefont {J.~M.}\ \bibnamefont {Parrondo}},\ }\href@noop {} {\bibfield  {journal} {\bibinfo  {journal} {Nature communications}\ }\textbf {\bibinfo {volume} {10}},\ \bibinfo {pages} {3542} (\bibinfo {year} {2019})}\BibitemShut {NoStop}%
\bibitem [{\citenamefont {Cocconi}\ \emph {et~al.}(2023)\citenamefont {Cocconi}, \citenamefont {Knight},\ and\ \citenamefont {Roberts}}]{CocconiKnightRoberts:2023}%
  \BibitemOpen
  \bibfield  {author} {\bibinfo {author} {\bibfnamefont {L.}~\bibnamefont {Cocconi}}, \bibinfo {author} {\bibfnamefont {J.}~\bibnamefont {Knight}},\ and\ \bibinfo {author} {\bibfnamefont {C.}~\bibnamefont {Roberts}},\ }\href@noop {} {\bibfield  {journal} {\bibinfo  {journal} {Phys. Rev. Lett.}\ }\textbf {\bibinfo {volume} {131}},\ \bibinfo {pages} {188301} (\bibinfo {year} {2023})}\BibitemShut {NoStop}%
\bibitem [{\citenamefont {Tailleur}\ and\ \citenamefont {Cates}(2008)}]{TailleurCates:2008}%
  \BibitemOpen
  \bibfield  {author} {\bibinfo {author} {\bibfnamefont {J.}~\bibnamefont {Tailleur}}\ and\ \bibinfo {author} {\bibfnamefont {M.~E.}\ \bibnamefont {Cates}},\ }\href {https://doi.org/10.1103/PhysRevLett.100.218103} {\bibfield  {journal} {\bibinfo  {journal} {Phys. Rev. Lett.}\ }\textbf {\bibinfo {volume} {100}},\ \bibinfo {pages} {218103} (\bibinfo {year} {2008})}\BibitemShut {NoStop}%
\bibitem [{\citenamefont {Solon}\ \emph {et~al.}(2015)\citenamefont {Solon}, \citenamefont {Cates},\ and\ \citenamefont {Tailleur}}]{SolonCatesTailleur:2015}%
  \BibitemOpen
  \bibfield  {author} {\bibinfo {author} {\bibfnamefont {A.~P.}\ \bibnamefont {Solon}}, \bibinfo {author} {\bibfnamefont {M.}~\bibnamefont {Cates}},\ and\ \bibinfo {author} {\bibfnamefont {J.}~\bibnamefont {Tailleur}},\ }\href@noop {} {\bibfield  {journal} {\bibinfo  {journal} {Eur. Phys. J. Spec. Top.}\ }\textbf {\bibinfo {volume} {224}},\ \bibinfo {pages} {1231} (\bibinfo {year} {2015})}\BibitemShut {NoStop}%
\bibitem [{\citenamefont {Martin}\ \emph {et~al.}(2021)\citenamefont {Martin}, \citenamefont {O'Byrne}, \citenamefont {Cates}, \citenamefont {Fodor}, \citenamefont {Nardini}, \citenamefont {Tailleur},\ and\ \citenamefont {Van~Wijland}}]{MartinETAL:2021}%
  \BibitemOpen
  \bibfield  {author} {\bibinfo {author} {\bibfnamefont {D.}~\bibnamefont {Martin}}, \bibinfo {author} {\bibfnamefont {J.}~\bibnamefont {O'Byrne}}, \bibinfo {author} {\bibfnamefont {M.~E.}\ \bibnamefont {Cates}}, \bibinfo {author} {\bibfnamefont {{\'E}.}~\bibnamefont {Fodor}}, \bibinfo {author} {\bibfnamefont {C.}~\bibnamefont {Nardini}}, \bibinfo {author} {\bibfnamefont {J.}~\bibnamefont {Tailleur}},\ and\ \bibinfo {author} {\bibfnamefont {F.}~\bibnamefont {Van~Wijland}},\ }\href@noop {} {\bibfield  {journal} {\bibinfo  {journal} {Phys. Rev. E}\ }\textbf {\bibinfo {volume} {103}},\ \bibinfo {pages} {032607} (\bibinfo {year} {2021})}\BibitemShut {NoStop}%
\bibitem [{\citenamefont {Knight}\ \emph {et~al.}(2025)\citenamefont {Knight}, \citenamefont {Kaveh},\ and\ \citenamefont {Pruessner}}]{KnightKavehPruessner:2025}%
  \BibitemOpen
  \bibfield  {author} {\bibinfo {author} {\bibfnamefont {J.}~\bibnamefont {Knight}}, \bibinfo {author} {\bibfnamefont {F.}~\bibnamefont {Kaveh}},\ and\ \bibinfo {author} {\bibfnamefont {G.}~\bibnamefont {Pruessner}},\ }\href@noop {} {\bibfield  {journal} {\bibinfo  {journal} {arXiv:2507.22199}\ } (\bibinfo {year} {2025})}\BibitemShut {NoStop}%
\bibitem [{\citenamefont {Dabelow}\ \emph {et~al.}(2021)\citenamefont {Dabelow}, \citenamefont {Bo},\ and\ \citenamefont {Eichhorn}}]{DabelowBoEichhorn:2021}%
  \BibitemOpen
  \bibfield  {author} {\bibinfo {author} {\bibfnamefont {L.}~\bibnamefont {Dabelow}}, \bibinfo {author} {\bibfnamefont {S.}~\bibnamefont {Bo}},\ and\ \bibinfo {author} {\bibfnamefont {R.}~\bibnamefont {Eichhorn}},\ }\href@noop {} {\bibfield  {journal} {\bibinfo  {journal} {Journal of Statistical Mechanics: Theory and Experiment}\ }\textbf {\bibinfo {volume} {2021}},\ \bibinfo {pages} {033216} (\bibinfo {year} {2021})}\BibitemShut {NoStop}%
\bibitem [{\citenamefont {Neri}(2022{\natexlab{b}})}]{Neri:2022}%
  \BibitemOpen
  \bibfield  {author} {\bibinfo {author} {\bibfnamefont {I.}~\bibnamefont {Neri}},\ }\href@noop {} {\bibfield  {journal} {\bibinfo  {journal} {J. Phys. A: Math. Theor.}\ }\textbf {\bibinfo {volume} {55}},\ \bibinfo {pages} {304005} (\bibinfo {year} {2022}{\natexlab{b}})}\BibitemShut {NoStop}%
\bibitem [{\citenamefont {Harris}(1963)}]{Harris:1963}%
  \BibitemOpen
  \bibfield  {author} {\bibinfo {author} {\bibfnamefont {T.~E.}\ \bibnamefont {Harris}},\ }\href@noop {} {\emph {\bibinfo {title} {The Theory of Branching Processes}}}\ (\bibinfo  {publisher} {Springer-Verlag},\ \bibinfo {address} {Berlin, Germany},\ \bibinfo {year} {1963})\BibitemShut {NoStop}%
\bibitem [{\citenamefont {Cameron}\ and\ \citenamefont {Tjhung}(2025)}]{CameronTjhung:2025}%
  \BibitemOpen
  \bibfield  {author} {\bibinfo {author} {\bibfnamefont {S.}~\bibnamefont {Cameron}}\ and\ \bibinfo {author} {\bibfnamefont {E.}~\bibnamefont {Tjhung}},\ }\href@noop {} {\bibfield  {journal} {\bibinfo  {journal} {Phys. Rev. E}\ }\textbf {\bibinfo {volume} {112}},\ \bibinfo {pages} {034109} (\bibinfo {year} {2025})}\BibitemShut {NoStop}%
\bibitem [{\citenamefont {Gibbs}(1873{\natexlab{a}})}]{gibbs1873graphical}%
  \BibitemOpen
  \bibfield  {author} {\bibinfo {author} {\bibfnamefont {J.~W.}\ \bibnamefont {Gibbs}},\ }\href@noop {} {\bibfield  {journal} {\bibinfo  {journal} {Transactions of Connecticut Academy of Arts and Sciences}\ ,\ \bibinfo {pages} {309}} (\bibinfo {year} {1873}{\natexlab{a}})}\BibitemShut {NoStop}%
\bibitem [{\citenamefont {Gibbs}(1873{\natexlab{b}})}]{gibbs1873method}%
  \BibitemOpen
  \bibfield  {author} {\bibinfo {author} {\bibfnamefont {J.~W.}\ \bibnamefont {Gibbs}},\ }\href@noop {} {\bibfield  {journal} {\bibinfo  {journal} {Transactions of Connecticut Academy of Arts and Sciences}\ ,\ \bibinfo {pages} {382}} (\bibinfo {year} {1873}{\natexlab{b}})}\BibitemShut {NoStop}%
\bibitem [{\citenamefont {Mruga{\l}a}(1978)}]{mrugala1978geometrical}%
  \BibitemOpen
  \bibfield  {author} {\bibinfo {author} {\bibfnamefont {R.}~\bibnamefont {Mruga{\l}a}},\ }\href@noop {} {\bibfield  {journal} {\bibinfo  {journal} {Reports on Mathematical Physics}\ }\textbf {\bibinfo {volume} {14}},\ \bibinfo {pages} {419} (\bibinfo {year} {1978})}\BibitemShut {NoStop}%
\bibitem [{\citenamefont {Weinhold}(1975)}]{weinhold1975metric}%
  \BibitemOpen
  \bibfield  {author} {\bibinfo {author} {\bibfnamefont {F.}~\bibnamefont {Weinhold}},\ }\href@noop {} {\bibfield  {journal} {\bibinfo  {journal} {The Journal of Chemical Physics}\ }\textbf {\bibinfo {volume} {63}},\ \bibinfo {pages} {2479} (\bibinfo {year} {1975})}\BibitemShut {NoStop}%
\bibitem [{\citenamefont {Ruppeiner}(1979)}]{ruppeiner1979thermodynamics}%
  \BibitemOpen
  \bibfield  {author} {\bibinfo {author} {\bibfnamefont {G.}~\bibnamefont {Ruppeiner}},\ }\href@noop {} {\bibfield  {journal} {\bibinfo  {journal} {Physical Review A}\ }\textbf {\bibinfo {volume} {20}},\ \bibinfo {pages} {1608} (\bibinfo {year} {1979})}\BibitemShut {NoStop}%
\bibitem [{\citenamefont {Salamon}\ and\ \citenamefont {Berry}(1983)}]{SalamonBerry1983}%
  \BibitemOpen
  \bibfield  {author} {\bibinfo {author} {\bibfnamefont {P.}~\bibnamefont {Salamon}}\ and\ \bibinfo {author} {\bibfnamefont {R.~S.}\ \bibnamefont {Berry}},\ }\href@noop {} {\bibfield  {journal} {\bibinfo  {journal} {Physical Review Letters}\ }\textbf {\bibinfo {volume} {51}},\ \bibinfo {pages} {1127} (\bibinfo {year} {1983})}\BibitemShut {NoStop}%
\bibitem [{\citenamefont {Crooks}(2007)}]{Crooks2007}%
  \BibitemOpen
  \bibfield  {author} {\bibinfo {author} {\bibfnamefont {G.~E.}\ \bibnamefont {Crooks}},\ }\href@noop {} {\bibfield  {journal} {\bibinfo  {journal} {Physical Review Letters}\ }\textbf {\bibinfo {volume} {99}} (\bibinfo {year} {2007})}\BibitemShut {NoStop}%
\bibitem [{\citenamefont {Sivak}\ and\ \citenamefont {Crooks}(2012)}]{SivakCrooks2012}%
  \BibitemOpen
  \bibfield  {author} {\bibinfo {author} {\bibfnamefont {D.~A.}\ \bibnamefont {Sivak}}\ and\ \bibinfo {author} {\bibfnamefont {G.~E.}\ \bibnamefont {Crooks}},\ }\href@noop {} {\bibfield  {journal} {\bibinfo  {journal} {Physical Review Letters}\ }\textbf {\bibinfo {volume} {108}} (\bibinfo {year} {2012})}\BibitemShut {NoStop}%
\bibitem [{\citenamefont {Aurell}\ \emph {et~al.}(2011)\citenamefont {Aurell}, \citenamefont {Mej{\'\i}a-Monasterio},\ and\ \citenamefont {Muratore-Ginanneschi}}]{aurell2011optimal}%
  \BibitemOpen
  \bibfield  {author} {\bibinfo {author} {\bibfnamefont {E.}~\bibnamefont {Aurell}}, \bibinfo {author} {\bibfnamefont {C.}~\bibnamefont {Mej{\'\i}a-Monasterio}},\ and\ \bibinfo {author} {\bibfnamefont {P.}~\bibnamefont {Muratore-Ginanneschi}},\ }\href@noop {} {\bibfield  {journal} {\bibinfo  {journal} {Physical review letters}\ }\textbf {\bibinfo {volume} {106}},\ \bibinfo {pages} {250601} (\bibinfo {year} {2011})}\BibitemShut {NoStop}%
\bibitem [{\citenamefont {Villani}\ \emph {et~al.}(2009)\citenamefont {Villani} \emph {et~al.}}]{villani2009optimal}%
  \BibitemOpen
  \bibfield  {author} {\bibinfo {author} {\bibfnamefont {C.}~\bibnamefont {Villani}} \emph {et~al.},\ }\href@noop {} {\emph {\bibinfo {title} {Optimal transport: old and new}}},\ Vol.\ \bibinfo {volume} {338}\ (\bibinfo  {publisher} {Springer},\ \bibinfo {year} {2009})\BibitemShut {NoStop}%
\bibitem [{\citenamefont {Jordan}\ \emph {et~al.}(1998)\citenamefont {Jordan}, \citenamefont {Kinderlehrer},\ and\ \citenamefont {Otto}}]{JKO1998}%
  \BibitemOpen
  \bibfield  {author} {\bibinfo {author} {\bibfnamefont {R.}~\bibnamefont {Jordan}}, \bibinfo {author} {\bibfnamefont {D.}~\bibnamefont {Kinderlehrer}},\ and\ \bibinfo {author} {\bibfnamefont {F.}~\bibnamefont {Otto}},\ }\href {https://doi.org/10.1137/S0036141096303359} {\bibfield  {journal} {\bibinfo  {journal} {SIAM Journal on Mathematical Analysis}\ }\textbf {\bibinfo {volume} {29}},\ \bibinfo {pages} {1} (\bibinfo {year} {1998})}\BibitemShut {NoStop}%
\bibitem [{\citenamefont {Grmela}\ and\ \citenamefont {{\"O}ttinger}(1997)}]{grmela1997dynamics}%
  \BibitemOpen
  \bibfield  {author} {\bibinfo {author} {\bibfnamefont {M.}~\bibnamefont {Grmela}}\ and\ \bibinfo {author} {\bibfnamefont {H.~C.}\ \bibnamefont {{\"O}ttinger}},\ }\href@noop {} {\bibfield  {journal} {\bibinfo  {journal} {Physical Review E}\ }\textbf {\bibinfo {volume} {56}},\ \bibinfo {pages} {6620} (\bibinfo {year} {1997})}\BibitemShut {NoStop}%
\bibitem [{\citenamefont {Espanol}(2023)}]{espanol2023statistical}%
  \BibitemOpen
  \bibfield  {author} {\bibinfo {author} {\bibfnamefont {P.}~\bibnamefont {Espanol}},\ }\href@noop {} {\bibfield  {journal} {\bibinfo  {journal} {The Journal of Chemical Physics}\ }\textbf {\bibinfo {volume} {159}} (\bibinfo {year} {2023})}\BibitemShut {NoStop}%
\bibitem [{\citenamefont {Beretta}(2014)}]{beretta2014steepest}%
  \BibitemOpen
  \bibfield  {author} {\bibinfo {author} {\bibfnamefont {G.~P.}\ \bibnamefont {Beretta}},\ }\href@noop {} {\bibfield  {journal} {\bibinfo  {journal} {Physical Review E}\ }\textbf {\bibinfo {volume} {90}},\ \bibinfo {pages} {042113} (\bibinfo {year} {2014})}\BibitemShut {NoStop}%
\bibitem [{\citenamefont {Reina}\ and\ \citenamefont {Zimmer}(2015)}]{reina2015entropy}%
  \BibitemOpen
  \bibfield  {author} {\bibinfo {author} {\bibfnamefont {C.}~\bibnamefont {Reina}}\ and\ \bibinfo {author} {\bibfnamefont {J.}~\bibnamefont {Zimmer}},\ }\href@noop {} {\bibfield  {journal} {\bibinfo  {journal} {Physical Review E}\ }\textbf {\bibinfo {volume} {92}},\ \bibinfo {pages} {052117} (\bibinfo {year} {2015})}\BibitemShut {NoStop}%
\bibitem [{\citenamefont {Dechant}\ \emph {et~al.}(2022)\citenamefont {Dechant}, \citenamefont {Sasa},\ and\ \citenamefont {Ito}}]{dechant2022geometric2}%
  \BibitemOpen
  \bibfield  {author} {\bibinfo {author} {\bibfnamefont {A.}~\bibnamefont {Dechant}}, \bibinfo {author} {\bibfnamefont {S.-i.}\ \bibnamefont {Sasa}},\ and\ \bibinfo {author} {\bibfnamefont {S.}~\bibnamefont {Ito}},\ }\href@noop {} {\bibfield  {journal} {\bibinfo  {journal} {Physical Review E}\ }\textbf {\bibinfo {volume} {106}},\ \bibinfo {pages} {024125} (\bibinfo {year} {2022})}\BibitemShut {NoStop}%
\bibitem [{\citenamefont {Otsubo}\ \emph {et~al.}(2020)\citenamefont {Otsubo}, \citenamefont {Ito}, \citenamefont {Dechant},\ and\ \citenamefont {Sagawa}}]{otsubo2020estimating}%
  \BibitemOpen
  \bibfield  {author} {\bibinfo {author} {\bibfnamefont {S.}~\bibnamefont {Otsubo}}, \bibinfo {author} {\bibfnamefont {S.}~\bibnamefont {Ito}}, \bibinfo {author} {\bibfnamefont {A.}~\bibnamefont {Dechant}},\ and\ \bibinfo {author} {\bibfnamefont {T.}~\bibnamefont {Sagawa}},\ }\href@noop {} {\bibfield  {journal} {\bibinfo  {journal} {Physical Review E}\ }\textbf {\bibinfo {volume} {101}},\ \bibinfo {pages} {062106} (\bibinfo {year} {2020})}\BibitemShut {NoStop}%
\bibitem [{\citenamefont {Movilla~Miangolarra}\ \emph {et~al.}(2021)\citenamefont {Movilla~Miangolarra}, \citenamefont {Taghvaei}, \citenamefont {Fu}, \citenamefont {Chen},\ and\ \citenamefont {Georgiou}}]{movilla2021energy}%
  \BibitemOpen
  \bibfield  {author} {\bibinfo {author} {\bibfnamefont {O.}~\bibnamefont {Movilla~Miangolarra}}, \bibinfo {author} {\bibfnamefont {A.}~\bibnamefont {Taghvaei}}, \bibinfo {author} {\bibfnamefont {R.}~\bibnamefont {Fu}}, \bibinfo {author} {\bibfnamefont {Y.}~\bibnamefont {Chen}},\ and\ \bibinfo {author} {\bibfnamefont {T.~T.}\ \bibnamefont {Georgiou}},\ }\href@noop {} {\bibfield  {journal} {\bibinfo  {journal} {Physical Review E}\ }\textbf {\bibinfo {volume} {104}},\ \bibinfo {pages} {044101} (\bibinfo {year} {2021})}\BibitemShut {NoStop}%
\bibitem [{\citenamefont {Dechant}\ and\ \citenamefont {Sakurai}(2019)}]{dechant2019thermodynamic}%
  \BibitemOpen
  \bibfield  {author} {\bibinfo {author} {\bibfnamefont {A.}~\bibnamefont {Dechant}}\ and\ \bibinfo {author} {\bibfnamefont {Y.}~\bibnamefont {Sakurai}},\ }\href@noop {} {\bibfield  {journal} {\bibinfo  {journal} {arXiv preprint arXiv:1912.08405}\ } (\bibinfo {year} {2019})}\BibitemShut {NoStop}%
\bibitem [{\citenamefont {Sabbagh}\ \emph {et~al.}(2024)\citenamefont {Sabbagh}, \citenamefont {Movilla~Miangolarra},\ and\ \citenamefont {Georgiou}}]{sabbagh2024wasserstein}%
  \BibitemOpen
  \bibfield  {author} {\bibinfo {author} {\bibfnamefont {R.}~\bibnamefont {Sabbagh}}, \bibinfo {author} {\bibfnamefont {O.}~\bibnamefont {Movilla~Miangolarra}},\ and\ \bibinfo {author} {\bibfnamefont {T.~T.}\ \bibnamefont {Georgiou}},\ }\href@noop {} {\bibfield  {journal} {\bibinfo  {journal} {Physical Review Research}\ }\textbf {\bibinfo {volume} {6}},\ \bibinfo {pages} {033308} (\bibinfo {year} {2024})}\BibitemShut {NoStop}%
\bibitem [{\citenamefont {Van~Vu}\ and\ \citenamefont {Hasegawa}(2021)}]{van2021geometrical}%
  \BibitemOpen
  \bibfield  {author} {\bibinfo {author} {\bibfnamefont {T.}~\bibnamefont {Van~Vu}}\ and\ \bibinfo {author} {\bibfnamefont {Y.}~\bibnamefont {Hasegawa}},\ }\href@noop {} {\bibfield  {journal} {\bibinfo  {journal} {Physical Review Letters}\ }\textbf {\bibinfo {volume} {126}},\ \bibinfo {pages} {010601} (\bibinfo {year} {2021})}\BibitemShut {NoStop}%
\bibitem [{\citenamefont {Dechant}(2022)}]{dechant2022minimum}%
  \BibitemOpen
  \bibfield  {author} {\bibinfo {author} {\bibfnamefont {A.}~\bibnamefont {Dechant}},\ }\href@noop {} {\bibfield  {journal} {\bibinfo  {journal} {Journal of Physics A: Mathematical and Theoretical}\ }\textbf {\bibinfo {volume} {55}},\ \bibinfo {pages} {094001} (\bibinfo {year} {2022})}\BibitemShut {NoStop}%
\bibitem [{\citenamefont {Van~Vu}\ and\ \citenamefont {Saito}(2023{\natexlab{a}})}]{van2023thermodynamic}%
  \BibitemOpen
  \bibfield  {author} {\bibinfo {author} {\bibfnamefont {T.}~\bibnamefont {Van~Vu}}\ and\ \bibinfo {author} {\bibfnamefont {K.}~\bibnamefont {Saito}},\ }\href@noop {} {\bibfield  {journal} {\bibinfo  {journal} {Physical Review X}\ }\textbf {\bibinfo {volume} {13}},\ \bibinfo {pages} {011013} (\bibinfo {year} {2023}{\natexlab{a}})}\BibitemShut {NoStop}%
\bibitem [{\citenamefont {Yoshimura}\ \emph {et~al.}(2023)\citenamefont {Yoshimura}, \citenamefont {Kolchinsky}, \citenamefont {Dechant},\ and\ \citenamefont {Ito}}]{yoshimura2023housekeeping}%
  \BibitemOpen
  \bibfield  {author} {\bibinfo {author} {\bibfnamefont {K.}~\bibnamefont {Yoshimura}}, \bibinfo {author} {\bibfnamefont {A.}~\bibnamefont {Kolchinsky}}, \bibinfo {author} {\bibfnamefont {A.}~\bibnamefont {Dechant}},\ and\ \bibinfo {author} {\bibfnamefont {S.}~\bibnamefont {Ito}},\ }\href@noop {} {\bibfield  {journal} {\bibinfo  {journal} {Physical Review Research}\ }\textbf {\bibinfo {volume} {5}},\ \bibinfo {pages} {013017} (\bibinfo {year} {2023})}\BibitemShut {NoStop}%
\bibitem [{\citenamefont {Kolchinsky}\ \emph {et~al.}(2026)\citenamefont {Kolchinsky}, \citenamefont {Dechant}, \citenamefont {Yoshimura},\ and\ \citenamefont {Ito}}]{kolchinsky2026generalized}%
  \BibitemOpen
  \bibfield  {author} {\bibinfo {author} {\bibfnamefont {A.}~\bibnamefont {Kolchinsky}}, \bibinfo {author} {\bibfnamefont {A.}~\bibnamefont {Dechant}}, \bibinfo {author} {\bibfnamefont {K.}~\bibnamefont {Yoshimura}},\ and\ \bibinfo {author} {\bibfnamefont {S.}~\bibnamefont {Ito}},\ }\href {https://doi.org/10.1103/r48t-dghl} {\bibfield  {journal} {\bibinfo  {journal} {Physical Review Research}\ }\textbf {\bibinfo {volume} {8}},\ \bibinfo {pages} {023025} (\bibinfo {year} {2026})}\BibitemShut {NoStop}%
\bibitem [{\citenamefont {Nagayama}\ \emph {et~al.}(2025{\natexlab{a}})\citenamefont {Nagayama}, \citenamefont {Yoshimura}, \citenamefont {Kolchinsky},\ and\ \citenamefont {Ito}}]{nagayama2025geometric}%
  \BibitemOpen
  \bibfield  {author} {\bibinfo {author} {\bibfnamefont {R.}~\bibnamefont {Nagayama}}, \bibinfo {author} {\bibfnamefont {K.}~\bibnamefont {Yoshimura}}, \bibinfo {author} {\bibfnamefont {A.}~\bibnamefont {Kolchinsky}},\ and\ \bibinfo {author} {\bibfnamefont {S.}~\bibnamefont {Ito}},\ }\href@noop {} {\bibfield  {journal} {\bibinfo  {journal} {Physical Review Research}\ }\textbf {\bibinfo {volume} {7}},\ \bibinfo {pages} {033011} (\bibinfo {year} {2025}{\natexlab{a}})}\BibitemShut {NoStop}%
\bibitem [{\citenamefont {Yoshimura}\ \emph {et~al.}(2025)\citenamefont {Yoshimura}, \citenamefont {Maekawa}, \citenamefont {Nagayama},\ and\ \citenamefont {Ito}}]{yoshimura2025force}%
  \BibitemOpen
  \bibfield  {author} {\bibinfo {author} {\bibfnamefont {K.}~\bibnamefont {Yoshimura}}, \bibinfo {author} {\bibfnamefont {Y.}~\bibnamefont {Maekawa}}, \bibinfo {author} {\bibfnamefont {R.}~\bibnamefont {Nagayama}},\ and\ \bibinfo {author} {\bibfnamefont {S.}~\bibnamefont {Ito}},\ }\href@noop {} {\bibfield  {journal} {\bibinfo  {journal} {Physical Review Research}\ }\textbf {\bibinfo {volume} {7}},\ \bibinfo {pages} {013244} (\bibinfo {year} {2025})}\BibitemShut {NoStop}%
\bibitem [{\citenamefont {Nagayama}\ \emph {et~al.}(2025{\natexlab{b}})\citenamefont {Nagayama}, \citenamefont {Yoshimura},\ and\ \citenamefont {Ito}}]{nagayama2025infinite}%
  \BibitemOpen
  \bibfield  {author} {\bibinfo {author} {\bibfnamefont {R.}~\bibnamefont {Nagayama}}, \bibinfo {author} {\bibfnamefont {K.}~\bibnamefont {Yoshimura}},\ and\ \bibinfo {author} {\bibfnamefont {S.}~\bibnamefont {Ito}},\ }\href@noop {} {\bibfield  {journal} {\bibinfo  {journal} {Physical Review Research}\ }\textbf {\bibinfo {volume} {7}},\ \bibinfo {pages} {013307} (\bibinfo {year} {2025}{\natexlab{b}})}\BibitemShut {NoStop}%
\bibitem [{\citenamefont {Maes}(2021)}]{maes2021local}%
  \BibitemOpen
  \bibfield  {author} {\bibinfo {author} {\bibfnamefont {C.}~\bibnamefont {Maes}},\ }\href@noop {} {\bibfield  {journal} {\bibinfo  {journal} {SciPost Physics Lecture Notes}\ ,\ \bibinfo {pages} {032}} (\bibinfo {year} {2021})}\BibitemShut {NoStop}%
\bibitem [{\citenamefont {Mehta}\ \emph {et~al.}(2016)\citenamefont {Mehta}, \citenamefont {Lang},\ and\ \citenamefont {Schwab}}]{mehta2016landauer}%
  \BibitemOpen
  \bibfield  {author} {\bibinfo {author} {\bibfnamefont {P.}~\bibnamefont {Mehta}}, \bibinfo {author} {\bibfnamefont {A.~H.}\ \bibnamefont {Lang}},\ and\ \bibinfo {author} {\bibfnamefont {D.~J.}\ \bibnamefont {Schwab}},\ }\href@noop {} {\bibfield  {journal} {\bibinfo  {journal} {Journal of Statistical Physics}\ }\textbf {\bibinfo {volume} {162}},\ \bibinfo {pages} {1153} (\bibinfo {year} {2016})}\BibitemShut {NoStop}%
\bibitem [{\citenamefont {Cao}\ and\ \citenamefont {Liang}(2025)}]{cao2025stochastic}%
  \BibitemOpen
  \bibfield  {author} {\bibinfo {author} {\bibfnamefont {Y.}~\bibnamefont {Cao}}\ and\ \bibinfo {author} {\bibfnamefont {S.}~\bibnamefont {Liang}},\ }\href@noop {} {\bibfield  {journal} {\bibinfo  {journal} {Quantitative Biology}\ }\textbf {\bibinfo {volume} {13}},\ \bibinfo {pages} {e75} (\bibinfo {year} {2025})}\BibitemShut {NoStop}%
\bibitem [{\citenamefont {Ciliberto}\ \emph {et~al.}(2010)\citenamefont {Ciliberto}, \citenamefont {Joubaud},\ and\ \citenamefont {Petrosyan}}]{CilibertoJoubaudPetrosyan2010}%
  \BibitemOpen
  \bibfield  {author} {\bibinfo {author} {\bibfnamefont {S.}~\bibnamefont {Ciliberto}}, \bibinfo {author} {\bibfnamefont {S.}~\bibnamefont {Joubaud}},\ and\ \bibinfo {author} {\bibfnamefont {A.}~\bibnamefont {Petrosyan}},\ }\href {https://doi.org/10.1088/1742-5468/2010/12/p12003} {\bibfield  {journal} {\bibinfo  {journal} {Journal of Statistical Mechanics: Theory and Experiment}\ }\textbf {\bibinfo {volume} {2010}},\ \bibinfo {pages} {P12003} (\bibinfo {year} {2010})}\BibitemShut {NoStop}%
\bibitem [{\citenamefont {Milo}\ and\ \citenamefont {Phillips}(2015)}]{milo2015cell}%
  \BibitemOpen
  \bibfield  {author} {\bibinfo {author} {\bibfnamefont {R.}~\bibnamefont {Milo}}\ and\ \bibinfo {author} {\bibfnamefont {R.}~\bibnamefont {Phillips}},\ }\href@noop {} {\emph {\bibinfo {title} {Cell biology by the numbers}}}\ (\bibinfo  {publisher} {Garland Science},\ \bibinfo {year} {2015})\BibitemShut {NoStop}%
\bibitem [{\citenamefont {Foster}\ \emph {et~al.}(2023)\citenamefont {Foster}, \citenamefont {Bae}, \citenamefont {Lemma}, \citenamefont {Zheng}, \citenamefont {Ireland}, \citenamefont {Chandrakar}, \citenamefont {Boros}, \citenamefont {Dogic}, \citenamefont {Needleman},\ and\ \citenamefont {Vlassak}}]{foster2023dissipation}%
  \BibitemOpen
  \bibfield  {author} {\bibinfo {author} {\bibfnamefont {P.~J.}\ \bibnamefont {Foster}}, \bibinfo {author} {\bibfnamefont {J.}~\bibnamefont {Bae}}, \bibinfo {author} {\bibfnamefont {B.}~\bibnamefont {Lemma}}, \bibinfo {author} {\bibfnamefont {J.}~\bibnamefont {Zheng}}, \bibinfo {author} {\bibfnamefont {W.}~\bibnamefont {Ireland}}, \bibinfo {author} {\bibfnamefont {P.}~\bibnamefont {Chandrakar}}, \bibinfo {author} {\bibfnamefont {R.}~\bibnamefont {Boros}}, \bibinfo {author} {\bibfnamefont {Z.}~\bibnamefont {Dogic}}, \bibinfo {author} {\bibfnamefont {D.~J.}\ \bibnamefont {Needleman}},\ and\ \bibinfo {author} {\bibfnamefont {J.~J.}\ \bibnamefont {Vlassak}},\ }\href@noop {} {\bibfield  {journal} {\bibinfo  {journal} {Proceedings of the National Academy of Sciences}\ }\textbf {\bibinfo {volume} {120}},\ \bibinfo {pages} {e2207662120} (\bibinfo {year} {2023})}\BibitemShut {NoStop}%
\bibitem [{\citenamefont {England}(2013)}]{england2013statistical}%
  \BibitemOpen
  \bibfield  {author} {\bibinfo {author} {\bibfnamefont {J.~L.}\ \bibnamefont {England}},\ }\href@noop {} {\bibfield  {journal} {\bibinfo  {journal} {The Journal of chemical physics}\ }\textbf {\bibinfo {volume} {139}} (\bibinfo {year} {2013})}\BibitemShut {NoStop}%
\bibitem [{\citenamefont {Lynn}\ \emph {et~al.}(2021)\citenamefont {Lynn}, \citenamefont {Cornblath}, \citenamefont {Papadopoulos}, \citenamefont {Bertolero},\ and\ \citenamefont {Bassett}}]{lynn2021broken}%
  \BibitemOpen
  \bibfield  {author} {\bibinfo {author} {\bibfnamefont {C.~W.}\ \bibnamefont {Lynn}}, \bibinfo {author} {\bibfnamefont {E.~J.}\ \bibnamefont {Cornblath}}, \bibinfo {author} {\bibfnamefont {L.}~\bibnamefont {Papadopoulos}}, \bibinfo {author} {\bibfnamefont {M.~A.}\ \bibnamefont {Bertolero}},\ and\ \bibinfo {author} {\bibfnamefont {D.~S.}\ \bibnamefont {Bassett}},\ }\href@noop {} {\bibfield  {journal} {\bibinfo  {journal} {Proceedings of the National Academy of Sciences}\ }\textbf {\bibinfo {volume} {118}},\ \bibinfo {pages} {e2109889118} (\bibinfo {year} {2021})}\BibitemShut {NoStop}%
\bibitem [{\citenamefont {Nartallo-Kaluarachchi}\ \emph {et~al.}(2026)\citenamefont {Nartallo-Kaluarachchi}, \citenamefont {Kringelbach}, \citenamefont {Deco}, \citenamefont {Lambiotte},\ and\ \citenamefont {Goriely}}]{nartallo2026nonequilibrium}%
  \BibitemOpen
  \bibfield  {author} {\bibinfo {author} {\bibfnamefont {R.}~\bibnamefont {Nartallo-Kaluarachchi}}, \bibinfo {author} {\bibfnamefont {M.}~\bibnamefont {Kringelbach}}, \bibinfo {author} {\bibfnamefont {G.}~\bibnamefont {Deco}}, \bibinfo {author} {\bibfnamefont {R.}~\bibnamefont {Lambiotte}},\ and\ \bibinfo {author} {\bibfnamefont {A.}~\bibnamefont {Goriely}},\ }\href@noop {} {\bibfield  {journal} {\bibinfo  {journal} {Physics Reports}\ }\textbf {\bibinfo {volume} {1152}},\ \bibinfo {pages} {1} (\bibinfo {year} {2026})}\BibitemShut {NoStop}%
\bibitem [{\citenamefont {Murashita}\ \emph {et~al.}(2014)\citenamefont {Murashita}, \citenamefont {Funo},\ and\ \citenamefont {Ueda}}]{murashita2014nonequilibrium}%
  \BibitemOpen
  \bibfield  {author} {\bibinfo {author} {\bibfnamefont {Y.}~\bibnamefont {Murashita}}, \bibinfo {author} {\bibfnamefont {K.}~\bibnamefont {Funo}},\ and\ \bibinfo {author} {\bibfnamefont {M.}~\bibnamefont {Ueda}},\ }\href@noop {} {\bibfield  {journal} {\bibinfo  {journal} {Physical Review E}\ }\textbf {\bibinfo {volume} {90}},\ \bibinfo {pages} {042110} (\bibinfo {year} {2014})}\BibitemShut {NoStop}%
\bibitem [{\citenamefont {Pal}\ \emph {et~al.}(2021)\citenamefont {Pal}, \citenamefont {Reuveni},\ and\ \citenamefont {Rahav}}]{pal2021thermodynamic}%
  \BibitemOpen
  \bibfield  {author} {\bibinfo {author} {\bibfnamefont {A.}~\bibnamefont {Pal}}, \bibinfo {author} {\bibfnamefont {S.}~\bibnamefont {Reuveni}},\ and\ \bibinfo {author} {\bibfnamefont {S.}~\bibnamefont {Rahav}},\ }\href@noop {} {\bibfield  {journal} {\bibinfo  {journal} {Physical Review Research}\ }\textbf {\bibinfo {volume} {3}},\ \bibinfo {pages} {013273} (\bibinfo {year} {2021})}\BibitemShut {NoStop}%
\bibitem [{\citenamefont {Busiello}\ \emph {et~al.}(2020)\citenamefont {Busiello}, \citenamefont {Gupta},\ and\ \citenamefont {Maritan}}]{busiello2020entropy}%
  \BibitemOpen
  \bibfield  {author} {\bibinfo {author} {\bibfnamefont {D.~M.}\ \bibnamefont {Busiello}}, \bibinfo {author} {\bibfnamefont {D.}~\bibnamefont {Gupta}},\ and\ \bibinfo {author} {\bibfnamefont {A.}~\bibnamefont {Maritan}},\ }\href@noop {} {\bibfield  {journal} {\bibinfo  {journal} {Physical Review Research}\ }\textbf {\bibinfo {volume} {2}},\ \bibinfo {pages} {023011} (\bibinfo {year} {2020})}\BibitemShut {NoStop}%
\bibitem [{\citenamefont {Kolchinsky}\ and\ \citenamefont {Wolpert}(2021)}]{kolchinsky2021dependence}%
  \BibitemOpen
  \bibfield  {author} {\bibinfo {author} {\bibfnamefont {A.}~\bibnamefont {Kolchinsky}}\ and\ \bibinfo {author} {\bibfnamefont {D.~H.}\ \bibnamefont {Wolpert}},\ }\href@noop {} {\bibfield  {journal} {\bibinfo  {journal} {Physical Review E}\ }\textbf {\bibinfo {volume} {104}},\ \bibinfo {pages} {054107} (\bibinfo {year} {2021})}\BibitemShut {NoStop}%
\bibitem [{\citenamefont {Ge}\ and\ \citenamefont {Qian}(2016)}]{ge2016mesoscopic}%
  \BibitemOpen
  \bibfield  {author} {\bibinfo {author} {\bibfnamefont {H.}~\bibnamefont {Ge}}\ and\ \bibinfo {author} {\bibfnamefont {H.}~\bibnamefont {Qian}},\ }\href@noop {} {\bibfield  {journal} {\bibinfo  {journal} {Physical Review E}\ }\textbf {\bibinfo {volume} {94}},\ \bibinfo {pages} {052150} (\bibinfo {year} {2016})}\BibitemShut {NoStop}%
\bibitem [{\citenamefont {Datta}\ \emph {et~al.}(2022)\citenamefont {Datta}, \citenamefont {Pietzonka},\ and\ \citenamefont {Barato}}]{Datta22}%
  \BibitemOpen
  \bibfield  {author} {\bibinfo {author} {\bibfnamefont {A.}~\bibnamefont {Datta}}, \bibinfo {author} {\bibfnamefont {P.}~\bibnamefont {Pietzonka}},\ and\ \bibinfo {author} {\bibfnamefont {A.~C.}\ \bibnamefont {Barato}},\ }\href {https://doi.org/10.1103/PhysRevX.12.031034} {\bibfield  {journal} {\bibinfo  {journal} {Phys. Rev. X}\ }\textbf {\bibinfo {volume} {12}},\ \bibinfo {pages} {031034} (\bibinfo {year} {2022})}\BibitemShut {NoStop}%
\bibitem [{\citenamefont {Freitas}\ and\ \citenamefont {Esposito}(2022)}]{freitas2022emergent}%
  \BibitemOpen
  \bibfield  {author} {\bibinfo {author} {\bibfnamefont {J.~N.}\ \bibnamefont {Freitas}}\ and\ \bibinfo {author} {\bibfnamefont {M.}~\bibnamefont {Esposito}},\ }\href@noop {} {\bibfield  {journal} {\bibinfo  {journal} {Nature Communications}\ }\textbf {\bibinfo {volume} {13}},\ \bibinfo {pages} {5084} (\bibinfo {year} {2022})}\BibitemShut {NoStop}%
\bibitem [{\citenamefont {Feitosa}\ and\ \citenamefont {Menon}(2004)}]{FeitosaMenon2004}%
  \BibitemOpen
  \bibfield  {author} {\bibinfo {author} {\bibfnamefont {K.}~\bibnamefont {Feitosa}}\ and\ \bibinfo {author} {\bibfnamefont {N.}~\bibnamefont {Menon}},\ }\href {https://doi.org/10.1103/physrevlett.92.164301} {\bibfield  {journal} {\bibinfo  {journal} {Physical Review Letters}\ }\textbf {\bibinfo {volume} {92}},\ \bibinfo {pages} {164301} (\bibinfo {year} {2004})}\BibitemShut {NoStop}%
\bibitem [{\citenamefont {Joubaud}\ \emph {et~al.}(2012)\citenamefont {Joubaud}, \citenamefont {Lohse},\ and\ \citenamefont {van~der Meer}}]{JoubaudLohseVanDerMeer2012}%
  \BibitemOpen
  \bibfield  {author} {\bibinfo {author} {\bibfnamefont {S.}~\bibnamefont {Joubaud}}, \bibinfo {author} {\bibfnamefont {D.}~\bibnamefont {Lohse}},\ and\ \bibinfo {author} {\bibfnamefont {D.}~\bibnamefont {van~der Meer}},\ }\href {https://doi.org/10.1103/physrevlett.108.210604} {\bibfield  {journal} {\bibinfo  {journal} {Physical Review Letters}\ }\textbf {\bibinfo {volume} {108}},\ \bibinfo {pages} {210604} (\bibinfo {year} {2012})}\BibitemShut {NoStop}%
\bibitem [{\citenamefont {Lagoin}\ \emph {et~al.}(2022)\citenamefont {Lagoin}, \citenamefont {Crauste-Thibierge},\ and\ \citenamefont {Naert}}]{LagoinCrausteThibiergeNaert2022}%
  \BibitemOpen
  \bibfield  {author} {\bibinfo {author} {\bibfnamefont {M.}~\bibnamefont {Lagoin}}, \bibinfo {author} {\bibfnamefont {C.}~\bibnamefont {Crauste-Thibierge}},\ and\ \bibinfo {author} {\bibfnamefont {A.}~\bibnamefont {Naert}},\ }\href {https://doi.org/10.1103/physrevlett.129.120606} {\bibfield  {journal} {\bibinfo  {journal} {Physical Review Letters}\ }\textbf {\bibinfo {volume} {129}},\ \bibinfo {pages} {120606} (\bibinfo {year} {2022})}\BibitemShut {NoStop}%
\bibitem [{\citenamefont {Puglisi}\ \emph {et~al.}(2005)\citenamefont {Puglisi}, \citenamefont {Visco}, \citenamefont {Barrat}, \citenamefont {Trizac},\ and\ \citenamefont {van Wijland}}]{PuglisiEtAl2005}%
  \BibitemOpen
  \bibfield  {author} {\bibinfo {author} {\bibfnamefont {A.}~\bibnamefont {Puglisi}}, \bibinfo {author} {\bibfnamefont {P.}~\bibnamefont {Visco}}, \bibinfo {author} {\bibfnamefont {A.}~\bibnamefont {Barrat}}, \bibinfo {author} {\bibfnamefont {E.}~\bibnamefont {Trizac}},\ and\ \bibinfo {author} {\bibfnamefont {F.}~\bibnamefont {van Wijland}},\ }\href {https://doi.org/10.1103/physrevlett.95.110202} {\bibfield  {journal} {\bibinfo  {journal} {Physical Review Letters}\ }\textbf {\bibinfo {volume} {95}},\ \bibinfo {pages} {110202} (\bibinfo {year} {2005})}\BibitemShut {NoStop}%
\bibitem [{\citenamefont {Maes}(2020)}]{maes2020frenesy}%
  \BibitemOpen
  \bibfield  {author} {\bibinfo {author} {\bibfnamefont {C.}~\bibnamefont {Maes}},\ }\href@noop {} {\bibfield  {journal} {\bibinfo  {journal} {Physics Reports}\ }\textbf {\bibinfo {volume} {850}},\ \bibinfo {pages} {1} (\bibinfo {year} {2020})}\BibitemShut {NoStop}%
\bibitem [{\citenamefont {Di~Terlizzi}\ and\ \citenamefont {Baiesi}(2019)}]{di2019kinetic}%
  \BibitemOpen
  \bibfield  {author} {\bibinfo {author} {\bibfnamefont {I.}~\bibnamefont {Di~Terlizzi}}\ and\ \bibinfo {author} {\bibfnamefont {M.}~\bibnamefont {Baiesi}},\ }\href@noop {} {\bibfield  {journal} {\bibinfo  {journal} {Journal of Physics A: Mathematical and Theoretical}\ }\textbf {\bibinfo {volume} {52}},\ \bibinfo {pages} {02LT03} (\bibinfo {year} {2019})}\BibitemShut {NoStop}%
\bibitem [{\citenamefont {Van~Vu}\ and\ \citenamefont {Saito}(2023{\natexlab{b}})}]{van2023topological}%
  \BibitemOpen
  \bibfield  {author} {\bibinfo {author} {\bibfnamefont {T.}~\bibnamefont {Van~Vu}}\ and\ \bibinfo {author} {\bibfnamefont {K.}~\bibnamefont {Saito}},\ }\href@noop {} {\bibfield  {journal} {\bibinfo  {journal} {Physical review letters}\ }\textbf {\bibinfo {volume} {130}},\ \bibinfo {pages} {010402} (\bibinfo {year} {2023}{\natexlab{b}})}\BibitemShut {NoStop}%
\bibitem [{\citenamefont {Krishnamurthy}\ \emph {et~al.}(2016)\citenamefont {Krishnamurthy}, \citenamefont {Ghosh}, \citenamefont {Chatterji}, \citenamefont {Ganapathy},\ and\ \citenamefont {Sood}}]{krishnamurthy2016micrometre}%
  \BibitemOpen
  \bibfield  {author} {\bibinfo {author} {\bibfnamefont {S.}~\bibnamefont {Krishnamurthy}}, \bibinfo {author} {\bibfnamefont {S.}~\bibnamefont {Ghosh}}, \bibinfo {author} {\bibfnamefont {D.}~\bibnamefont {Chatterji}}, \bibinfo {author} {\bibfnamefont {R.}~\bibnamefont {Ganapathy}},\ and\ \bibinfo {author} {\bibfnamefont {A.}~\bibnamefont {Sood}},\ }\href@noop {} {\bibfield  {journal} {\bibinfo  {journal} {Nature Physics}\ }\textbf {\bibinfo {volume} {12}},\ \bibinfo {pages} {1134} (\bibinfo {year} {2016})}\BibitemShut {NoStop}%
\bibitem [{\citenamefont {Bebon}\ \emph {et~al.}(2025)\citenamefont {Bebon}, \citenamefont {Robinson},\ and\ \citenamefont {Speck}}]{Bebon25}%
  \BibitemOpen
  \bibfield  {author} {\bibinfo {author} {\bibfnamefont {R.}~\bibnamefont {Bebon}}, \bibinfo {author} {\bibfnamefont {J.~F.}\ \bibnamefont {Robinson}},\ and\ \bibinfo {author} {\bibfnamefont {T.}~\bibnamefont {Speck}},\ }\href {https://doi.org/10.1103/PhysRevX.15.021050} {\bibfield  {journal} {\bibinfo  {journal} {Phys. Rev. X}\ }\textbf {\bibinfo {volume} {15}},\ \bibinfo {pages} {021050} (\bibinfo {year} {2025})}\BibitemShut {NoStop}%
\bibitem [{\citenamefont {Andrae}\ \emph {et~al.}(2010)\citenamefont {Andrae}, \citenamefont {Cremer}, \citenamefont {Reichenbach},\ and\ \citenamefont {Frey}}]{Andrae2010}%
  \BibitemOpen
  \bibfield  {author} {\bibinfo {author} {\bibfnamefont {B.}~\bibnamefont {Andrae}}, \bibinfo {author} {\bibfnamefont {J.}~\bibnamefont {Cremer}}, \bibinfo {author} {\bibfnamefont {T.}~\bibnamefont {Reichenbach}},\ and\ \bibinfo {author} {\bibfnamefont {E.}~\bibnamefont {Frey}},\ }\href {https://doi.org/10.1103/PhysRevLett.104.218102} {\bibfield  {journal} {\bibinfo  {journal} {Phys. Rev. Lett.}\ }\textbf {\bibinfo {volume} {104}},\ \bibinfo {pages} {218102} (\bibinfo {year} {2010})}\BibitemShut {NoStop}%
\bibitem [{\citenamefont {Qian}(2014)}]{Qian2014}%
  \BibitemOpen
  \bibfield  {author} {\bibinfo {author} {\bibfnamefont {H.}~\bibnamefont {Qian}},\ }\href {https://doi.org/https://doi.org/10.1007/s40484-014-0028-4} {\bibfield  {journal} {\bibinfo  {journal} {Quantitative Biology}\ }\textbf {\bibinfo {volume} {2}},\ \bibinfo {pages} {47} (\bibinfo {year} {2014})},\ \Eprint {https://arxiv.org/abs/https://onlinelibrary.wiley.com/doi/pdf/10.1007/s40484-014-0028-4} {https://onlinelibrary.wiley.com/doi/pdf/10.1007/s40484-014-0028-4} \BibitemShut {NoStop}%
\bibitem [{\citenamefont {Rao}\ and\ \citenamefont {Leibler}(2022)}]{Rao22}%
  \BibitemOpen
  \bibfield  {author} {\bibinfo {author} {\bibfnamefont {R.}~\bibnamefont {Rao}}\ and\ \bibinfo {author} {\bibfnamefont {S.}~\bibnamefont {Leibler}},\ }\href {https://doi.org/10.1073/pnas.2112083119} {\bibfield  {journal} {\bibinfo  {journal} {Proceedings of the National Academy of Sciences}\ }\textbf {\bibinfo {volume} {119}},\ \bibinfo {pages} {e2112083119} (\bibinfo {year} {2022})},\ \Eprint {https://arxiv.org/abs/https://www.pnas.org/doi/pdf/10.1073/pnas.2112083119} {https://www.pnas.org/doi/pdf/10.1073/pnas.2112083119} \BibitemShut {NoStop}%
\bibitem [{\citenamefont {Tom\'e}\ \emph {et~al.}(2023)\citenamefont {Tom\'e}, \citenamefont {Fiore},\ and\ \citenamefont {de~Oliveira}}]{Tome23}%
  \BibitemOpen
  \bibfield  {author} {\bibinfo {author} {\bibfnamefont {T.}~\bibnamefont {Tom\'e}}, \bibinfo {author} {\bibfnamefont {C.~E.}\ \bibnamefont {Fiore}},\ and\ \bibinfo {author} {\bibfnamefont {M.~J.}\ \bibnamefont {de~Oliveira}},\ }\href {https://doi.org/10.1103/PhysRevE.107.064135} {\bibfield  {journal} {\bibinfo  {journal} {Phys. Rev. E}\ }\textbf {\bibinfo {volume} {107}},\ \bibinfo {pages} {064135} (\bibinfo {year} {2023})}\BibitemShut {NoStop}%
\bibitem [{\citenamefont {Hawthorne}\ \emph {et~al.}(2023)\citenamefont {Hawthorne}, \citenamefont {Harunari}, \citenamefont {de~Oliveira},\ and\ \citenamefont {Fiore}}]{Hawthorne23}%
  \BibitemOpen
  \bibfield  {author} {\bibinfo {author} {\bibfnamefont {F.}~\bibnamefont {Hawthorne}}, \bibinfo {author} {\bibfnamefont {P.~E.}\ \bibnamefont {Harunari}}, \bibinfo {author} {\bibfnamefont {M.~J.}\ \bibnamefont {de~Oliveira}},\ and\ \bibinfo {author} {\bibfnamefont {C.~E.}\ \bibnamefont {Fiore}},\ }\bibfield  {journal} {\bibinfo  {journal} {Entropy}\ }\textbf {\bibinfo {volume} {25}},\ \href {https://doi.org/10.3390/e25081230} {10.3390/e25081230} (\bibinfo {year} {2023})\BibitemShut {NoStop}%
\bibitem [{\citenamefont {Oliveira}\ \emph {et~al.}(2024)\citenamefont {Oliveira}, \citenamefont {Wang}, \citenamefont {Dong}, \citenamefont {Du}, \citenamefont {Fiore}, \citenamefont {Vilela},\ and\ \citenamefont {Stanley}}]{OLIVEIRA2024114694}%
  \BibitemOpen
  \bibfield  {author} {\bibinfo {author} {\bibfnamefont {I.~V.}\ \bibnamefont {Oliveira}}, \bibinfo {author} {\bibfnamefont {C.}~\bibnamefont {Wang}}, \bibinfo {author} {\bibfnamefont {G.}~\bibnamefont {Dong}}, \bibinfo {author} {\bibfnamefont {R.}~\bibnamefont {Du}}, \bibinfo {author} {\bibfnamefont {C.~E.}\ \bibnamefont {Fiore}}, \bibinfo {author} {\bibfnamefont {A.~L.}\ \bibnamefont {Vilela}},\ and\ \bibinfo {author} {\bibfnamefont {H.~E.}\ \bibnamefont {Stanley}},\ }\href {https://doi.org/https://doi.org/10.1016/j.chaos.2024.114694} {\bibfield  {journal} {\bibinfo  {journal} {Chaos, Solitons \& Fractals}\ }\textbf {\bibinfo {volume} {181}},\ \bibinfo {pages} {114694} (\bibinfo {year} {2024})}\BibitemShut {NoStop}%
\bibitem [{\citenamefont {Irisarri}\ \emph {et~al.}(2025)\citenamefont {Irisarri}, \citenamefont {Trigal}, \citenamefont {Toral},\ and\ \citenamefont {Manzano}}]{irisarri2025stochasticthermodynamicssocialimitation}%
  \BibitemOpen
  \bibfield  {author} {\bibinfo {author} {\bibfnamefont {L.}~\bibnamefont {Irisarri}}, \bibinfo {author} {\bibfnamefont {L.}~\bibnamefont {Trigal}}, \bibinfo {author} {\bibfnamefont {R.}~\bibnamefont {Toral}},\ and\ \bibinfo {author} {\bibfnamefont {G.}~\bibnamefont {Manzano}},\ }\href {https://arxiv.org/abs/2511.14006} {\bibinfo {title} {Stochastic thermodynamics of social imitation beyond energetics}} (\bibinfo {year} {2025}),\ \Eprint {https://arxiv.org/abs/2511.14006} {arXiv:2511.14006 [physics.soc-ph]} \BibitemShut {NoStop}%
\bibitem [{\citenamefont {Zanin}\ \emph {et~al.}(2020)\citenamefont {Zanin}, \citenamefont {Güntekin}, \citenamefont {Aktürk}, \citenamefont {Hanoğlu},\ and\ \citenamefont {Papo}}]{Zanin2020}%
  \BibitemOpen
  \bibfield  {author} {\bibinfo {author} {\bibfnamefont {M.}~\bibnamefont {Zanin}}, \bibinfo {author} {\bibfnamefont {B.}~\bibnamefont {Güntekin}}, \bibinfo {author} {\bibfnamefont {T.}~\bibnamefont {Aktürk}}, \bibinfo {author} {\bibfnamefont {L.}~\bibnamefont {Hanoğlu}},\ and\ \bibinfo {author} {\bibfnamefont {D.}~\bibnamefont {Papo}},\ }\bibfield  {journal} {\bibinfo  {journal} {Frontiers in Physiology}\ }\textbf {\bibinfo {volume} {Volume 10 - 2019}},\ \href {https://doi.org/10.3389/fphys.2019.01619} {10.3389/fphys.2019.01619} (\bibinfo {year} {2020})\BibitemShut {NoStop}%
\bibitem [{\citenamefont {Maldonado}\ and\ \citenamefont {Merino-Negrete}(2024)}]{maldonado2024irreversibility}%
  \BibitemOpen
  \bibfield  {author} {\bibinfo {author} {\bibfnamefont {C.}~\bibnamefont {Maldonado}}\ and\ \bibinfo {author} {\bibfnamefont {N.}~\bibnamefont {Merino-Negrete}},\ }\href@noop {} {\bibfield  {journal} {\bibinfo  {journal} {Physica A: Statistical Mechanics and its Applications}\ }\textbf {\bibinfo {volume} {637}},\ \bibinfo {pages} {129584} (\bibinfo {year} {2024})}\BibitemShut {NoStop}%
\bibitem [{\citenamefont {Holehouse}(2026)}]{holehouse2026quantifying}%
  \BibitemOpen
  \bibfield  {author} {\bibinfo {author} {\bibfnamefont {J.}~\bibnamefont {Holehouse}},\ }\href@noop {} {\bibfield  {journal} {\bibinfo  {journal} {npj Complexity}\ }\textbf {\bibinfo {volume} {3}},\ \bibinfo {pages} {10} (\bibinfo {year} {2026})}\BibitemShut {NoStop}%
\bibitem [{\citenamefont {Tononi}\ \emph {et~al.}(1994)\citenamefont {Tononi}, \citenamefont {Sporns},\ and\ \citenamefont {Edelman}}]{tononi1994measure}%
  \BibitemOpen
  \bibfield  {author} {\bibinfo {author} {\bibfnamefont {G.}~\bibnamefont {Tononi}}, \bibinfo {author} {\bibfnamefont {O.}~\bibnamefont {Sporns}},\ and\ \bibinfo {author} {\bibfnamefont {G.~M.}\ \bibnamefont {Edelman}},\ }\href@noop {} {\bibfield  {journal} {\bibinfo  {journal} {Proceedings of the National Academy of Sciences}\ }\textbf {\bibinfo {volume} {91}},\ \bibinfo {pages} {5033} (\bibinfo {year} {1994})}\BibitemShut {NoStop}%
\bibitem [{\citenamefont {Castelló}\ \emph {et~al.}(2006)\citenamefont {Castelló}, \citenamefont {Eguíluz},\ and\ \citenamefont {Miguel}}]{Castello2006}%
  \BibitemOpen
  \bibfield  {author} {\bibinfo {author} {\bibfnamefont {X.}~\bibnamefont {Castelló}}, \bibinfo {author} {\bibfnamefont {V.~M.}\ \bibnamefont {Eguíluz}},\ and\ \bibinfo {author} {\bibfnamefont {M.~S.}\ \bibnamefont {Miguel}},\ }\href {https://doi.org/10.1088/1367-2630/8/12/308} {\bibfield  {journal} {\bibinfo  {journal} {New Journal of Physics}\ }\textbf {\bibinfo {volume} {8}},\ \bibinfo {pages} {308} (\bibinfo {year} {2006})}\BibitemShut {NoStop}%
\bibitem [{\citenamefont {Nowak}\ \emph {et~al.}(2022)\citenamefont {Nowak}, \citenamefont {Grabisch},\ and\ \citenamefont {Sznajd-Weron}}]{Nowak22}%
  \BibitemOpen
  \bibfield  {author} {\bibinfo {author} {\bibfnamefont {B.}~\bibnamefont {Nowak}}, \bibinfo {author} {\bibfnamefont {M.}~\bibnamefont {Grabisch}},\ and\ \bibinfo {author} {\bibfnamefont {K.}~\bibnamefont {Sznajd-Weron}},\ }\href {https://doi.org/10.1103/PhysRevE.105.054314} {\bibfield  {journal} {\bibinfo  {journal} {Phys. Rev. E}\ }\textbf {\bibinfo {volume} {105}},\ \bibinfo {pages} {054314} (\bibinfo {year} {2022})}\BibitemShut {NoStop}%
\bibitem [{\citenamefont {Siedlecki}\ \emph {et~al.}(2016)\citenamefont {Siedlecki}, \citenamefont {Szwabi{\'n}ski},\ and\ \citenamefont {Weron}}]{siedlecki2016interplay}%
  \BibitemOpen
  \bibfield  {author} {\bibinfo {author} {\bibfnamefont {P.}~\bibnamefont {Siedlecki}}, \bibinfo {author} {\bibfnamefont {J.}~\bibnamefont {Szwabi{\'n}ski}},\ and\ \bibinfo {author} {\bibfnamefont {T.}~\bibnamefont {Weron}},\ }\href@noop {} {\bibfield  {journal} {\bibinfo  {journal} {arXiv preprint arXiv:1603.07556}\ } (\bibinfo {year} {2016})}\BibitemShut {NoStop}%
\bibitem [{\citenamefont {Kirman}(1993)}]{Kirman93}%
  \BibitemOpen
  \bibfield  {author} {\bibinfo {author} {\bibfnamefont {A.}~\bibnamefont {Kirman}},\ }\href {https://doi.org/10.2307/2118498} {\bibfield  {journal} {\bibinfo  {journal} {The Quarterly Journal of Economics}\ }\textbf {\bibinfo {volume} {108}},\ \bibinfo {pages} {137} (\bibinfo {year} {1993})},\ \Eprint {https://arxiv.org/abs/https://academic.oup.com/qje/article-pdf/108/1/137/5361523/108-1-137.pdf} {https://academic.oup.com/qje/article-pdf/108/1/137/5361523/108-1-137.pdf} \BibitemShut {NoStop}%
\bibitem [{\citenamefont {Liu}\ \emph {et~al.}(2026)\citenamefont {Liu}, \citenamefont {Datta},\ and\ \citenamefont {Barato}}]{liu2026cyclic}%
  \BibitemOpen
  \bibfield  {author} {\bibinfo {author} {\bibfnamefont {S.}~\bibnamefont {Liu}}, \bibinfo {author} {\bibfnamefont {A.}~\bibnamefont {Datta}},\ and\ \bibinfo {author} {\bibfnamefont {A.}~\bibnamefont {Barato}},\ }\href@noop {} {\bibfield  {journal} {\bibinfo  {journal} {Journal of Physics A: Mathematical and Theoretical}\ }\textbf {\bibinfo {volume} {59}},\ \bibinfo {pages} {105005} (\bibinfo {year} {2026})}\BibitemShut {NoStop}%
\bibitem [{\citenamefont {Khodabandehlou}\ \emph {et~al.}(2026)\citenamefont {Khodabandehlou}, \citenamefont {Maes},\ and\ \citenamefont {Rold{\'a}n}}]{khodabandehlou2026bringing}%
  \BibitemOpen
  \bibfield  {author} {\bibinfo {author} {\bibfnamefont {F.}~\bibnamefont {Khodabandehlou}}, \bibinfo {author} {\bibfnamefont {C.}~\bibnamefont {Maes}},\ and\ \bibinfo {author} {\bibfnamefont {{\'E}.}~\bibnamefont {Rold{\'a}n}},\ }\href@noop {} {\bibfield  {journal} {\bibinfo  {journal} {arXiv preprint arXiv:2602.15127}\ } (\bibinfo {year} {2026})}\BibitemShut {NoStop}%
\bibitem [{\citenamefont {Falasco}\ and\ \citenamefont {Esposito}(2025)}]{Falasco25}%
  \BibitemOpen
  \bibfield  {author} {\bibinfo {author} {\bibfnamefont {G.}~\bibnamefont {Falasco}}\ and\ \bibinfo {author} {\bibfnamefont {M.}~\bibnamefont {Esposito}},\ }\href {https://doi.org/10.1103/RevModPhys.97.015002} {\bibfield  {journal} {\bibinfo  {journal} {Rev. Mod. Phys.}\ }\textbf {\bibinfo {volume} {97}},\ \bibinfo {pages} {015002} (\bibinfo {year} {2025})}\BibitemShut {NoStop}%
\end{thebibliography}%
\end{document}